\documentclass[twocolumn]{aastex62}
\usepackage[utf8]{inputenc}
\DeclareUnicodeCharacter{2212}{-}% support older LaTeX versions
\usepackage{graphicx}
\graphicspath{{./}{Figures}}
\usepackage{amsmath}
\usepackage{afterpage}
\usepackage{enumitem}
\usepackage{xcolor}
\usepackage{float}
\usepackage{xspace} 
\usepackage{url}
\usepackage{fontawesome}
\setenumerate{itemsep=0mm}%noitemsep}
\usepackage{wrapfig}
\usepackage{placeins}

\usepackage{amsmath}
\usepackage{amssymb}
\usepackage{xspace}
\usepackage{xifthen}
\usepackage{eso-pic}

\newcommand{\Gaia}{{\it Gaia}\xspace}

\definecolor{forestgreen}{HTML}{228B22}
\definecolor{urlblue}{HTML}{000000}

\mathchardef\mhyphen="2D

\newlength{\dhatheight}

\newcommand{\code}[1]{\texttt{#1}\xspace}

\newcommand{\unit}[1]{\ensuremath{\mathrm{\,#1}}\xspace}

\newcommand{\degree}{\ensuremath{{}^{\circ}}\xspace}

\newcommand{\kpc}{\unit{kpc}}

\newcommand{\Msun}{\unit{M_\odot}}

\newcommand{\Lsun}{\unit{L_\odot}}

\newcommand{\secref}[1]{Section~\ref{sec:#1}}

\newcommand{\tabref}[1]{Table~\ref{tab:#1}}

\newcommand{\figref}[1]{Figure~\ref{fig:#1}}

\newcommand{\bandvar}[2][]{%
  \ifthenelse{\isempty{#1}}{\var{#2}}{\var{#2\_#1}}%
}

\newcommand{\ra}{{\ensuremath{\alpha_{2000}}}\xspace}
\newcommand{\dec}{{\ensuremath{\delta_{2000}}}\xspace}

\newcommand{\PA}{\ensuremath{\mathrm{P.A.}}\xspace}

\newcommand{\HEALPix}{\code{HEALPix}}
\newcommand{\healpix}{\HEALPix}

\newcommand{\emcee}{\code{emcee}}
\newcommand{\ugali}{\code{ugali}}

\newcommand{\var}[1]{\ensuremath{\texttt{\MakeUppercase{#1}}}\xspace}

\providecommand\physrep{\ref@jnl{Phys.~Rep.}}%
\providecommand\apjs{\ref@jnl{ApJS}}%
\providecommand{\jcap}{\ref@jnl{JCAP}}%
\reportnum{\footnotesize FERMILAB-PUB-24-0359-LDRD-PPD}

\shorttitle{Discovery and Spectroscopic Confirmation of Aquarius~III}

\shortauthors{DELVE Collaboration}

\begin{document}

\reportnum{\footnotesize}

\title{Discovery and Spectroscopic Confirmation of Aquarius~III: A Low-Mass Milky Way Satellite Galaxy}

\correspondingauthor{William Cerny}
\email{william.cerny@yale.edu}

\author[0000-0003-1697-7062]{W.~Cerny}
\affiliation{Department of Astronomy, Yale University, New Haven, CT 06520, USA}

\author[0000-0002-7155-679X]{A.~Chiti}
\affiliation{Department of Astronomy and Astrophysics, University of Chicago, Chicago, IL 60637, USA}
\affiliation{Kavli Institute for Cosmological Physics, University of Chicago, Chicago, IL 60637, USA}

\author[0000-0002-7007-9725]{M.~Geha}
\affiliation{Department of Astronomy, Yale University, New Haven, CT 06520, USA}

\author[0000-0001-9649-4815]{B.~Mutlu-Pakdil}
\affiliation{Department of Physics and Astronomy, Dartmouth College, Hanover, NH 03755, USA}

\author[0000-0001-8251-933X]{A.~Drlica-Wagner}
\affiliation{Fermi National Accelerator Laboratory, P.O.\ Box 500, Batavia, IL 60510, USA}
\affiliation{Kavli Institute for Cosmological Physics, University of Chicago, Chicago, IL 60637, USA}
\affiliation{Department of Astronomy and Astrophysics, University of Chicago, Chicago, IL 60637, USA}

\author[0000-0003-0478-0473]{C.~Y.~Tan}
\affiliation{Department of Physics, University of Chicago, Chicago, IL 60637, USA}
\affiliation{Kavli Institute for Cosmological Physics, University of Chicago, Chicago, IL 60637, USA}

\author[0000-0002-6904-359X]{M.~Adam\'ow}	
\affiliation{Center for Astrophysical Surveys, National Center for Supercomputing Applications, 1205 West Clark St., Urbana, IL 61801, USA}

 \author[0000-0002-6021-8760]{A.~B.~Pace}
\affiliation{McWilliams Center for Cosmology, Carnegie Mellon University, 5000 Forbes Ave, Pittsburgh, PA 15213, USA}
\affiliation{Department of Astronomy, University of Virginia, 530 McCormick Road, Charlottesville, VA 22904 USA}
\author[0000-0002-4733-4994]{J.~D.~Simon}
\affiliation{Observatories of the Carnegie Institution for Science, 813 Santa Barbara St., Pasadena, CA 91101, USA}		

\author[0000-0003-4102-380X]{D.~J.~Sand}
\affiliation{Department of Astronomy/Steward Observatory, 933 North Cherry Avenue, Room N204, Tucson, AZ 85721-0065, USA}

\author[0000-0002-4863-8842]{A.~P.~Ji}
\affiliation{Department of Astronomy and Astrophysics, University of Chicago, Chicago, IL 60637, USA}
\affiliation{Kavli Institute for Cosmological Physics, University of Chicago, Chicago, IL 60637, USA}

\author[0000-0002-9110-6163]{T.~S.~Li}
\affiliation{Department of Astronomy and Astrophysics, University of Toronto, 50 St. George Street, Toronto ON, M5S 3H4, Canada}

\author[0000-0003-4341-6172]{A.~K.~Vivas}
\affiliation{Cerro Tololo Inter-American Observatory/NSF NOIRLab, Casilla 603, La Serena, Chile}

\author[0000-0002-5564-9873]{E.~F.~Bell}
\affiliation{Department of Astronomy, University of Michigan, Ann Arbor,
MI 48109, USA}

\author[0000-0002-3936-9628]{J.~L.~Carlin}
\affiliation{Rubin Observatory/AURA, 950 North Cherry Avenue, Tucson, AZ, 85719, USA}

\author[0000-0002-3690-105X]{J.~A.~Carballo-Bello}
\affiliation{Instituto de Alta Investigaci\'on, Universidad de Tarapac\'a, Casilla 7D, Arica, Chile}

\author[0000-0001-5143-1255]{A.~Chaturvedi}
\affiliation{Department of Physics, University of Surrey, Guildford GU2 7XH, UK}

\author[0000-0003-1680-1884]{Y.~Choi}
\affiliation{NSF NOIRLab, 950 N. Cherry Ave., Tucson, AZ 85719, USA}

\author[0000-0001-9775-9029]{A.~Doliva-Dolinsky}
\affiliation{Department of Chemistry and Physics, University of Tampa, 401 West Kennedy Boulevard, Tampa, FL 33606, USA}
\affiliation{Department of Physics and Astronomy, Dartmouth College, Hanover, NH 03755, USA}

\author[0000-0001-9852-9954]{O.~Y.~Gnedin}
\affiliation{Department of Astronomy, University of Michigan, Ann Arbor,
MI 48109, USA}

\author[0000-0002-9269-8287]{G.~Limberg}
\affiliation{Department of Astronomy \& Astrophysics, University of Chicago, 5640 S Ellis Avenue, Chicago, IL 60637, USA}
\affiliation{Kavli Institute for Cosmological Physics, University of Chicago, Chicago, IL 60637, USA}
\affiliation{Universidade de S\~ao Paulo, IAG, Departamento de Astronomia, SP 05508-090, S\~ao Paulo, Brazil}

\author[0000-0002-9144-7726]{C.~E.~Mart\'inez-V\'azquez}
\affiliation{International Gemini Observatory/NSF NOIRLab, 670 N. A'ohoku Place, Hilo, Hawai'i, 96720, USA}

\author[0000-0003-3519-4004]{S.~Mau}
\affiliation{Department of Physics, Stanford University, 382 Via Pueblo Mall, Stanford, CA 94305, USA}
\affiliation{Kavli Institute for Particle Astrophysics \& Cosmology, P.O.\ Box 2450, Stanford University, Stanford, CA 94305, USA}

\author[0000-0003-0105-9576]{G.~E.~Medina}
\affiliation{Department of Astronomy and Astrophysics, University of Toronto, 50 St. George Street, Toronto ON, M5S 3H4, Canada}

\author[0000-0001-9438-5228]{M.~Navabi}
\affiliation{Department of Physics, University of Surrey, Guildford GU2 7XH, UK}

\author[0000-0002-8282-469X]{N.~E.~D.~No\"el}
\affiliation{Department of Physics, University of Surrey, Guildford GU2 7XH, UK}

\author[0000-0003-4479-1265]{V.~M.~Placco}
\affiliation{NSF NOIRLab, 950 N. Cherry Ave., Tucson, AZ 85719, USA}

\author[0000-0001-5805-5766]{A.~H.~Riley}
\affiliation{Institute for Computational Cosmology, Department of Physics, Durham University, South Road, Durham DH1 3LE, UK}

\author[0000-0001-5107-8930]{Ian U.\ Roederer}
\affiliation{Department of Physics, North Carolina State University,
Raleigh, NC 27695, USA}

\author[0000-0003-1479-3059]{G.~S.~Stringfellow}
\affiliation{Center for Astrophysics and Space Astronomy, University of Colorado, 389 UCB, Boulder, CO 80309-0389, USA}

\author[0000-0003-4383-2969]{C.~R.~Bom}
\affiliation{Centro Brasileiro de Pesquisas F\'isicas, Rua Dr. Xavier Sigaud 150, 22290-180 Rio de Janeiro, RJ, Brazil}

\author[0000-0001-6957-1627]{P.~S.~Ferguson}
\affiliation{Department of Physics, University of Wisconsin-Madison, Madison, WI 53706, USA}

\author[0000-0001-5160-4486]{D.~J.~James}
\affiliation{ASTRAVEO LLC, PO Box 1668, Gloucester, MA 01931}
\affiliation{Applied Materials Inc., 35 Dory Road, Gloucester, MA 01930}

\author{D.~Mart\'{i}nez-Delgado}
\affiliation{Centro de Estudios de F\'isica del Cosmos de Arag\'on (CEFCA), Unidad 
Asociada al CSIC,
Plaza San Juan 1, 44001 Teruel, Spain}
\affiliation{ARAID Foundation, Avda. de Ranillas, 1-D, E-50018 Zaragoza, Spain}

\author[0000-0002-8093-7471]{P.~Massana}
\affiliation{NSF NOIRLab, Casilla 603, La Serena, Chile}

\author[0000-0002-1793-3689
]{D.~L.~Nidever}
\affiliation{Department of Physics, Montana State University, P.O. Box 173840, Bozeman, MT 59717-3840; NSF's National Optical-Infrared Astronomy Research Laboratory, 950 N. Cherry Ave., Tucson, AZ 85719, USA}
\affiliation{NSF NOIRLab, 950 N. Cherry Ave., Tucson, AZ 85719, USA}

\author[0000-0002-1594-1466]{J.~D.~Sakowska}
\affiliation{Department of Physics, University of Surrey, Guildford GU2 7XH, UK}

\author[0000-0003-3402-6164]{L.~Santana-Silva}
\affiliation{Centro Brasileiro de Pesquisas F\'isicas, Rua Dr. Xavier Sigaud 150, 22290-180 Rio de Janeiro, RJ, Brazil}

\author[0000-0001-5399-0114]{N.~F.~Sherman}
\affiliation{Institute for Astrophysical Research, Boston University, 725 Commonwealth Avenue,Boston, MA 02215, USA}

\author[0000-0002-9599-310X]{E.~J.~Tollerud}
\affiliation{Space Telescope Science Institute, 3700 San Martin Drive, Baltimore, MD 21218, USA}

\collaboration{(DELVE Collaboration)}

\begin{abstract}
We present the discovery of Aquarius~III, an ultra-faint Milky Way satellite galaxy identified in the second data release of the DECam Local Volume Exploration (DELVE) survey. Based on deeper follow-up imaging with DECam,  we find that Aquarius~III is a low-luminosity ($M_V = -2.5^{+0.3}_{-0.5}$; $L_V = 850^{+380}_{-260} \ L_{\odot}$), extended ($r_{1/2} = 41^{+9}_{-8}$~pc) stellar system located in the outer halo ($D_{\odot} = 85 \pm 4$~kpc). From medium-resolution Keck/DEIMOS  spectroscopy, we identify 11 member stars and measure a mean heliocentric radial velocity of $v_{\rm sys} = -13.1^{+1.0}_{-0.9} \ \rm km \ s^{-1}$ for the system and place an upper limit of $\sigma_v < 3.5 \rm \ km \ s^{-1}$ ($\sigma_v < 1.6 \rm \ km \ s^{-1}$) on its velocity dispersion at the 95\% (68\%)  credible level. Based on Calcium-Triplet-based metallicities of the six brightest red giant members, we find that Aquarius~III is very metal-poor ([Fe/H]$ = -2.61 \pm 0.21$) with a statistically-significant metallicity spread ($\sigma_{\rm [Fe/H]} = 0.46^{+0.26}_{-0.14}$ dex). We interpret this metallicity spread as strong evidence that the system is a dwarf galaxy as opposed to a star cluster. Combining our velocity measurement with \Gaia proper motions, we find that Aquarius~III is currently situated near its orbital pericenter in the outer halo ($r_{\rm peri} = 78 \pm 7$ kpc) and that it is plausibly on first infall onto the Milky Way. This orbital history likely precludes significant tidal disruption from the Galactic disk, notably unlike other satellites with comparably low velocity dispersion limits in the literature. Thus, if further velocity measurements confirm that its velocity dispersion is truly below $\sigma_v \lesssim 2 \rm \ km \ s^{-1}$, Aquarius~III may serve as a useful laboratory for probing galaxy formation physics in low-mass halos.
\end{abstract}

\keywords{dwarf galaxies; Local Group; photometric surveys}

%-------------------------------------------------------------------------------

\section{Introduction}
\label{sec:intro}
% \vdispupperlimit
Cosmological simulations of structure formation within the $\Lambda$ Cold Dark Matter ($\Lambda$CDM) paradigm strongly suggest that galaxy formation is hierarchical, with massive galaxies and their host dark matter halos forming from the continuous merging and accretion of relatively lower-mass systems \citep[e.g.,][]{1974ApJ...187..425P,1978MNRAS.183..341W,1991ApJ...379...52W}. A direct consequence of this bottom-up assembly process is the abundance of substructures around $L_*$ galaxies: galaxies like the Milky Way are expected to be surrounded by scores of accreted subhalos hosting low-mass dwarf galaxies in addition to many orders of magnitude more ``dark,'' very-low-mass subhalos with no luminous counterparts \citep[e.g.,][]{1999ApJ...524L..19M,2008MNRAS.391.1685S,2016ApJ...818...10G}. The density profiles, mass function, and radial distribution of these subhalos are sensitive probes of the nature of dark matter and its power spectrum on small scales \citep[see][for a review]{2017ARA&A..55..343B}. However, these subhalo properties are challenging to study directly in the absence of a luminous tracer. Thus, the low-mass satellite galaxies inhabiting these small-scale halos have long played a special role as  observationally-accessible windows into the elusive substructures surrounding the Milky Way and other nearby host galaxies.

\par At the turn of the 21st century, just eleven satellite galaxies of the Milky Way were known, raising concerns about the consistency of the observed satellite galaxy population with the subhalo population predicted by $\Lambda$CDM simulations \citep{10.1093/mnras/264.1.201,1999ApJ...522...82K, 1999ApJ...524L..19M}. The advent of the Sloan Digital Sky Survey (SDSS) in the early 2000s quickly changed the landscape, however, with the discovery of the first ``ultra-faint'' stellar systems: exceedingly faint, low-mass satellites beyond the detection limits of prior photographic surveys \citep[e.g.,][]{2005AJ....129.2692W,2005ApJ...626L..85W,2006ApJ...647L.111B,2006ApJ...650L..41Z}. Soon after, concerted efforts to spectroscopically characterize the internal velocity and metallicity distributions of these newly-discovered systems robustly established their nature as the most dark-matter-dominated, least chemically-enriched dwarf galaxies in the universe \citep{2005ApJ...630L.141K, 2006ApJ...650L..51M,2007ApJ...670..313S,2007MNRAS.380..281M}. In so doing, these studies pointed toward a reconciliation of the completeness-corrected satellite counts with the substructure predicted by $\Lambda$CDM \citep[e.g.,][]{2009ApJ...696.2179K,10.1111/j.1365-2966.2009.16031.x}.

\par Since the early years of SDSS, a succession of sensitive wide-field surveys has continued to drive the rapid discovery of ultra-faint dwarf galaxies in the Galactic halo \citep[e.g.,][]{Bechtol:2015,2015ApJ...813..109D,2015ApJ...805..130K,2015ApJ...808L..39K,2018MNRAS.479.5343K,2018MNRAS.475.5085T,2019PASJ...71...94H, 2022arXiv220912422C, 2023arXiv231106037G, 2023AJ....166...76S, 2023arXiv231110147S,2023arXiv231105439H} and throughout the Local Volume \citep[e.g.,][for recent examples]{2022MNRAS.509...16M,2022ApJ...935L..17S,2022ApJ...926...77M,PegasusW,2023arXiv230708738M,2023ApJ...957L...5J}. 
Paired with advances in numerical simulations \citep[e.g.,][]{2015MNRAS.453.1305W,2016ApJ...831..204R,2019ApJ...874...40M,2021ApJ...906...96A} and semi-analytical modelling \citep[e.g.,][]{2022MNRAS.516.3944M, 2023arXiv230813599A,2023ApJ...948...87W}, the large statistical sample of satellites built by these search efforts has not only largely alleviated concerns of tension with $\Lambda$CDM (\citealt{2017ARA&A..55..343B,2018PhRvL.121u1302K, 2020ApJ...893...47D,2022NatAs...6..897S}) but also has enabled wide-ranging and detailed tests of galaxy formation physics and the nature of dark matter \citep[e.g.,][]{2018MNRAS.473.2060J,2020ApJ...893...48N,2021PhRvL.126i1101N,2021JCAP...08..062N,2022ApJ...932..128M,2023arXiv230604674E}. Nonetheless, fully leveraging the constraining power of the ultra-faint dwarfs as physical laboratories will require a \textit{complete} census of these systems in the local universe as well as a complete accounting of their dynamical masses and chemistries through follow-up spectroscopy \citep[e.g.,][]{2019BAAS...51c.409S,2024arXiv240110318N}. Thus, the continued discovery and characterization of these extreme galaxies remains a central focus of ``near-field cosmology.''

\begin{figure*}
    \centering
    \includegraphics[width = .95\textwidth]{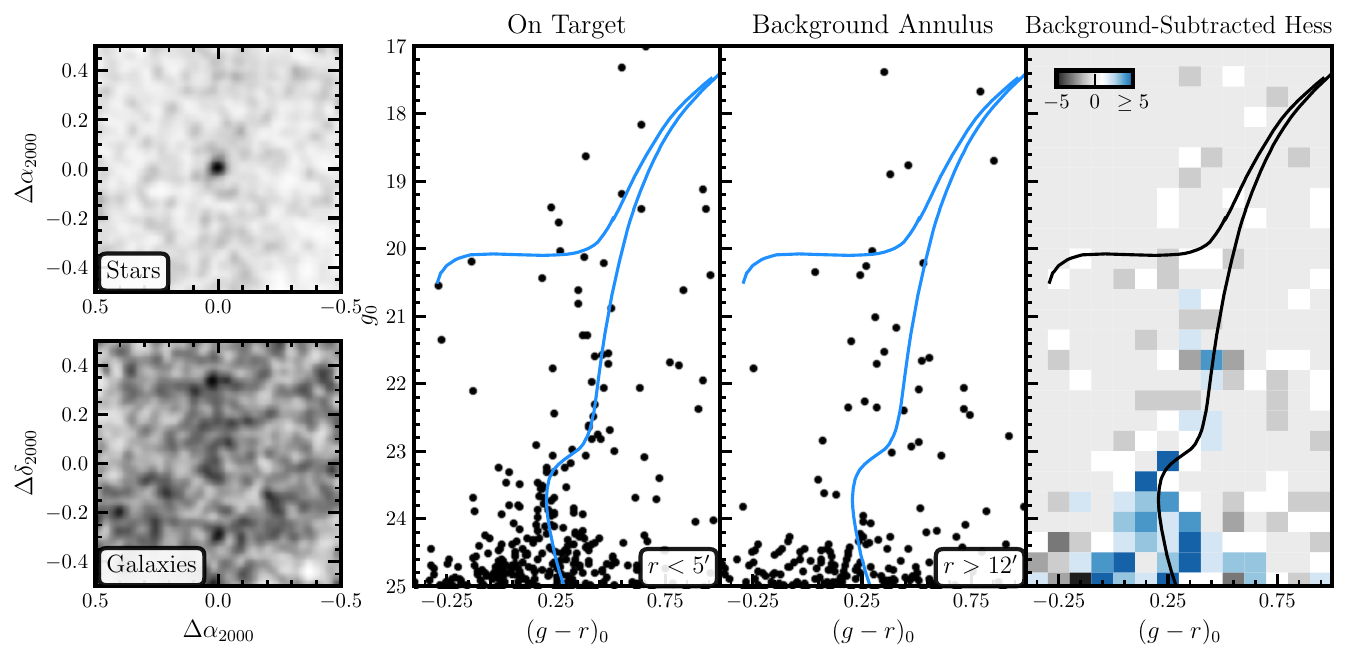}
    \caption{Diagnostic views of Aquarius~III based on the deeper DECam photometric catalog described in \secref{decamfollowup}. (Far left) Isochrone-filtered spatial distribution of stars (top subpanel) and galaxies (bottom subpanel) in a $1.0\degree \times \ 1.0\degree$ region centered on Aquarius~III. In both of these panels, a magnitude cut of $g_0 < 25$ has been applied and the (normalized) filtered density field has been smoothed with a Gaussian kernel. (Center Panels) Color--magnitude diagrams of stars within a $r = 5 \arcmin$ region (center-left) and in concentric, equal-area background annulus with inner radius $r = 12\arcmin$ (center-right). An old, metal-poor stellar isochrone with $\tau = 13.5$~Gyr, $Z=0.0001$ is shown as a solid blue curve. (Far right) Hess diagram constructed by subtracting the binned background CMD from the binned on-target CMD. This diagram clearly reveals the lower RGB, SGB, and MSTO of Aquarius~III while also making the paucity of brighter upper RGB stars apparent.}
    \label{fig:diagnostic}
\end{figure*}

\par Toward this broader goal, here we present the discovery of DELVE J2348$-$0329 (Aquarius~III), the latest entrant in the ongoing community census of ultra-faint Milky Way satellites with wide-field imaging surveys. In \secref{disco}, we describe our discovery of the satellite and our subsequent deeper follow-up imaging of the system with the 4m Blanco telescope / Dark Energy Camera. In \secref{prop}, we use this deeper imaging to characterize the satellite's morphology and stellar population.  Then, in \secref{dynamics}, we present Keck II/DEIMOS multi-object spectroscopy of resolved stars in the system from which we measure its intrinsic velocity and metallicity distribution. We then describe our exploration of its brightest stars' chemical abundances based on a Magellan/MagE longslit spectrum in \secref{mage}. Lastly, we discuss the implications of our measurements for the system's classification, orbital history, and dark matter halo mass in \secref{disc},  and we conclude in \secref{conc}.
\par Given our eventual determination that DELVE J2348$-$0329 is an ultra-faint dwarf galaxy as opposed to a star cluster, we follow the historical naming convention for Local Group satellite galaxies and hereafter refer to the system as Aquarius~III based on the constellation within which it resides.

\section{Discovery and Follow-up Imaging}
\label{sec:disco}
\subsection{Identification in DELVE DR2}
\par The DECam Local Volume Exploration survey (DELVE; \citealt{2021ApJS..256....2D}) is an ongoing campaign to uncover and characterize the satellite populations of the Milky Way, Magellanic Clouds, and several Local Volume hosts with the Dark Energy Camera (DECam; \citealt{2015AJ....150..150F}) on the 4m Blanco Telescope at Cerro Tololo Inter-American Observatory, Chile. Toward this goal, DELVE has assembled near-contiguous imaging of the southern celestial sky through a combination of 150+ nights of new DECam observations and an extensive reprocessing of archival community observations on the same instrument. The survey's most recent public data release, DELVE DR2 \citep{2022ApJS..261...38D}, includes more than $21,000\deg^2$ of high-Galactic-latitude sky coverage  including $\sim 17,000\deg^2$ of overlapping coverage in each of the $g,r,i,z$ bands.

\par In \citet{2023ApJ...953....1C}, we presented the primary results of our searches for ultra-faint Milky Way satellites over this wide-area dataset. These searches relied on the \texttt{simple} algorithm\footnote{\url{https://github.com/DarkEnergySurvey/simple}}, which applies a isochrone matched-filter approach in color--magnitude space (based on the algorithm described in \citealt{Bechtol:2015}) to identify overdensities of stars consistent with an old, metal-poor stellar population. From the thousands of candidates identified above our nominal significance threshold of $>5.5\sigma$, we identified a sample of seven especially promising candidates based on both diagnostic plots and on visual inspection of color images from the Legacy Surveys Data Releases 9 and 10 \citep{2019AJ....157..168D}. We presented our discovery and characterization of six of these systems in the aforementioned work but refrained from presenting the seventh candidate -- Aquarius~III -- as the available data were too shallow to robustly confirm its status as a real Milky Way satellite and to characterize its properties. Deeper follow-up imaging has since been obtained and offers clear confirmation of Aquarius~III as a \textit{bona-fide} new satellite, as we describe in the subsections below. We note that Aquarius~III is the last satellite we anticipate reporting based on searches over DELVE DR2's WIDE component. Ongoing and future searches will focus on the deeper, more homogeneous dataset provided by the upcoming DELVE DR3 (see \citealt{2024arXiv240800865T} for early results).

\subsection{Deeper Imaging with DECam}
\label{sec:decamfollowup}
We obtained dedicated follow-up exposures of Aquarius~III with DECam on the nights of 2022 July 28, 31 and 2023 July 14.  On each of the former two nights, we collected $\rm3\times300$s dithered $r$-band exposures, while on the latter night we collected $\rm3\times300$s $g$-band exposures and $\rm 3\times300$s $r$-band exposures. The majority of these exposures were taken in decent seeing conditions ($\sim 1.0$--$1.3\arcsec$) while Aquarius~III was at low airmass (sec(z) $\lesssim 1.2$); the exception was the first of these nights, when seeing was $\sim 1.4$--$1.5''$. Owing to the longer integration times compared to the existing data and relatively dark skies, these data typically achieved effective exposure times significantly longer than the previously-available data from public DECam surveys.
\par We processed these 12 new exposures, as well as all existing archival $g,r$-band exposures of the same field, using a pipeline similar to that used for the Dark Energy Survey (DES) Year 3 cosmology analyses \citep{2021ApJS..254...24S,10.1093/mnras/stab3055}. This processing is described in more detail by \citet{2024arXiv240800865T} in the context of early science results from DELVE Data Release 3; however, we highlight here that this pipeline relies on image-level coaddition for detection, followed by simultaneous fits to all individual images, to derive the full benefit from overlapping exposures of the same area. This is in contrast to the processing used for DELVE DR1 and DELVE DR2, which constructed multi-band catalogs by collating measurements made on individual exposures (as originally introduced by \citealt{2015ApJ...813..109D}). The combined effect of the new exposures and the coadd image processing improved the $10\sigma$~depth of our catalogs to $g_0, r_0 \sim 24.5$, representing a nearly one magnitude improvement in each band relative to the median depth of DELVE DR2 despite the relatively modest investment of additional exposure time.
\par For all analyses described below, we used a cleaned version of this catalog that retained only sources that passed the SourceExtractor cuts \texttt{FLAGS < 4} and \texttt{IMAFLAGS\_ISO} = 0 in each band (see e.g. \citealt{Abbott_2021} for a description of these parameters).\footnote{We make this deeper catalog available in an online repository associated with this work (see \secref{opendata})} We specifically made use of the ``Single Object Fitting'' (SOF) PSF magnitudes derived from the coadd processing, which we corrected for interstellar extinction using the maps of \citet{Schlegel:1998} with the recalibration from \citet{2011ApJ...737..103S}. 
Stars were separated from background galaxies using a morphological classifier (\texttt{EXTENDED\_CLASS}) based on the distribution of sources in the \texttt{BDF\_T}--\texttt{BDF\_S2N} plane, where \texttt{BDF\_T} is a parameter describing the best-fit pre-seeing bulge + disc model size \citep{10.1093/mnras/stab3055} and \texttt{BDF\_S2N}  is the associated signal-to-noise.
This classifier, which was developed for DES Y6, assigns sources an integer score from 0 to 4, with 0 being the most pointlike. For our analysis, we adopted a selection $0 <= \texttt{EXTENDED\_CLASS\_G} <= 2$ based on measurements derived from the $g$-band images but note that using different thresholds did not significantly influence our results.

\begin{figure*}
    \centering
    \includegraphics[width=\textwidth]{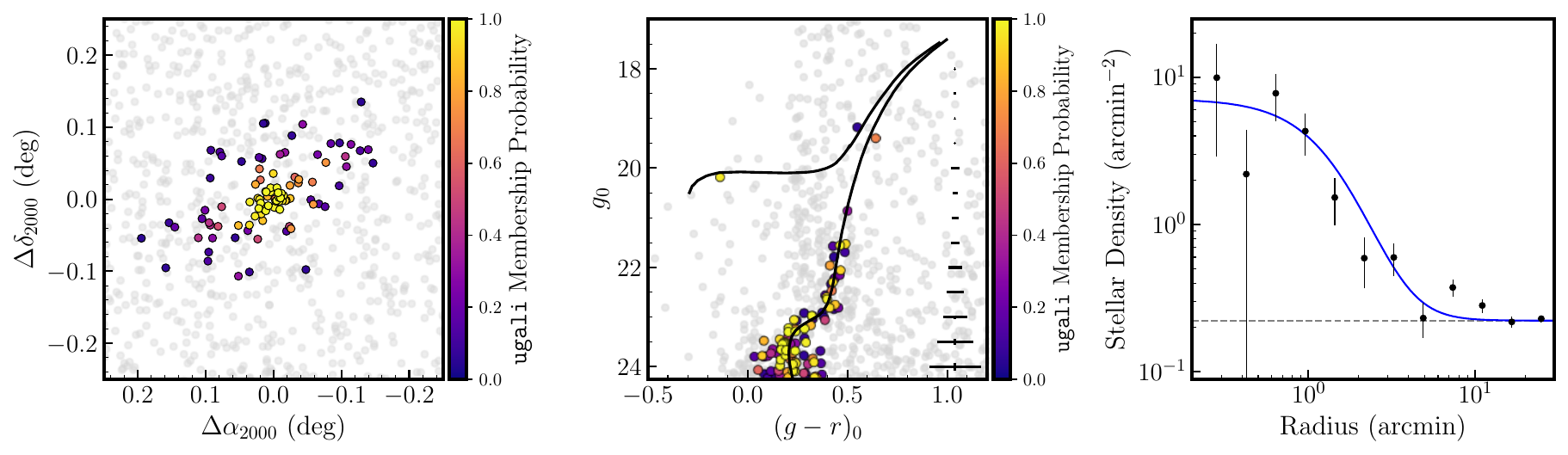}
    \caption{(Left) Spatial distribution of stars in a $0.5\degree \times 0.5\degree$ region centered on Aquarius~III. All stars are colored by their membership probabilities assigned by our \texttt{ugali} fit, which incorporates both spatial and color--magnitude information but does not include velocity or metallicity information from our spectroscopy. Stars with probabilities $p < 0.05$ are shown in grey for ease of visualization. (Center) CMD for sources in the lefthand panel, with the same coloring scheme. A PARSEC isochrone with age 13.5 Gyr and metallicity $Z=0.0001$, shifted to a distance modulus $(m-M)_0 = 19.66$, is shown as a solid black curve. (Right) Radial stellar density profile of isochrone-filtered stars derived from concentric circular bins (black points, with Poisson uncertainties shown) assuming the same magnitude cuts and star/galaxy separation criterion as used for the \texttt{ugali} fit. The best-fitting Plummer model with azimuthally-averaged angular half-light radius $r_h = 1.6\arcmin$ is shown in blue. }
    \label{fig:member}
\end{figure*}

\subsubsection{A Clearer Look at Aquarius~III}
In \figref{diagnostic}, we present several views of Aquarius~III based on the deeper DECam photometric catalog described above. In the leftmost panels, we depict the spatial distribution of stars and galaxies in a small region around the candidate system after filtering with an old, metal poor isochrone and smoothing the resulting density field with a Gaussian kernel. Comparing the density of stars within a $5\arcmin$ radius to that measured from an equal-area concentric background annulus with an inner radius $r = 12'$ away,  Aquarius~III stands out as a robust overdensity of stars with a Poisson significance of $\sim 7\sigma$. The detection of this as-yet unreported stellar system is further elucidated by its color--magnitude diagram (CMD; center-left panel), which clearly displays a distinct lower red giant branch (RGB), main-sequence turnoff (MSTO), and upper main sequence (MS) characteristic of an old, metal-poor halo stellar population. These features are not seen for the sample of stars in a concentric control annulus (center-right panel), as emphasized by a background-subtracted Hess diagram (far righthand panel). The MSTO feature that clearly appears in these panels was only marginally visible in the discovery CMD from DELVE DR2, emphasizing the importance of the deeper follow-up imaging for characterization of the putative stellar system.

\section{Structural and Stellar Population Properties }
\label{sec:prop}
We simultaneously fit Aquarius~III's structure and stellar population with the Ultra-faint Galaxy LIkelihood (\texttt{ugali}) toolkit\footnote{https://github.com/DarkEnergySurvey/ugali}, which implements the unbinned Poisson maximum-likelihood formalism described in Appendix C of \citet{2020ApJ...893...47D}. Aquarius~III was modelled with a \citet{1911MNRAS..71..460P} radial stellar density profile, and its $g,r$-band CMD was fit with a PARSEC stellar isochrone \citep[][Version 1.2S]{Bressan:2012,2014MNRAS.444.2525C,10.1093/mnras/stv1281}. The main free parameters of the radial profile model were Aquarius~III's centroid coordinates ($\alpha_{2000}$ and $\delta_{2000}$), angular elliptical half-light radius ($a_h$), ellipticity ($\epsilon$, defined as $\epsilon = 1 - \frac{b}{a}$), and position angle east of north ($\rm P.A.$). The isochrone age was fixed at $\tau = 13.5$~Gyr and the metallicity was fixed at $Z = 0.0001$ (the oldest age and lowest metallicity in our grid), and thus the only free parameter for the CMD component of the fit  was the distance modulus, $(m-M)_0$. Lastly, as an additional free parameter, \texttt{ugali} models the ``richness'' of the putative satellite which is defined as the total number of stars in the system above the hydrogen-burning limit \citep{2020ApJ...893...47D}.

\unskip
\begin{deluxetable*}{l c c c c}
\tablecolumns{5}
\tablewidth{0pt}
\tabletypesize{\small}
\tablecaption{\label{tab:properties} Properties of the Aquarius III Milky Way Satellite Galaxy}
\tablehead{\colhead{Parameter} & \colhead{Description} &  \colhead{Value} & \colhead{Units} & \colhead{Section}}
\startdata
\ra & Centroid Right Ascension &  $357.218^{+0.005}_{-0.004}$ & deg & \ref{sec:prop}\\
\dec & Centroid Declination &  $-3.489^{+0.004}_{-0.003}$ & deg & \ref{sec:prop} \\
$a_h$ & Elliptical Angular Half-Light Radius  &  $2.1^{+0.7}_{-0.5}$ & arcmin & \ref{sec:prop} \\
$a_{1/2}$ & Elliptical Physical Half-Light Radius & $51^{+16}_{-12}$  & pc & \ref{sec:prop}  \\
$r_h$ & Azimuthally-Averaged Angular Half-Light Radius &  $1.6^{+0.4}_{-0.3}$ & arcmin & \ref{sec:prop} \\
$r_{1/2}$ & Azimuthally-Averaged Physical Half-Light Radius &  $41^{+9}_{-8}$ & pc & \ref{sec:prop}  \\
$\epsilon$ & Ellipticity & $0.47^{+0.14}_{-0.28}$ & ... & \ref{sec:prop}\\
\PA  & Position Angle of Major Axis (East of North) & $119^{+17}_{-11}$ & deg & \ref{sec:prop} \\
$(m-M)_0$  & Distance Modulus & $19.66 \pm 0.11\tablenotemark{a}$ & mag & \ref{sec:prop}\\
$D_{\odot}$  & Heliocentric Distance & $85 \pm 4\tablenotemark{a} $ & kpc & \ref{sec:prop} \\
$M_V$ & Absolute $V$-band Magnitude & $-2.5^{+0.3}_{-0.5}$ & mag  & \ref{sec:prop}  \\
$L_V$ & $V$-band Luminosity & $850^{+380}_{-260}$ & $\Lsun$ & \ref{sec:prop} \\
$M_{*}$ & Stellar Mass (assming $M_*/L_V = 2$) &  $1700^{+760}_{-520}$ & $\Msun$ & \ref{sec:prop} \\
$E(B-V)$ & Mean Reddening ($r < 5\arcmin)$ & 0.04 & mag & \ref{sec:prop} \\
\hline
$N_{\rm spec}$ &  Number of Spectroscopic Members &   11 & ... & \ref{sec:members} \\ 
$v_{\rm sys}$ & Mean Heliocentric Radial Velocity & $-13.1^{+1.0}_{-0.9}$  & $\rm km s^{-1}$ & \ref{sec:dynamics}\\
% $v_{\rm GSR}$ & Radial Velocity in the Galactic Standard of Rest & $129^{+9}_{-10}$  & $\rm km s^{-1}$ & \ref{sec:dynamics}\\
$\sigma_v$ & Velocity Dispersion\tablenotemark{b}, uniform prior  $0 < \sigma_v < 10$ & $<$ 3.5 & $\rm km s^{-1}$   & \ref{sec:vdisp} \\
$\sigma_v$ & Velocity Dispersion, prior: $\lvert \log_{10}\sigma_v\rvert < 1$ & $< 2.1$  & $\rm km s^{-1}$   & \ref{sec:vdisp} \\
$M_{1/2}$ & Dynamical Mass within $r_{1/2}$ & $< 5.1 \times 10^{5}$ \rm & $\Msun$ & \ref{sec:dynmass}\\ 
$M_{1/2}/L_{V,1/2}$ & Mass-to-Light Ratio within $r_{1/2}$ & $<$ 1300 & $\Msun/\Lsun$ & \ref{sec:dynmass}\\ 
$\log_{10} J(0.5\degree)$ & $J$-factor within a solid angle of radius $0.5\degree$ & $<$ 17.8 & GeV$^2$cm$^{−5}$ & \ref{sec:jfactor} \\
\hline
$[\rm Fe/H]_{\rm spec}$ &  Mean Spectroscopic Metallicity & $-2.61 \pm 0.21$& dex & \ref{sec:mdisp}\\
$\sigma_{[\rm Fe/H]}$ &  Metallicity Dispersion among Spectroscopic Members & $0.46^{+0.26}_{-0.14}$ & dex & \ref{sec:mdisp}\\
\hline 
$\mu_{\alpha *}$ & Proper Motion in Right Ascension & $1.01 \pm 0.25$ & mas yr$^{-1}$ & \ref{sec:orbit}\\ % cos(\delta)
$\mu_{\delta}$ & Proper Motion in Declination & $-0.10 \pm 0.20$ & mas yr$^{-1}$& \ref{sec:orbit}\\
$r_{\rm GC}$ & Galactocentric Distance & $86 \pm 4$ & \kpc &\ref{sec:orbit}\\
$r_{\rm apo}$ & Orbital Apocenter & Unconstrained & \kpc & \ref{sec:orbit}\\
$r_{\rm peri}$ & Orbital Pericenter & $78 \pm 7$ & $\kpc$ & \ref{sec:orbit}\\
$e$ & Orbital Eccentricity & Unconstrained & ... & \ref{sec:orbit}\\
$L_Z$ & Angular Momentum about the Galactocentric $Z$ Axis &  $13^{+3}_{-5}$ & $\rm 10^3 \ \text{kpc km s}^{-1}$ & \ref{sec:orbit}\\
\hline 
\enddata
\tablecomments{The velocity dispersion posterior peaked near zero, and we therefore quote upper limits for $\sigma_v$, $M_{1/2}$, $M_{1/2}/L_{V,1/2}$, and $\log_{10} J(0.5\degree)$ at the 95\% credible level.}
\tablenotetext{a}{We assume a  $\pm 0.1$ mag systematic uncertainty  on the distance modulus to account for uncertainties in isochrone modeling following \citealt{2015ApJ...813..109D}. This systematic term has been propagated to the quoted heliocentric distance in kiloparsecs as well as to our physical size measurements. }
\tablenotetext{b}{We recommend the use of this velocity dispersion limit derived assuming the uniform prior on 0 $< \sigma_v/(\rm \ km \ s^{-1}) < 10$. We use this estimate when deriving the dynamical mass and mass-to-light ratio reported in the subsequent rows.}
\end{deluxetable*}
\unskip

\unskip

\par We derived posterior probability distributions for each of the seven free model parameters using the affine-invariant Monte Carlo Markov Chain (MCMC) sampler \texttt{emcee} \citep{Foreman-Mackey:2013}, for which we used 80 walkers each taking 15000 steps with the first 3000 steps discarded as burn-in. In \tabref{properties}, we report the resulting estimates of the model parameters and their uncertainties where the uncertainties were derived in most cases from the highest density interval containing the peak and 68\% of the marginalized posterior distribution. We also report a number of additional quantities derived from these posteriors, including Aquarius~III's azimuthally-averaged angular and physical half-light radii ($r_h$ and $r_{1/2}$, respectively, where $r_h = a_h \sqrt{1-\epsilon}$), absolute magnitude ($M_V$, derived using the formalism from \citealt{2008ApJ...684.1075M}), $V$-band luminosity ($L_V \equiv 10^{0.4(4.83-M_V)})$, and stellar mass ($M_*$, calculated from $L_V$ assuming a stellar-mass-to-light ratio of two). 

\par All of these results were specifically derived from a fit using our deeper DECam catalog with an assumed magnitude limit of $g_0, r_0 = 24.25$ and both the isochrone age and metallicity fixed. We estimate the formal $S/N = 10$ magnitude limit of our deeper DECam catalog to be $g_0, r_0 \approx 24.5$, but chose to use a more conservative limit for our analysis to derive measurements in a regime where the star/galaxy classification and photometric uncertainties are better controlled.

\subsection{Summary of Structural Fit Results}
\par We find that Aquarius~III is a low luminosity ($M_V = -2.5^{+0.3}_{-0.5}$, $L_V = 850^{+380}_{-260} \ L_{\odot}$) Milky Way satellite with a CMD that is fit closely by the oldest, most metal-poor isochrone in our PARSEC grid. The satellite's structure is well fit by a \citet{1911MNRAS..71..460P} radial stellar density profile with an elliptical half-light radius $a_{h} = 2.1^{+0.7}_{-0.5}\arcmin$ ($a_{1/2} = 51^{+16}_{-12}$~pc) and a moderate ellipticity ($\epsilon = 0.47^{+0.14}_{-0.28}$). The corresponding azimuthally-averaged half-radius is $r_h = 1.6^{+0.4}_{-0.3}\arcmin$ ($r_{1/2} = 41^{+9}_{-8}$~pc). The total number of observed (photometric) members above our nominal magnitude limit, calculated by summing over the membership probabilities computed by \texttt{ugali}, is $\Sigma p_{i,obs} \sim 56$ stars. 
\par We visualize all of these results in \figref{member}, where we show the spatial distribution and CMD of stars in the system in the left and center panels, respectively, as well as its radial profile in the righthand panel. Stars in the former two panels are colored by their membership probabilities from our \texttt{ugali} fit, which incorporates both spatial and color--magnitude information but no spectroscopic information.

\section{Stellar Kinematics and Metallicities from Keck/DEIMOS Spectroscopy}
\label{sec:dynamics}
\subsection{ Observations and Data Reduction}

To characterize Aquarius~III's mean velocity, internal kinematics, and stellar metallicities, we obtained medium-resolution, multi-object spectroscopy of stars in Aquarius~III with the DEep Imaging Multi Object Spectrograph (DEIMOS; \citealt{2003SPIE.4841.1657F}) on the Keck~II telescope at the W.M. Keck Observatory on Maunakea, Hawai'i.
Following numerous past studies of Milky Way satellites with DEIMOS \citep[e.g.,][]{2007MNRAS.380..281M,2007ApJ...670..313S}, we used the 1200G grating with the OG550 order-blocking filter. This configuration provides near-continuous coverage over the wavelength range $6500 \rm \ \AA$ to $9000 \rm \ \AA$ at a resolution of $\mathcal{R} \sim 6500$. This wavelength range contains a number of strong stellar absorption features including $\rm H\alpha$, the Calcium Triplet (CaT), and the Mg I $\lambda$ 8807 $\rm \ \AA$ line, in addition to the strong telluric A-band feature at $\sim 7600 \rm \ \AA $.
\par On the night of 2023 October 5, we collected $\rm 1\times1200s$, $\rm 7\times1800s$, and $\rm 1\times1500s$ exposures for a total shutter-open exposure time of 15300s (4.25 hours), all in clear conditions.\footnote{The mean Modified Julian Day (MJD) of these exposures was 60223.40527.} We used a single multi-object mask comprised of slits of width $0.7\arcsec$ and minimum length 4.5\arcsec. Targets for this mask were drawn primarily from the probabilistic member catalog provided by a preliminary \ugali fit (see \secref{prop}) as well as from an additional pool of targets drawn from the DESI Legacy Imaging Surveys Data Release 10 \citep{2019AJ....157..168D}. XeNeArKr arcs and internal quartz flats were taken at the beginning of the night; this is sufficient for precise wavelength calibration thanks to DEIMOS' excellent stability and active flexure compensation system.

\par The raw DEIMOS spectra were reduced using a lightly modified version of the official Keck-supported data reduction pipeline implemented within the \texttt{PypeIt} framework \citep{2020JOSS....5.2308P}. \texttt{PypeIt} reduces the eight DEIMOS CCDs as four separate mosaic images each containing a red and blue chip and automatically performs flat-fielding, sky subtraction, and spectral extraction followed by wavelength calibration based on the calibration arc frames.  For the reductions used here, we disabled \texttt{PypeIt}'s default flexure corrections and heliocentric velocity corrections, and we instead determined linear flexure corrections for each reduced 1D spectrum during our velocity measurement procedure described below.\\ 
\vspace{-2em}
\subsection{Velocity Measurements}
We measured line-of-sight velocities of stars using an in-development version of the \texttt{DMOST} package (M. Geha et al., in prep.),\footnote{\url{https://github.com/marlageha/dmost}} a dedicated measurement pipeline for observations made with DEIMOS' 1200G grating. \texttt{DMOST} measures stellar velocities by forward-modelling the 1D spectrum of a given star with both a stellar template from the PHOENIX stellar atmosphere library  \citep{2013A&A...553A...6H} and a telluric absorption spectrum from TelFit \citep{2014AJ....148...53G}. The stellar spectrum template is selected from a coadded spectrum derived from all exposures of a given source, while the telluric template is selected based on a fit to the highest-$S/N$ sources on a given mask (for each exposure) and is assumed to be representative across all sources on the mask. After these templates have been selected, velocities are determined on an exposure-by-exposure basis by minimizing the $\chi^2$ of the best-fit template against the observed spectrum from each exposure. This is carried out through a MCMC fit simultaneously constraining both the radial velocity of a given star as well as a linear wavelength shift of the telluric spectrum needed to correct for the miscentering of stars within their slits (see \citealt{2007ApJ...663..960S}). If no individual exposures yielded a velocity measurement (as commonly occurs for the very faintest sources), \texttt{DMOST} instead derives the velocity from the coadded spectrum across all exposures.
\par The final radial velocity of a given star is calculated from an inverse-variance-weighted average across the measurements from individual exposures that had well-behaved (nearly Gaussian) posteriors. The associated statistical error is taken as the standard deviation across exposures. Lastly, a total velocity error that includes the contribution of systematic effects was calculated by scaling this statistical error by a factor of 1.4 and adding an additional $1.1 \rm \ km \ s^{-1}$ in quadrature, i.e., $\epsilon_{v, \rm tot}^2 = \sqrt{1.4\epsilon_{\rm v,stat}^2 + 1.1^2}$. Here, the scaling term encapsulates the signal-to-noise-dependent component of the systematic error while the fixed term represents an uncertainty floor. This systematic error prescription was derived based on the repeatability of velocity measurements across hundreds of DEIMOS masks. The total uncertainties from this procedure have been validated by comparing stellar velocities against public radial velocity data from large-scale spectroscopic surveys (M. Geha et al., in prep).\footnote{We emphasize that this systematic error prescription is not just instrument-dependent: it is \textit{pipeline-dependent}. Thus, this prescription is not generally applicable to all DEIMOS analyses nor is it expected to match other prescriptions in the literature. }

\subsection{Equivalent Width Measurements}

\par \texttt{DMOST} also measures the Equivalent Widths (EWs) of the CaT lines of stellar sources from their coadded 1D spectra. For this work, each of the CaT lines were modelled with a Gaussian-plus-Lorentzian profile (for stars at $S/N > 15$ per spectral pixel) or a Gaussian profile (for stars at $S/N < 15/$pixel). The profile model parameters were derived through a non-linear least-squares fit using \texttt{scipy} \citep{2020SciPy-NMeth}, and we integrated the resulting fits to get the EW of each line. The statistical error on the EW of each line was then derived analytically from the fit errors. Lastly, a total CaT EW error was computed by summing the EW uncertainties of the three individual lines in quadrature and then further adding a $0.2\rm \ \AA$ systematic uncertainty floor in quadrature. Analogously to the velocity uncertainties, this systematic uncertainty floor was derived based on the repeatability of total EW measurements across masks and validated against spectroscopic metallicities from large-scale public surveys.

\begin{figure*}
    \includegraphics[width = \textwidth]{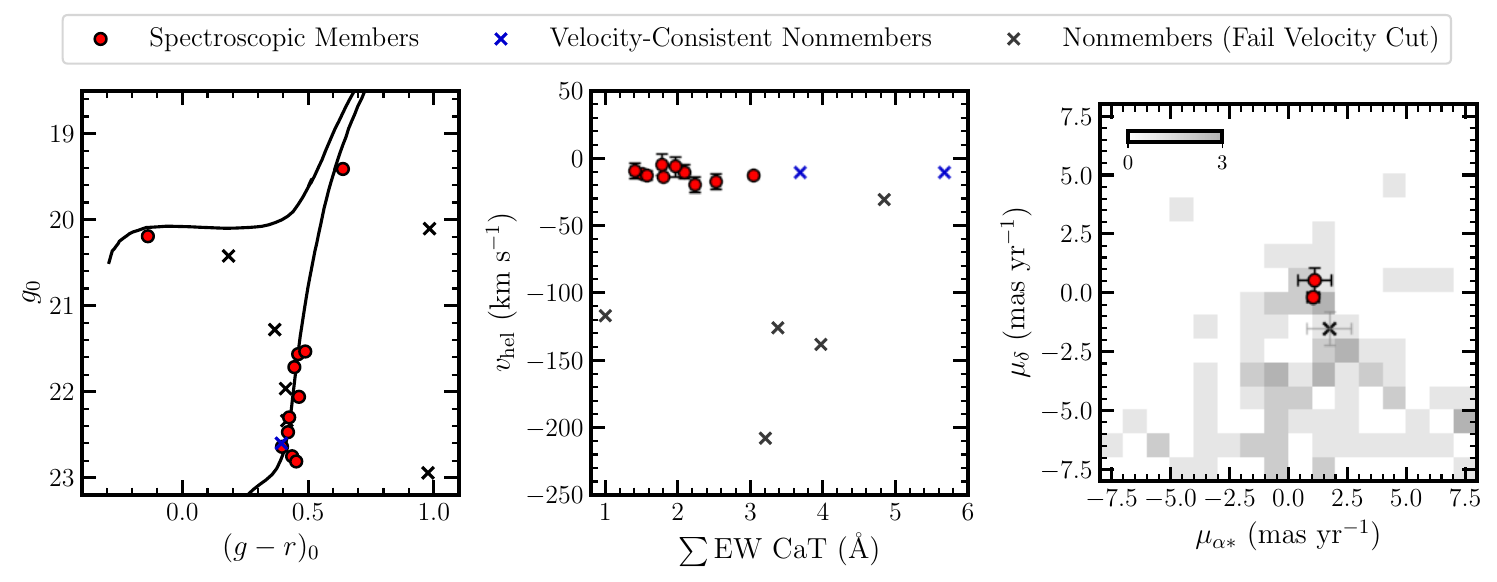}
    \caption{\label{fig:specmembers} Three views of our Keck/DEIMOS spectroscopic sample for the Aquarius~III field. In each panel, our sample of confirmed members are shown as red filled circles while nonmembers are shown as blue and black crosses. (Left) CMD showing just the spectroscopic sample. One velocity-consistent non-member is redder and falls outside the axis limits shown here. (Center) Radial velocity in the heliocentric frame ($v_{\rm hel.}$) vs. summed equivalent width of the Calcium Triplet lines ($\Sigma$~EW CaT). No CaT EW errorbars are shown for ease of visualization. The suspected members of Aquarius~III cluster in the velocity range $-20 < v_{\rm hel.}/(\rm km \ s^{-1}) < 0$ and all share low CaT EWs. Note that the BHB star has been excluded from this central panel as its CaT EW is not well-measured due to the strong Paschen absorption in its spectrum. (Right) \Gaia proper motions of the spectroscopic sample overlaid over a 2D histogram of the proper motions of all stars within a $10\arcmin$ radius. Only the brightest two spectroscopic members have reported proper motions from \Gaia. These two members' proper motions are closely consistent with each other and form the basis of our measurement for the Aquarius~III system.}
\end{figure*}

\par From these CaT EW measurements, we derived [Fe/H] measurements for all candidate RGB stars assuming the luminosity-dependent EW--[Fe/H] calibration from \citet{2013MNRAS.434.1681C}. 
We specifically adopted the form of the calibration that requires the absolute $V$-band magnitude of each source as an input, which we estimated by transforming our DECam $g,r$ photometry using the piecewise relations derived for DES DR2 \citep[Appendix B of][]{Abbott_2021} and subtracting our derived distance modulus of $(m-M)_0 = 19.66 \pm 0.11$. 
Posterior distributions for the metallicity of a given star were constructed through Monte Carlo sampling from the error distributions on the total equivalent widths $\Sigma \rm EW$ (including the assumed 0.2 $\rm \AA$ systematic error), the $g,r$-band photometry\footnote{We add a 0.02 mag uncertainty floor in quadrature to the photometric errors in each filter to account for zeropoint uncertainties.}, Aquarius~III's distance modulus, and the coefficients on the \citet{2013MNRAS.434.1681C} relation; we assumed Gaussian errors in all cases. Our final metallicity measurements were then derived based on the median and 16th/84th percentiles of the resultant posteriors.
\par Lastly, we also used \texttt{DMOST} to measure the EW of the Mg I $\lambda 8807 \rm \ \AA$ absorption line. The strength of this line is correlated with stellar surface gravity and thus its EW is a useful discriminant for separating foreground main-sequence stars from red giants such as those expected in a halo dwarf galaxy \citep{2012A&A...539A.123B}. We performed a simple Gaussian fit to this line through a procedure like that used for the CaT and integrated to get the EW. The errors on the Mg I EW measurements have not been extensively validated and here we opted to only use these measurements to retroactively check that our member sample did not include any interloper main-sequence stars with large Mg I EWs.

\begin{deluxetable*}{ccccccccccc}
\tablecaption{Basic Properties of Stars Observed with Keck/DEIMOS}
\tablehead{StarName & RA & DEC & $g_0$ & $r_0$ & S/N & $v_{\rm hel.}$ & $\Sigma$ EW CaT & [Fe/H] & Member & Type}
\startdata
\textit{Gaia} DR3 2447566690779941504 & $357.216$ & $-3.488$ & $19.41$ & $18.77$ & $78.1$ & $-14.5 \pm 1.2$  & $1.80 \pm 0.21$ & $-3.05 \pm 0.11$ & True & RGB \\
\textit{Gaia} DR3 2447566656420203008 & $357.239$ & $-3.489$ & $20.19$ & $20.33$ & $25.7$ & $-11.1 \pm 1.9$  & --- & ---   & True & BHB \\
Aqu III J234851.34$-$032925.89 & $357.214$ & $-3.491$ & $21.56$ & $21.10$ & $17.0$ & $-12.9 \pm 2.2$  & $3.03 \pm 0.28$ & $-1.95 \pm 0.13$ & True & RGB \\
Aqu III J234849.41$-$032917.06 & $357.206$ & $-3.488$ & $21.53$ & $21.04$ & $15.5$ & $-12.5 \pm 2.0$  & $1.52 \pm 0.29$ & $-2.81 \pm 0.15$ & True & RGB \\
Aqu III J234851.13$-$032916.12 & $357.213$ & $-3.488$ & $21.71\tablenotemark{a}$ & $21.27\tablenotemark{a}$ & $12.6$ & $-13.6 \pm 3.2$  & $1.56 \pm 0.30$ & $-2.74 \pm 0.15$ & True & RGB \\
Aqu III J234848.22$-$032927.00 & $357.201$ & $-3.491$ & $22.06$ & $21.60$ & $9.4$ & $-10.9 \pm 4.9$  & $2.09 \pm 0.34$ & $-2.34 \pm 0.17$ & True & RGB \\
Aqu III J234855.22$-$032840.90 & $357.230$ & $-3.478$ & $22.30$ & $21.87$ & $7.2$ & $-10.3 \pm 5.5$  & $1.41 \pm 0.37$ & $-2.74 \pm 0.19$ & True & RGB \\
Aqu III J234838.64$-$032753.36 & $357.161$ & $-3.465$ & $22.47$ & $22.05$ & $6.5$ & $-19.9 \pm 5.6$  & $2.23 \pm 0.48$ & $-2.17 \pm 0.22$ & True & RGB \\
Aqu III J234850.46$-$032929.86 & $357.210$ & $-3.492$ & $22.64$ & $22.24$ & $5.3$ & $-17.8 \pm 5.9$  & $2.52 \pm 0.58$ & $-1.97 \pm 0.28$ & True & RGB \\
Aqu III J234846.51$-$032915.26 & $357.194$ & $-3.488$ & $22.75$ & $22.31$ & $5.0$ & $-7.1 \pm 6.9$  & $1.96 \pm 0.51$ & $-2.27 \pm 0.25$ & True & RGB \\
Aqu III J234901.54$-$033156.18 & $357.256$ & $-3.532$ & $22.81$ & $22.36$ & $4.2$ & $-5.4 \pm 8.0$  & $1.78 \pm 0.56$ & $-2.39 \pm 0.27$ & True & RGB \\
\hline
Aqu III J234856.79$-$033032.90 & $357.237$ & $-3.509$ & $22.60$ & $22.21$ & $19.0$ & $-11.2 \pm 1.5$ & $5.66 \pm 0.27$ & --- & False & --- \\
\textit{Gaia} DR3 2447566690779941120 & $357.225$ & $-3.491$ & $20.79$ & $19.47$ & $13.1$ & $-11.5 \pm 1.6$ & $3.69 \pm 0.32$ & --- & False & --- \\
\hline
\textit{Gaia} DR3 2447567137456807936 & $357.123$ & $-3.497$ & $20.10$ & $19.12$ & $67.6$ & $-30.8 \pm 1.1$ & $4.85 \pm 0.31$ & --- & False & --- \\
\textit{Gaia} DR3 2447566656420202624 & $357.248$ & $-3.491$ & $20.42$ & $20.24$ & $28.1$ & $-250.8 \pm 1.8$ & --- & --- & False & BHB \\
Aqu III J234910.87$-$033040.95 & $357.295$ & $-3.511$ & $21.28$ & $20.91$ & $14.1$ & $-207.9 \pm 2.3$ & $3.21 \pm 0.31$ & --- & False & --- \\
Aqu III J234832.38$-$032729.57 & $357.135$ & $-3.458$ & $22.94$ & $21.97$ & $12.5$ & $-117.0 \pm 1.2$ & $1.00 \pm 1.48$ & --- & False & --- \\
Aqu III J234846.59$-$032810.30 & $357.194$ & $-3.470$ & $21.96$ & $21.55$ & $9.6$ & $-125.9 \pm 3.3$ & $3.38 \pm 0.40$ & --- & False & --- \\
Aqu III J234817.17$-$032618.61 & $357.072$ & $-3.439$ & $22.34$ & $21.92$ & $5.7$ & $-138.3 \pm 5.7$ & $3.97 \pm 0.61$ & --- & False & --- \\
\enddata
\tablecomments{Stars are separated by membership category with the confirmed members first, followed by the two velocity-consistent non-members, and lastly, the six non-members. Celestial coordinate positions (RA, DEC) and extinction-corrected magnitudes are taken from our deeper DECam photometric catalog.  The quoted signal-to-noise ($S/N$) reported here relate to the DEIMOS spectroscopy. Note that metallicities derived from spectra below $S/N$ = 7 may be less reliable and thus were  excluded from our metallicity dispersion fit.} 
 \tablenotetext{a}{This star was found to have inconsistent $r-$band photometry between our deeper DECam catalog and Legacy Surveys DR10, such that it would have been rejected in the former dataset by a CMD selection. Given the star's spectrum, velocity,  central position, and lack of obvious time variability, we choose to adopt the LS DR10 $r$-band magnitude and therefore consider this star is indeed a true member.}
 \label{tab:membertable}
 \end{deluxetable*}

\subsection{Stellar Membership}
\label{sec:members}
\par The measurement procedures described above yielded a sample of 19 stars with reliable velocity measurements. We report the basic properties for these 19 stars in \tabref{membertable}, including their positions, apparent magnitudes, velocities, metallicities, and the signal-to-noise ($S/N$) of their coadded DEIMOS spectra. 
At a glance, the signal-to-noise ratios of stars in our sample range from $S/N$ = 78/pixel for the brightest star ($g_0 \sim 19.4$) down to $S/N = 4$/pixel for the faintest ($g_0 \sim 22.8$). The full range of velocity errors is $1.2-8.0 \rm \ km \ s^{-1}$, with measurements at $S/N \gtrsim 15$/pixel generally being dominated by the systematic uncertainty component. The CaT EW uncertainties span $\sim 0.2-0.6 \rm \ \AA$; however, we opted to exclude EW measurements derived from spectra at $S/N < 7$/pixel from our analysis. In practice, this means we considered only stars with CaT EW measurements uncertainties less than $0.4 \ \rm \AA$.
\par In \figref{specmembers}, we visualize the properties of all 19 stars in a CMD (left panel), the $v_{\rm hel} - \rm CaT \ EW$ plane (center panel), and in a proper motion vector-point diagram (right panel) derived using measurements available from \Gaia Data Release 3 \citep{2023A&A...674A...1G}. 
As seen in the center panel, the velocity distribution of this 19-star sample includes a conspicuous excess of 13 stars in the velocity range $-20\rm  \ km \ s^{-1} \lesssim  \textit{v}_{\rm hel.} \lesssim 0 \rm \ km \ s^{-1} $. Of these 13 stars, 12 are consistent with the best-fit isochrone from our \ugali fit. The remaining star, \Gaia DR3 2447566690779941120, is $\sim 0.8$ mag redder than the best-fit isochrone and we therefore rejected it from membership despite its consistent velocity. We then further rejected one of the 12 velocity-consistent, isochrone-consistent stars (J234856.79$-$033032.90) because its implied metallicity, [Fe/H] = $-0.44 \pm 0.14$, would be inconsistent with the remaining member candidates and the composite metallicity distribution of known ultra-faint dwarf galaxy stars (see e.g., Figure 11 of \citealt{2023ApJ...958..167F}). This star also exhibited a noticeably higher Mg I EW of than the remaining member candidates, $\rm 0.48 \pm 0.17 \AA$ (statistical error only), further suggesting its nature as a foreground main-sequence contaminant.
\par After these selections, we were left with a sample of eleven stars which we regard as clear members of Aquarius III including ten red giant branch (RGB) stars and one blue horizontal branch (BHB) star. Within the sample of ten plausible RGB member stars with similar velocities,  all were found to have a Mg I EW measurement consistent with being giants according to the \citet{2012A&A...539A.123B} criterion at the $< 1.5\sigma$ level.  In addition, the candidate BHB star displays the broad Paschen absorption lines expected from a BHB star of its temperature. Thus, we found no reason to exclude any stars and we regard all eleven of these velocity-selected stars as likely members of Aquarius~III.

\subsection{Velocity and Velocity Dispersion}
\label{sec:vdisp}
\begin{figure*}
    \centering
    \includegraphics[width = .48\textwidth]{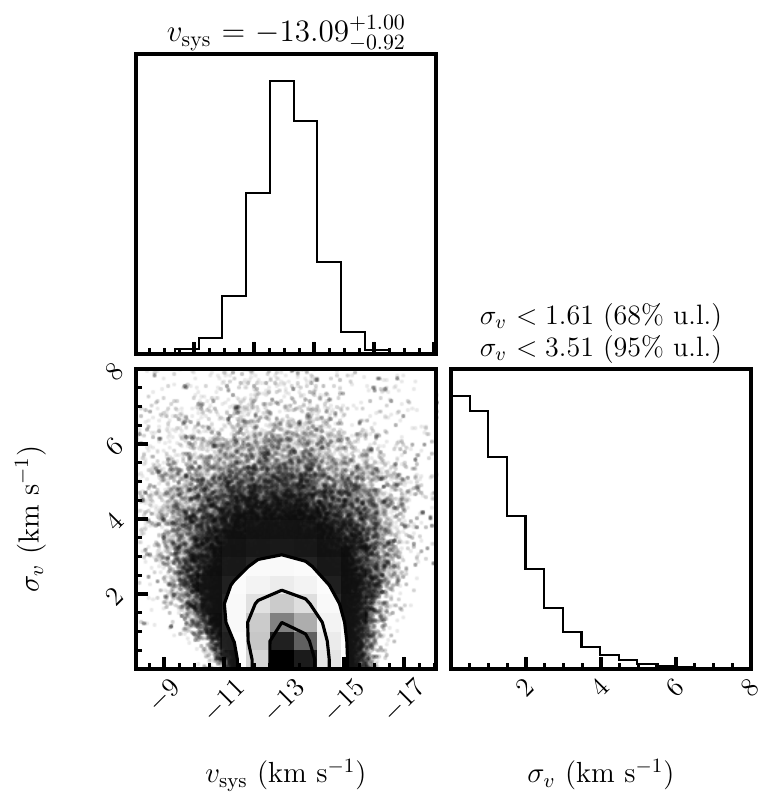}
    \includegraphics[width = .49\textwidth]{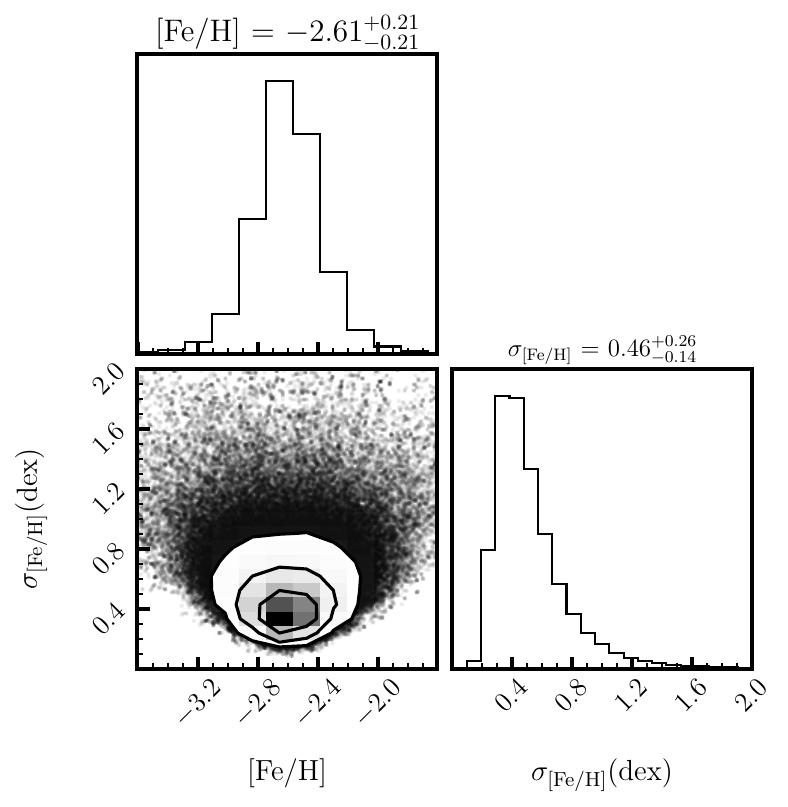}
    \caption{(Left) Posterior distributions for Aquarius~III's systemic mean velocity ($v_{\rm sys}$) and velocity dispersion ($\sigma_v$), derived from all eleven member stars assuming a uniform prior of $0 < \sigma_v/ (\rm \ km \ s ^{-1}) < 10$. 
    The velocity dispersion is unresolved with an upper limit of $3.5 \rm \ km \ s^{-1}$ at the 95\% credible level. (Right) The equivalent posterior distributions for Aquarius~III's systemic metallicity ([Fe/H]) and metallicity dispersion ($\sigma_{\rm [Fe/H]}$), derived from the six RGB member stars. The metallicity dispersion is clearly non-zero with a 95\% (99.5\%) credible lower limit of 0.25 (0.19) dex.}
    \label{fig:cornerplots}
\end{figure*}

We derived estimates of Aquarius~III's systemic radial velocity $(v_{\rm sys}$) and velocity dispersion ($\sigma_v$) through a simple two-parameter fit assuming the likelihood from \citet{2006AJ....131.2114W}. The observed velocity distribution was modelled as a Gaussian distribution with mean $v_{\rm sys}$ and a dispersion constituted both by an intrinsic component $\sigma_{v}$ and a component associated with the observational errors. We then performed a Bayesian fit assuming a default uniform prior on the velocity dispersion of $0 < \sigma_v/(\rm km \ s^{-1}) < 10$. Posterior probability distributions for each parameter were derived through MCMC sampling with \texttt{emcee}; for this sampling, we used 100 walkers each taking 6000 steps with the first 1000 steps for each walker discarded as burn-in. The resulting posteriors are shown in the left side of \figref{cornerplots}.

\par From the median and 16th/84th percentile of the marginalized posterior distribution, we found a systemic velocity of $v_{\rm sys} = -13.1^{+1.0}_{-0.9} \rm \ km \ s^{-1}$ for Aquarius~III. For the velocity dispersion, $\sigma_v$, the MCMC sampling produced marginalized posterior distributions with a mode approaching the lower boundary of our velocity dispersion prior ($\sigma_v = 0$). We are therefore only able to place an upper limit on the dispersion, which we find to be $\sigma_v < 3.5 \rm \ km \ s^{-1}$ (95\% credible upper limit) for our default prior. This limit strengthens to $\sigma_v < 2.1 \rm \ km \ s^{-1}$ at the 95\% credible level if we instead adopt a log-uniform prior of $-1 < \log_{10}(\sigma_v) < 1$.\footnote{For the logarithmic prior, the MCMC sampling was performed over $\log_{10}(\sigma$). We then recasted the samples to linear $\sigma_v$ estimates in order to quote upper limits.} We adopt the first of these as our nominal measurement because it most accurately reflects our prior belief on the range of possibility velocity dispersions of the system. Moreover, we view this as the most conservative choice for the sake of interpreting Aquarius~III's seemingly low mass (see \secref{cold}).

\subsection{Dynamical Mass and Mass-to-Light Ratio}
\label{sec:dynmass}
\par We proceeded to place an upper limit on Aquarius~III's dynamical mass within the half-light radius ($M_{1/2}$) using the mass estimator from \citet{2010MNRAS.406.1220W},
\[ M_{1/2} = 930 \ M_{\odot} \left(\frac{\sigma_v}{\rm \ km \ s^{-1}}\right)^2 \left(\frac{r_{1/2}}{\rm pc}\right), \]
under the assumption that the system is in dynamical equilibrium. We constructed the posterior distribution of $M_{1/2}$ by directly Monte Carlo sampling from the posterior distributions for $r_{1/2}$ and $\sigma_v$ without replacement and transforming them according to this relation. This yielded an upper limit on Aquarius~III's dynamical mass within the half-light radius of $M_{1/2} < 5.1 \times 10^{5} \ M_{\odot}$ at the 95\% credible level. 
Adopting a luminosity within the half-light radius of $L_{1/2} \equiv 0.5L_V = 425_{-130}^{+190} \ L_{\odot}$, this dynamical mass implies an upper-limit on the mass-to-light ratio within $r_{1/2}$ of $M_{1/2} / L_{1/2} < 1300 \ M_{\odot}/L_{\odot}$ (at the 95\% credible level).  

\par
\begin{figure*}[htp!]
    \centering
    \includegraphics[width = \textwidth]{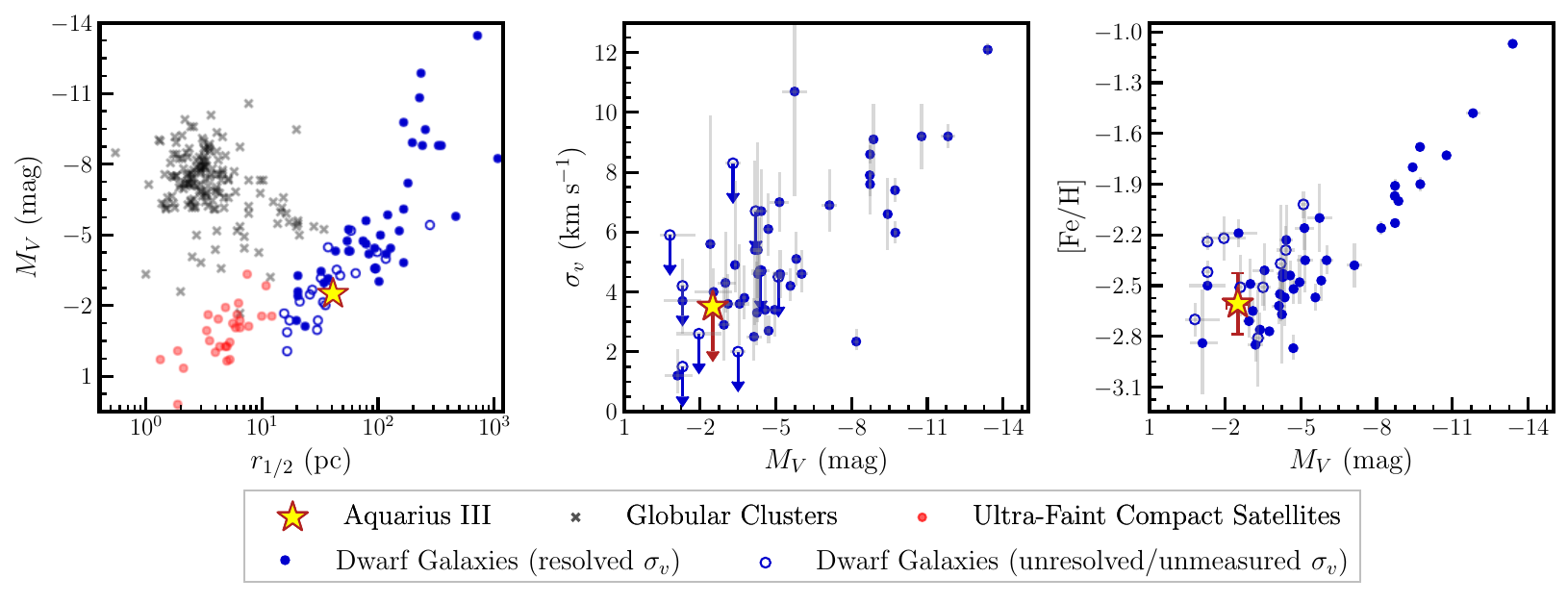}
    \caption{(Left) Absolute magnitude vs. half-light radius for the known population of Milky Way satellite galaxies (excluding the LMC, SMC, and Sagittarius), ``classical'' globular clusters, and ultra-faint, compact systems generally presumed to be star clusters. (Center) Velocity dispersion vs. absolute magnitude for the population of known Milky Way satellite galaxies for which spectroscopic measurements are available (blue circles; excluding the LMC, SMC, and Sagittarius) and for Aquarius~III (yellow star). Upper limits are denoted by downward arrows. Aquarius~III's velocity dispersion upper limit is comparable to the upper limits observed for other faint satellite galaxies. (Right) Spectroscopic metallicity vs. absolute magnitude for the same population of satellite galaxies. Aquarius~III falls comfortably near the $M_V$-[Fe/H] sequence delineated by the population of known Milky Way satellites. A complete reference list for the underlying measurements is provided in Appendix \ref{sec:references}.}
    \label{fig:pop}
\end{figure*}

 Taken at face value, this mass-to-light ratio limit suggests that Aquarius~III is consistent with having a substantial amount of dark matter or none at all. We therefore conclude that the available kinematic data do not clearly distinguish whether the system is a dark-matter dominated dwarf galaxy or a baryon-dominated, self-gravitating star cluster. This being said, we are able to eventually conclude that Aquarius~III is a dark-matter-dominated dwarf galaxy based on its size and metallicity distribution (see \figref{pop} and \secref{classification}). In this light, our relatively strong limits on Aquarius~III's velocity dispersion and total mass may position the galaxy as a useful laboratory for studying galaxy formation in low-mass halos (see \secref{cold}). 

\subsection{Metallicity and Metallicity Dispersion}
\label{sec:mdisp}
We also derived estimates Aquarius~III's mean metallicity ([Fe/H]) and metallicity dispersion ($\sigma_{\rm [Fe/H]}$) through a nearly identical Bayesian fit to that used for the velocity and velocity dispersion. The observed metallicity distribution of the RGB stars was modelled as a Gaussian with both an intrinsic component and a component associated with observational errors. We assumed a weak uniform prior on Aquarius~III's metallicity dispersion of $0 < \sigma_{\rm [Fe/H]} < 2$ and derived posterior probability distributions for each parameter with \texttt{emcee} using 100 walkers taking 6000 steps (with the first 1000 discarded as burn-in).
\par For our primary measurement, we limited our metallicity sample to the six RGB stars with $S/N > 7$/pixel spectra. The resulting posterior distributions for this sample are shown in the righthand panel of \figref{cornerplots}. From the 16th/84th percentile of the marginalized 1D posteriors, we find a mean metallicity of $\rm [Fe/H] = -2.61 \pm 0.21$ and a metallicity dispersion of $\sigma_{\rm [Fe/H]} = 0.46^{+0.26}_{-0.14}$. If we opted to revise our minimum $S/N$ threshold for metallicities to $S/N > 10$/pixel ($S/N > 4$/pixel), we find [Fe/H] = $-2.64^{+0.35}_{-0.34}$ and $\sigma_{\rm [Fe/H]} =  0.64^{+0.48}_{-0.24}$ ([Fe/H] = $-2.47 \pm 0.14$ and $\sigma_{\rm [Fe/H]} = 0.39^{+0.15}_{-0.10}$). In short, these tests demonstrate our metallicity results are fairly insensitive to the exact signal-to-noise cutoff we applied.

\par Our primary metallicity and metallicity dispersion constraints are subject to the strong caveat that they are derived from a sample of only six stars. Thus, their magnitudes should be interpreted cautiously. What is nearly certain, however, is that our measured metallicity dispersion is nonzero: the posterior probability distribution derived from the MCMC suggests a lower limit of $\sigma_{\rm [Fe/H]} > 0.25$  at the 95\% credible level, clearly indicating that measurement uncertainty alone cannot explain the observed spread. A resolved dispersion persists even if either the most metal-poor star or the most metal-rich star is excluded. This resolved metallicity spread provides strong evidence in favor of a dwarf galaxy classification for Aquarius~III (see \secref{classification}).

\section{Elemental Abundances from Magellan/MagE Spectroscopy}
\label{sec:mage}
\subsection{Observations and Data Reduction}
The brightest star in Aquarius~III (\Gaia DR3 2447566690779941504 in \tabref{membertable}) is sufficiently bright so as to allow for more detailed spectroscopic investigation of its chemical abundances. On the night of 2023 October 14, we obtained a longslit spectrum of this star with the Magellan Echellete (MagE) spectrograph. We used the 1$\farcs$0 slit, which provides a resolution of $\mathcal{R} \sim 4100$ over a wavelength range of $\sim 3000 \ \rm \AA $--$11000 \rm  \ \AA$. We collected $2\rm \times 2400$s exposures and $1\rm \times 1800$s exposure, for a total integration time of 6600s (1.8~hrs). ThAr frames were collected after each science exposure, and Xe-flash flats and quartz flats were taken at the beginning and end of the night. 
Each exposure was reduced individually using the MagE pipeline provided in the Carnegie Python Distribution (\texttt{CarPy}; \citealt{2000ApJ...531..159K,2003PASP..115..688K}) and subsequently normalized, and the three spectra were then coadded with inverse-variance weighting. 
\par Information-rich wavelength regions of the resultant coadded spectrum are shown in \figref{magespectrum}, including the $\sim 3850$--$4400 \rm \ \AA$ spectral region covering the Ca II  H\&K lines and CH $G$-band as well as the $5100$--$5300 \rm \ \AA$ spectral region featuring the Mg I b triplet. Even from visual inspection alone, the strength and width of these features support the classification of the star as a cool, luminous, metal-poor, and likely carbon-enhanced K giant. In the following subsections, we formalize this interpretation through an automated spectral fit as well as through synthesis of several key lines/features.

\subsection{Stellar Parameters, Metallicity, and $\alpha$-abundance }
\par We performed a fit to the star's radial velocity, effective temperature, surface gravity, iron abundance, and $\alpha$-element abundance from each MagE spectrum using \texttt{Payne4MagE}\footnote{\url{https://github.com/yupengyao/Payne4MagE}}. \texttt{Payne4MagE} is an instrument-specific wrapper for The Payne \citep{2019ApJ...879...69T}, which is a neural network-based emulator designed for constructing synthetic stellar spectra given a fixed set of labels (stellar parameters and abundances). For our application here, we simultaneously fit the spectrum from each of the three exposures over the restricted wavelength range of  $4700 \ \rm \AA$ -- $6700 \rm \ \AA$. This intentionally selected for a region of the spectrum redward of the $\rm CH$ $G$-band, which was necessary to avoid biasing the fit since the emulator used here was trained on models limited to  $\rm [C/Fe] = 0.$ Wavelengths redder than $6700 \rm \ \AA$ in our spectrum primarily consisted of either information-poor continuum or absorption lines not matched well by most spectral synthesis models (e.g., the CaT), and thus we avoided this regime as well. In the same vein, we also masked several strong absorption lines within the main fit region of $\sim 4700 \ \rm \AA - 6700$ (including $\rm H\alpha$ and $\rm H\beta$).
\par The best-fitting spectral model from the joint \texttt{Payne4MagE} fit to the three exposures was that of a cool giant with $T_{\rm eff} = 4900 \pm 130 \rm \ K$, $\log(g)/(\rm cm \ s^{-2}) = 1.9 \pm 0.37$, [Fe/H] = $ -3.2 \pm 0.11$, $[\alpha/\rm Fe] = 0.59 \pm 0.07$, where the reported uncertainties were derived from a Gaussian approximation to the empirical covariance matrix.\footnote{The effective line broadening of the MagE spectrum, which includes the contribution of both the instrumental resolution and the contribution of macroturbulence, was found to be $\sim~20 \rm \ km \ s ^{-1}$.} The iron abundance derived from the MagE spectrum and \texttt{Payne4MagE} is in excellent agreement with the EW-based CaT metallicity derived from our DEIMOS spectrum of the same star ([Fe/H] $= -3.04 \pm 0.10$). As a third and final independent estimate of the metallicity, we computed the ``KP index'' of the Ca II K line at 3933.7 $\rm \AA$ \citep{1990AJ....100..849B} following a similar procedure to \citet{2018ApJ...856..142C}. This equivalent-width-based calibration yielded an estimate of [Fe/H] = $-2.89 \pm 0.26$,  consistent with both other estimates described above. Collectively, these measurements demonstrate that Aquarius~III's brightest red giant member is at the boundary of the extremely metal poor (EMP) regime.

\subsection{Carbon and Barium Abundance}
A significant fraction of the extremely metal-poor stars in the Milky Way halo and in ultra-faint dwarf galaxies exhibit an enhancement in carbon (e.g., $\gtrsim 20$-$50\%$ below [Fe/H] = $-3.0$ have $\rm [C/Fe] > 0.7-1.0$; \citealt{2006ApJ...652L..37L,2013AJ....146..132L,2014ApJ...797...21P,2022MNRAS.515.4082A}). Within the ultra-faint dwarfs in particular, this carbon enhancement is often paired with a deficiency in neutron-capture elements -- the so-called CEMP-no pattern \citep[e.g.,][]{2005ARA&A..43..531B} The CEMP-no stars in UFDs may be the descendants of faint Population III supernova, and thus these stars offer an insightful window into the chemical evolution of their host galaxies at early times \citep[e.g.,][]{2021MNRAS.502....1J}. On the other hand, $\sim 80\%$ of halo CEMP stars outside known dwarf galaxies are CEMP-$s$ stars thought to be the products of mass transfer from an AGB star  binary companion  \citep[e.g.,][]{2005ARA&A..43..531B,2007ApJ...655..492A}.
\par To test whether Aquarius~III's brightest star is carbon-enhanced, we compared our observed MagE spectrum against synthetic spectra generated using the Julia-based spectral synthesis package \texttt{Korg} \citep{2023AJ....165...68W,2024AJ....167...83W}. \texttt{Korg} interpolates from a grid of MARCS model atmospheres and generates synthetic spectra under the assumption of 1D local thermodynamic equilibrium. For our purposes, we adopted a model atmosphere with the temperature and surface gravity from the \texttt{Payne4Mage} fit, [M/H] = $-2.5$ (the lowest available in  \texttt{Korg}'s MARCS atmosphere grid at the time of our analysis), and [$\alpha$/Fe] = +0.6. We then synthesized spectra with varied carbon abundances assuming the same  $T_{\rm eff}$, log($g$), and a marginally higher iron abundance than derived above, [Fe/H] = $-2.8$, which was found to better match the strength of the observed iron lines. The best-fitting carbon abundance was found to be a model with [C/Fe] $= +1.4$ for the star assuming this higher metallicity. We then added a 0.08 dex evolutionary correction based on the online calculator associated with \citet{2014ApJ...797...21P}\footnote{\url{https://vplacco.pythonanywhere.com/}} to account for the surface carbon depletion associated with CN cycling on the RGB. Adopting a conservative of uncertainty of $\pm 0.3$~dex motivated by our use of a model atmosphere at a higher metallicity, our final measurement is $\rm [C/Fe] = 1.48 \pm 0.3$. We therefore conclude that the  star is consistent with a CEMP classification. 

\begin{figure*}
    \centering
    \includegraphics[width = \textwidth]{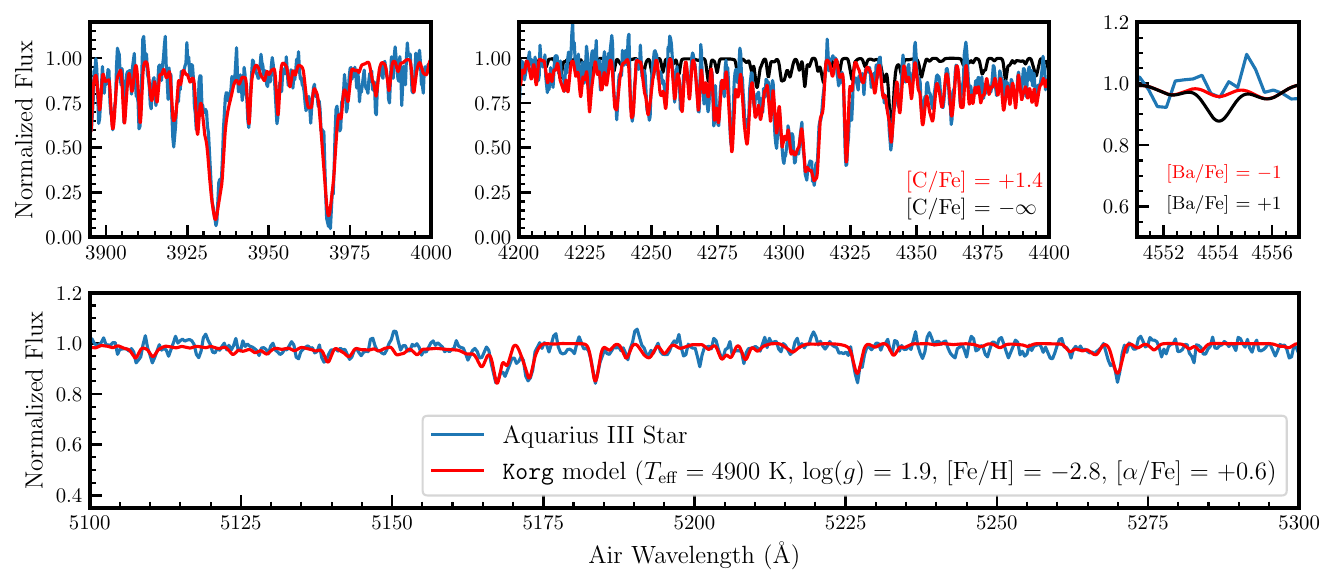}
    \caption{Important wavelength regions of our MagE spectrum for the brightest star in Aquarius~III (\Gaia DR3 2447566690779941504) compared against a spectral synthesis model from \texttt{Korg}. In each panel, the observed spectrum is shown in blue while a matched synthetic spectrum, smoothed to $R = 4100$, is shown in red. (Top Row, Left Panel) The Ca II H $\&$ K lines. (Top Row, Center Panel) The CH $G$-band, compared to a model with [C/Fe] = +1.4 (red) and no Carbon (black).  (Top Row, Right) Narrow wavelength range covering the strong Ba II $\lambda \rm  4554 \ \AA$ resonance line. A model with [Ba/Fe] = +1 is shown in black. Although the data is noisy, this Ba-enhanced model is clearly disfavored by the non-detection of this line.   (Bottom panel) The wavelength range $5100$--$5300 \rm \ \AA$, which most notably includes the Mg I b triplet as well as a number of iron lines. }
    \label{fig:magespectrum}
\end{figure*}

In an attempt to distinguish between the CEMP-s and CEMP-no scenarios for this star, we next explored whether it was possible to measure Barium (an $s$-process element) from the Ba II $\lambda \rm  4554 \ \AA$ line or Strontium (a predominantly $r$-process element for metal-poor stars) from the Sr II $\lambda 4077 \rm \ \AA$ line. We ultimately concluded that a reliable quantitative measurement could not be made from either line due to the low resolution and somewhat low $S/N$ of our MagE spectrum. This being said, we see no evidence for strong absorption at the expected wavelength for any of the strong Ba lines covered by our spectrum (see \figref{magespectrum} for the Ba II $\lambda \rm  4554 \ \AA$ resonance line). Comparison to \texttt{Korg} models with varied barium abundances allows us to rule out  [Ba/Fe] $> +1$ -- excluding the CEMP-s possibility for this star. This lack of $s$-process enhancement therefore favors a CEMP-no classification for this star.
\par In summary, our MagE spectrum indicates that the brightest star in Aquarius~III is a carbon-enhanced, metal poor star at the boundary of the extremely metal poor regime ([Fe/H] $\approx -3$) that is most consistent with a CEMP-no classification. Our results derived here from spectral synthesis should be regarded as indicative as opposed to a precise quantitative analysis, and a higher-resolution, higher-$S/N$ spectrum will be necessary to further ascertain this star's nature.

\section{Discussion}
\label{sec:disc}

\subsection{Classification of Aquarius~III as a Dwarf Galaxy}
\label{sec:classification}
Recent discoveries of ultra-faint systems in the Milky Way halo have generally fallen into two broad categories: extended, dark-matter-dominated ultra-faint dwarf galaxies and compact, low-luminosity systems generally presumed to be baryon-dominated ultra-faint star clusters. We find that Aquarius~III is more consistent with the former class of satellites on the basis of its size and metallicity distribution. 
\par In detail, Aquarius~III's physical size of $r_{1/2} = 41^{+9}_{-8}$~pc is larger than all known Milky Way globular clusters  (see left panel of \figref{pop});  even if we adopted the lower bound of our $1\sigma$ credible interval the only comparably extended clusters fall several magnitudes brighter and at much higher surface brightness. With respect to its metallicity distribution, Aquarius~III's metallicity dispersion of $\sigma_{\rm [Fe/H]} = 0.46^{+0.26}_{-0.14}$~dex is clearly non-zero. This not only points to multiple generations of star formation in the system but also suggests that Aquarius~III inhabits a dark matter halo with a gravitational potential deep enough to retain supernova ejecta \citep{2012AJ....144...76W}. This statement is not in tension with our finding of a low velocity dispersion, as dynamical masses within the half-light radius as large as $5.1 \times 10^5 M_{\odot}$ are permitted by the current kinematic observations (at the 95\% credible level). Lastly, but perhaps least persuasively, we note that Aquarius~III's mean metallicity is closely consistent with the expectation from the luminosity-metallicity relation for dwarf galaxies \citep{2013ApJ...779..102K}, and the metallicity of its brightest star -- as measured through several independent techniques -- is more metal poor than any known star in any known intact globular cluster.  This star's combination of a low barium abundance and strong carbon enhancement also matches the enrichment pattern commonly seen in very-metal-poor ultra-faint dwarf galaxy stars \citep{2019ApJ...870...83J}.

\subsection{Proper Motion and Orbit}
\label{sec:orbit}

The brightest two spectroscopically-confirmed members in Aquarius~III each have a proper motion measurement reported in \Gaia Data Release 3 \citep{2016A&A...595A...1G,2023A&A...674A...1G}, enabling us to measure the galaxy's systemic proper motion.\footnote{For completeness, the \Gaia DR3 proper motions for these two stars are $(1.03\pm0.27,-0.20\pm0.21) \rm \ mas \ yr^{-1} \rm \ and \ (1.07\pm0.70,	0.48\pm0.55)   \ mas \ yr^{-1}$. Both stars have high-quality astrometric solutions as quantified by the \texttt{fidelity\_v2} classifier introduced by \citet{2022MNRAS.510.2597R}.} We performed a simple two-parameter MCMC fit constraining the Aquarius~III's systemic proper motion in right ascension and declination ($\mu_{\alpha*}$ and $\mu_{\delta}$, respectively) based on the likelihood presented in Equations 3 and 4 of \citet{2019ApJ...875...77P}. We applied no priors and assumed no intrinsic proper motion dispersion in either component. Posterior probability distributions for each component were derived using \texttt{emcee}, from which we estimated $\mu_{\alpha *} = 1.01 \pm 0.25 \rm \ mas \ yr^{-1}$ and $ \mu_{\delta} = -0.10 \pm 0.20 \rm \ mas \ yr^{-1}$.  These estimates and uncertainties account for the covariance between the proper motion components for a given star; however, we neglected the $\mathcal{O}(0.02 \rm \ mas \ yr^{-1}$) spatially-covariant systematic errors discussed by \citet{2021A&A...649A...2L} as they are subdominant. 

\par With complete 6D phase-space information in-hand, we integrated 5000 realizations of Aquarius-III's orbit using the \texttt{galpy} Python package \citep{2015ApJS..216...29B}. Initial conditions for these realizations were generated by sampling directly from the posterior probability distributions on each parameter ($\alpha, \delta, D_{\odot}, \mu_{\alpha}, \mu_{\delta}, v_{\rm hel.}$).  For each set of initial conditions, we rewound Aquarius~III's orbit for the last 3 Gyr in the static, axisymmetric \citet{2017MNRAS.465...76M} potential model. This is a six-component model including a bulge, thin and a thick stellar disk, an atomic and molecular gaseous disk, and a Navarro-Frenk-White dark matter halo \citep{1997ApJ...490..493N}, all summing to a total virial mass of $1.3 \times 10^{12} M_{\odot}$.  At the conclusion of each integration, we recorded Aquarius~III's orbital properties including its orbital eccentricity ($e$) apocentric and pericentric radii ($r_{\rm apo}$ and $r_{\rm peri}$,  respectively), total orbital energy per unit mass ($E_{\rm tot}$), and the $Z$ component of its angular momentum ($L_Z$). Throughout, we adopted a right-handed Galactocentric coordinate frame with the Solar distance from the Galactic center and the corresponding circular velocity set to the 
properties of the best-fitting potential model from \citet{2017MNRAS.465...76M}, namely $R_0 = 8.21$~kpc and $v_{\rm circ} = 233.1 \rm \ km \ s^{-1}$. We further assumed the Solar peculiar motion about the Local Standard of Rest from \citet{2010MNRAS.403.1829S}.

\begin{figure*}
    \centering
    \includegraphics[width=\textwidth]{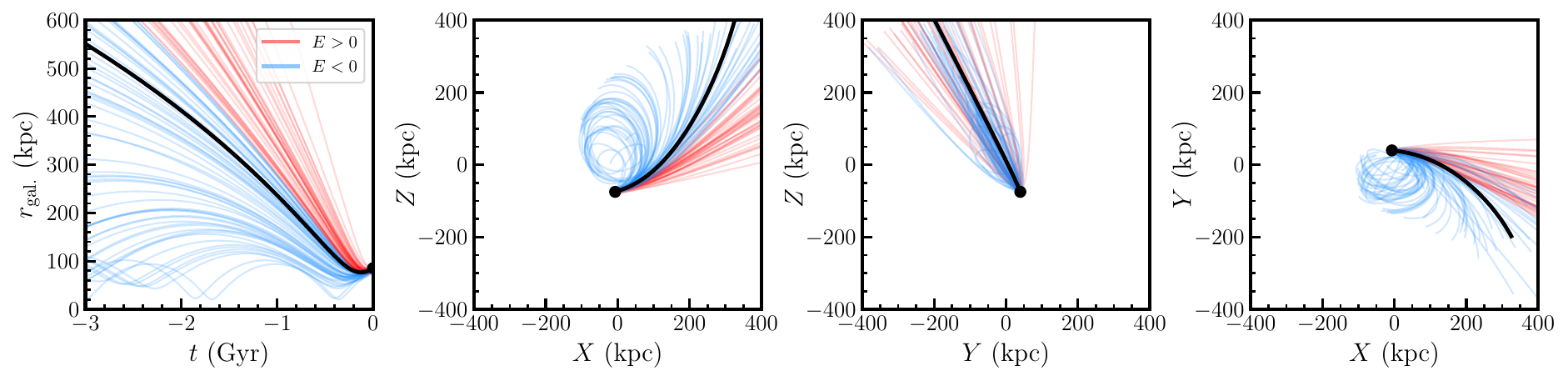}
    \caption{Projections of Aquarius~III's orbit for the last 3 Gyr in the \citet{2017MNRAS.465...76M} Milky Way potential.  In each panel, the solid black curve depicts a ``fiducial'' orbit assuming the median positions and velocities in \tabref{properties}, and the blue (red) orbits depict bound (unbound) realizations randomly sampled from our parent sample of 5000 realizations (100 random samples in total).  Aquarius~III's current position is depicted in each panel by a black point. (Left) Galactocentric radius $(r_{\rm gal.})$ vs. time ($t$), where $t=0$ denotes the present-day. (Center Left) Galactocentric $Z$ vs $X$, i.e., an edge on view of disk plane. (Center Right)  Galactocentric $Z$ vs $Y$. (Right)  Galactocentric $Y$ vs $X$, i.e., the plane parallel to the disk. Note that we have truncated the axis limits in all four panels for sake of visualization.}
    \label{fig:orbit}
\end{figure*}

\par In \figref{orbit}, we display four different representations of Aquarius~III's orbital history for the last 3 Gyr. Each panel shows 100 randomly-drawn realizations where realizations
corresponding to bound orbits $(E_{\rm tot} < 0)$  are shown in blue and realizations corresponding to unbound orbits ($E_{\rm tot} > 0$) are shown in red. These represent $\sim 61\%$ and $\sim 39\%$ of cases, respectively. In black, we also depict a ``fiducial'' orbit corresponding to the case wherein Aquarius~III's present-day phase-space properties were set to be exactly equal to the best-fit values quoted in \tabref{properties}. Broadly,  our results favor a scenario in which Aquarius~III is orbiting retrograde with respect to the Milky Way disk ($L_Z = (13^{+3}_{-5}) \times  \rm 10^3 \ \text{kpc km s}^{-1}$, in our right-handed coordinate frame) and has passed its orbital pericenter within the last $\sim 250$~Myr. Aquarius~III's Galactocentric radius at this recent pericentric passage is relatively tightly constrained to $r_{\rm peri} = 78 \pm 7$~kpc. By contrast, its orbital apocenter is very poorly constrained: of the 61\% of orbits that are bound, about half have no apocentric passages within the last $3$~Gyr while the remainder are distributed across a wide range of possible apocenters.  
\par Although our fiducial orbit suggests Aquarius III is a bound satellite, these results are fully consistent with a scenario in which Aquarius~III is on first infall onto the Milky Way, as is believed to the case for a significant fraction of the known ultra-faint dwarfs \citep{2018ApJ...863...89S, 2021ApJ...922...93H}. The possibility of first infall led us to explore whether Aquarius~III is consistent with having accreted with the Large and Small Magellanic Clouds (LMC and SMC, respectively), which themselves are believed to be on their first infall \citep{2010ApJ...721L..97B,2013ApJ...764..161K}. This hypothesis is supported by Aquarius~III's projected proximity to the trailing arm of the Magellanic Stream's HI gas component and its 3D proximity to stars comprising the trailing portion of the recently reported ``Magellanic Stellar Stream'' \citep[MSS;][]{2023ApJ...956..110C}. However, despite this positional similarity, we found that Aquarius~III's 3D kinematics are inconsistent with a Magellanic association: its retrograde orbit immediately disfavors an association, and more convincingly, its orbital angular momentum about the Galactocentric $Y$-axis is clearly inconsistent with that of the claimed MSS debris stars ($L_Y = (24^{+7}_{-8}) \times 10^{3} \rm \ kpc \ km \ s^{-1}$, compared to the selection  $L_Y < 5 \times 10^3 \rm \ kpc \ km \ s^{-1}$ from \citealt{2023ApJ...956..110C}). This kinematic inconsistency manifests most clearly in the center-right panel of \figref{orbit}, where Aquarius~III's orbit has it approaching the Milky Way disk from above the disk plane (positive $Z$), in contrast to the Clouds which are infalling toward the Milky Way from below the disk plane (negative $Z$). We conclude that an association between Aquarius~III and the LMC/SMC is highly improbable. 
\par This all being said, we caution that our proper motion measurement is quite uncertain owing to both the small number of stars in Aquarius~III with \Gaia proper motions (just two) as well as the large proper motion uncertainties on each of these stars.  Furthermore, we note that orbit integrations performed based on uniformly sampling from a proper motion measurement posterior distribution with large errors can bias orbital history inferences toward the case of an eccentric orbit with the satellite near its pericenter (see Section 5.2 of \citealt{2022MNRAS.511.2610C}). This is because these uniformly-drawn samples will favor values for the tangential velocity that are \textit{larger} in magnitude than the true tangential velocity, whereas these possible orbits are in reality far less likely to be ``sampled'' from the host halo's distribution function than bound, low-velocity orbits \citep{2022MNRAS.511.2610C}. A proper treatment involving weighting samples by the distribution function is beyond the scope of this work, and for now we simply conclude that improved proper motion measurements (e.g., from future \textit{Gaia} data releases) will provide a more precise and accurate constraint on Aquarius~III's orbital history. As we describe in the following section, a secure determination of Aquarius~III's orbit will be important for interpreting its cold velocity dispersion.
\subsection{A Kinematically Cold, Tidally-Unscathed Dark Matter Halo for Aquarius~III?}
\label{sec:cold}
One consequence of Aquarius~III's plausible first-infall orbital history and large pericenter distance ($r_{\rm peri} = 78 \pm 7$~kpc) is that the galaxy very likely has not experienced significant tidal mass loss induced by the Milky Way disk. This permits the possibility that Aquarius~III's low present-day halo mass is a reasonably faithful tracer of its halo mass at the time of its formation, i.e., that it formed in a very-low-mass dark matter halo. This would be a stronger statement than has been previously possible with other faint dwarf satellites with strong velocity dispersion limits -- Segue~2, Tucana~III,  and Draco~II ($\sigma_v < $\ [2.6, 1.5, 5.9] $\rm \ km \ s^{-1}$, respectively) --  all of which display signatures of mass loss or disruption and are situated in the inner halo \citep{2013ApJ...770...16K,  2017ApJ...838...11S, 2018MNRAS.480.2609L}. Likewise, three additional ultra-faint dwarfs with strong velocity dispersion upper limits, namely Grus~II, Triangulum II, and Tucana~V  ($\sigma_v < [3.4, 2.0, 3.1] \rm \ km \ s^{-1}$; \citealt{2020ApJ...892..137S,2022MNRAS.514.1706B,2024ApJ...968...21H}) each have orbital pericenters that leave them vulnerable to the disk's tidal influence, even if direct evidence for disruption has not yet been observed. In each of these cases -- barring Aquarius~III -- tidal mass loss induced by the Milky Way disk remains a viable explanation for the low observed velocity dispersions. Indeed, for Tucana III this is a near certainty: $N$-body simulations suggest that extreme tidal dark matter loss in satellites precedes substantial stellar mass loss \citep{2008ApJ...673..226P}, and thus the presence of clear tidal tails in Tucana~III's stellar component \citep{2016ApJ...833L...5D,2018ApJ...866...22L} points to severe mass loss in its dark matter component. In short, then, Aquarius~III stands alone among the known Milky Way satellite population in that it appears both very kinematically cold and tidally unscathed by the Milky Way disk. 
\par  The combination of these properties potentially makes Aquarius III an interesting laboratory for studying galaxy formation physics in low-mass dark matter halos.  If its velocity dispersion is confirmed to be $\sigma_v \lesssim 2 \rm \ km \ s^{-1}$ based on additional kinematic observations, Aquarius~III's implied  halo mass will likely fall in the low $10^8 \ M_{\odot}$ regime. The galaxy occupation fraction in this regime (and at lower halo masses) is sensitive to the available channels for gas cooling and to the timing of cosmic reionization \citep[e.g.,][]{1992MNRAS.256P..43E,2002ApJ...572L..23S,2020MNRAS.498.4887B,2023MNRAS.524.2290N}, and whether these halos are expected to exist in the first place is sensitive to the nature of dark matter. Put succinctly, confirmation of a galaxy that formed in such a low-mass halo would almost certainly require the need for $\rm H_2$ cooling and would disfavor cosmological models with early reionization \citep{2022MNRAS.516.3944M,2023arXiv230813599A}. Moreover, such a galaxy would disfavor warm dark matter particle models which suppress the small-scale power spectrum at masses approaching $\sim 10^8 M_{\odot}$ (e.g., warm dark matter particles with masses $\lesssim  $5~keV). Similar conclusions have previously been drawn at the satellite population level based on abundance matching \citep[e.g.,][]{2018MNRAS.473.2060J,2021JCAP...08..062N,2021PhRvL.126i1101N} but the robust confirmation of an individual non-disrupting galaxy in this halo mass range would be a novel and clean confirmation of these constraints. 
\par One caveat to the above is that it remains possible that Aquarius III has lost mass through ``preprocessing'' in a \textit{different} group environment prior to infall onto the Milky Way \citep[see e.g.,][for useful context]{10.1111/j.1365-2966.2009.15507.x,2013MNRAS.432..336W,2014MNRAS.442..406H,2019MNRAS.483..235J}, which would complicate the interpretation of its low velocity dispersion. Illustratively, \citet{2015ApJ...807...49W} studied Milky Way/M31-like pairs in the ELVIS simulations and report that $\sim 50\%$ of satellites of comparable stellar mass to Aquarius III resided in another massive host halo prior to infall. These groups of galaxies/subhalos are commonly disrupted at time of infall \citep{2015ApJ...807...49W}, and thus we are fundamentally limited in our ability to assess whether preprocessing has occurred for Aquarius III.
 \par In any case, Aquarius III's status as one of the Milky Way's least massive confirmed satellite~galaxies yet identified makes it an appealing target for future follow-up. Specifically, higher-precision radial velocity measurements -- ideally from an expanded member sample -- are a critical next step before any actual constraints can be made based on Aquarius~III's stellar kinematics. Expanding the sample of spectroscopic members will be challenging given that our current Keck/DEIMOS sample already nearly reaches the base of the RGB at $g_0 \sim 23$. However, even a second equal-depth epoch of Keck/DEIMOS observations covering the same members may be sufficient to place a strong enough dispersion limit to settle whether Aquarius~III truly is an unprecedentedly-low-mass galaxy. A second, high precision epoch would also enable the identification and removal of short-period spectroscopic binaries, which could be either inflating or deflating the velocity dispersion constraint derived from our current single-epoch dataset \citep[e.g.,][]{2010ApJ...722L.209M,2010ApJ...721.1142M,2022ApJ...939....3P,2023ApJ...956...91W}

\subsection{(Limits on) Astrophysical $J$-Factor}
\label{sec:jfactor}
The Milky Way's ultra-faint dwarf satellites are excellent targets in the search for gamma-ray emission associated with annihilating dark matter owing to their high dark matter densities and relatively minimal baryonic components \citep{2015PhRvL.115w1301A, 2015PhRvD..91h3535G,2023arXiv231104982M,2024PhRvD.109j3007B}. The astrophysical component of the expected signal from dark matter annihilation is quantified in the J-factor ($J$), which represents the integral over the line of sight of the dark matter density squared,
\[ J(\theta) = \underset{\mathrm{l.o.s}}{\iint }\, \rho _{\mathrm{DM}}^2 (r) \, \mathrm{d}\ell \mathrm{d}\Omega\]
within a solid angle $\Omega$ of angular radius $\theta$.  The standard method for computing the $J$-factor is to infer the dark matter density profile by comparing solutions to the spherical Jeans equations to the measured velocity dispersion \citep[e.g.,][]{Bonnivard2015MNRAS.453..849B, GeringerSameth2015ApJ...801...74G, 10.1093/mnras/sty2839}. To apply this approach to Aquarius III, we followed the Jeans modelling implementation  described by \citet{10.1093/mnras/sty2839}, which assumes that the velocity anisotropy was constant with radius. Taking a prior on the maximum circular velocity of $v_{\rm max} > 1 \rm \ km \ s^{-1}$, this yielded an estimate for Aquarius~III's $J$-factor within a solid angle of radius $\theta = 0.5\deg$ (hereafter $J(0.5\degree))$ of  $\log10(J(0.5\degree)/(\rm GeV^2 \ cm^{-5}) < 17.7$. As a simple check on this result, we also estimated Aquarius~III's $J$-factor using the empirical scaling relation from \citet{10.1093/mnras/sty2839},
% \small
\[ \frac{J(0.5\degree)}{\rm GeV^2 \ cm^{-5}} \approx 10^{17.87} \left(\frac{\sigma_v}{5 \rm \ km \ s^{-1}}\right)^4 \left(\frac{D_{\odot}}{100 \rm \ kpc}\right)^{-2} \left(\frac{r_{1/2}}{100 \rm \ pc}\right)^{-1}.\] 
% \normalsize 
Neglecting uncertainties on this relation itself, this calculation yielded an upper limit on Aquarius~III's logarithmic $J$-factor of $\log10(J(0.5)/(\rm GeV^2 \ cm^{-5}) < 17.8 $ in good agreement with the full Jeans modelling treatment. 
\par Owing to Aquarius~III's small velocity dispersion and large distance, this upper limit on the galaxy's estimated $J$-factor is quite stringent and places Aquarius~III in the bottom third of ultra-faint dwarf galaxy $J$-factors (see e.g., Fig. 11 of \citealt{2024ApJ...961..234H}).  We therefore conclude that Aquarius~III will not meaningfully contribute to indirect-detection constraints even within the context of stacked analyses.

\section{Summary and Conclusions}
\label{sec:conc}
We have presented the discovery of Aquarius~III, a low-luminosity Milky Way halo satellite identified in DECam imaging data processed by DELVE. Follow-up imaging from DECam clearly established it as a \textit{bona fide} ultra-faint stellar system in the outer Milky Way halo. Multi-object spectroscopy from Keck/DEIMOS then enabled measurement of its stellar kinematics and metallicity distribution from a parent sample of 11 member stars, affirming its nature as a metal-poor dwarf galaxy - albeit a low-mass one with no currently-resolvable velocity dispersion. An additional longslit spectrum of Aquarius~III's brightest star from Magellan/MagE revealed it to be carbon-enhanced, $\alpha$-element enhanced, extremely metal-poor star worthy of follow-up with higher-resolution spectroscopy. \Gaia proper motions for Aquarius~III's two brightest stars allowed us to explore its orbital history, from which we concluded that the galaxy is on a retrograde orbit having recently passed its pericenter in the outer halo. Lastly, we synthesized the sum total of these measurements to argue that Aquarius~III may represent a clean example of a galaxy that formed in a halo at $M_{\rm peak} \sim 10^8 M_{\odot}$ -- a hypothesis that can be tested with additional spectroscopic observations.

\par Looking to the future, the unprecedentedly deep, wide-field data from the upcoming Vera C. Rubin Observatory and its Legacy Survey of Space and Time will yield a deluge of dozens to hundreds of faint Local Group satellites in the southern sky \citep[e.g.,][]{2014ApJ...795L..13H,2018MNRAS.479.2853N, 2022MNRAS.516.3944M}. Aquarius~III is, in many ways, emblematic of the Milky Way dwarf satellites that Rubin will discover in great number: low-luminosity, distant galaxies with at most a few dozen red giant branch stars (see e.g., Figures 8 and 9 of \citealt{2019ARA&A..57..375S}). Spectroscopically characterizing these systems is costly - here, for example, requiring a half night on a 10m-class telescope - yet these data are indispensible for measuring their dynamical masses and metallicities and thereby solidly establishing their classifications as dwarf galaxies. We therefore emphasize that capitalizing on these discoveries from Rubin will require a major investment of time on large-aperture telescopes for multi-object spectroscopy. Only with these data can we maximize these galaxies' utility as laboratories for testing models of galaxy formation and the small-scale structure predictions of $\Lambda$CDM cosmology.

\section{Data Availability}
\label{sec:opendata}
All code and data products associated with this work are archived on Zenodo and will be made available at time of publication (or upon request to the corresponding author before that time). This repository includes our deeper DECam catalog derived from image-level coadds, our DEIMOS spectroscopic catalog including both members and non-members and an expanded range of columns, our fully reduced and normalized MagE spectrum, and all MCMC chains produced by our analyses here. The raw DEIMOS spectra are available through the Keck Observatory archive.\footnote{\url{koa.ipac.caltech.edu}}

\section{Acknowledgments}

\par This work was partially supported by the National Science Foundation under Grant No. AST-2108168 and AST-2307126. This material is based upon work supported by the National Science Foundation Graduate Research Fellowship Program under Grant No. DGE-2139841. Any opinions, findings, and conclusions or recommendations expressed in this material are those of the author(s)
and do not necessarily reflect the views of the National Science Foundation.  This project is partially supported by the NASA Fermi Guest Investigator Program Cycle 9 No. 91201. This work is partially supported by Fermilab LDRD project
L2019-011.
\par W.C. gratefully acknowledges support from a Gruber Science Fellowship at Yale University. DJS acknowledges support from NSF grants AST-1821967, 1813708 and AST-2205863.
JLC acknowledges support from NSF grant AST-1816196. G.E.M. acknowledges support from the University of Toronto Arts \& Science Post-doctoral Fellowship program, the Dunlap Institute, and the Natural Sciences and Engineering Research Council of Canada (NSERC) through grant RGPIN-2022-04794. JAC-B acknowledges support from FONDECYT Regular N 1220083. C.E.M.-V. is supported by the international Gemini Observatory, a program of NSF NOIRLab, which is managed by the Association of Universities for Research in Astronomy (AURA) under a cooperative agreement with the U.S. National Science Foundation, on behalf of the Gemini partnership of Argentina, Brazil, Canada, Chile, the Republic of Korea, and the United States of America. A.B.P. acknowledges support from NSF grant AST-1813881. I.U.R.\ acknowledges support from the NASA Astrophysics Data Analysis Program, grant 80NSSC21K0627, and the NSF, grant AST~2205847.
\par Some of the data presented herein were obtained at Keck Observatory, which is a private 501(c)3 non-profit organization operated as a scientific partnership among the California Institute of Technology, the University of California, and the National Aeronautics and Space Administration. The Observatory was made possible by the generous financial support of the W. M. Keck Foundation. The authors wish to recognize and acknowledge the very significant cultural role and reverence that the summit of Maunakea has always had within the Native Hawaiian community. We are most fortunate to have the opportunity to conduct observations from this mountain. This research has also made use of the Keck Observatory Archive (KOA), which is operated by the W. M. Keck Observatory and the NASA Exoplanet Science Institute (NExScI), under contract with the National Aeronautics and Space Administration.
\par This paper includes data gathered with the 6.5 meter Magellan Telescopes located at Las Campanas Observatory, Chile. 
This project used data obtained with the Dark Energy Camera, which was constructed by the Dark Energy Survey (DES) collaboration.
Funding for the DES Projects has been provided by 
the DOE and NSF (USA),   
MISE (Spain),   
STFC (UK), 
HEFCE (UK), 
NCSA (UIUC), 
KICP (U. Chicago), 
CCAPP (Ohio State), 
MIFPA (Texas A\&M University),  
CNPQ, 
FAPERJ, 
FINEP (Brazil), 
MINECO (Spain), 
DFG (Germany), 
and the collaborating institutions in the Dark Energy Survey, which are
Argonne Lab, 
UC Santa Cruz, 
University of Cambridge, 
CIEMAT-Madrid, 
University of Chicago, 
University College London, 
DES-Brazil Consortium, 
University of Edinburgh, 
ETH Z{\"u}rich, 
Fermilab, 
University of Illinois, 
ICE (IEEC-CSIC), 
IFAE Barcelona, 
Lawrence Berkeley Lab, 
LMU M{\"u}nchen, and the associated Excellence Cluster Universe, 
University of Michigan, 
NSF's National Optical-Infrared Astronomy Research Laboratory, 
University of Nottingham, 
Ohio State University, 
OzDES Membership Consortium
University of Pennsylvania, 
University of Portsmouth, 
SLAC National Lab, 
Stanford University, 
University of Sussex, 
and Texas A\&M University.

Based on observations at NSF Cerro Tololo Inter-American Observatory, NSF NOIRLab (NOIRLab Prop. ID 2019A-0305; PI: Drlica-Wagner), which is managed by the Association of Universities for Research in Astronomy (AURA) under a cooperative agreement with the U.S. National Science Foundation.

The Legacy Surveys consist of three individual and complementary projects: the Dark Energy Camera Legacy Survey (DECaLS; Proposal ID 2014B-0404; PIs: David Schlegel and Arjun Dey), the Beijing-Arizona Sky Survey (BASS; NOAO Prop. ID 2015A-0801; PIs: Zhou Xu and Xiaohui Fan), and the Mayall z-band Legacy Survey (MzLS; Prop. ID 2016A-0453; PI: Arjun Dey). DECaLS, BASS and MzLS together include data obtained, respectively, at the Blanco telescope, Cerro Tololo Inter-American Observatory, NSF’s NOIRLab; the Bok telescope, Steward Observatory, University of Arizona; and the Mayall telescope, Kitt Peak National Observatory, NOIRLab. Pipeline processing and analyses of the data were supported by NOIRLab and the Lawrence Berkeley National Laboratory (LBNL). The Legacy Surveys project is honored to be permitted to conduct astronomical research on Iolkam Du’ag (Kitt Peak), a mountain with particular significance to the Tohono O’odham Nation.

NOIRLab is operated by the Association of Universities for Research in Astronomy (AURA) under a cooperative agreement with the National Science Foundation. LBNL is managed by the Regents of the University of California under contract to the U.S. Department of Energy.

BASS is a key project of the Telescope Access Program (TAP), which has been funded by the National Astronomical Observatories of China, the Chinese Academy of Sciences (the Strategic Priority Research Program “The Emergence of Cosmological Structures” Grant \# XDB09000000), and the Special Fund for Astronomy from the Ministry of Finance. The BASS is also supported by the External Cooperation Program of Chinese Academy of Sciences (Grant \# 114A11KYSB20160057), and Chinese National Natural Science Foundation (Grant \# 12120101003, \# 11433005).

The Legacy Survey team makes use of data products from the Near-Earth Object Wide-field Infrared Survey Explorer (NEOWISE), which is a project of the Jet Propulsion Laboratory/California Institute of Technology. NEOWISE is funded by the National Aeronautics and Space Administration.

The Legacy Surveys imaging of the DESI footprint is supported by the Director, Office of Science, Office of High Energy Physics of the U.S. Department of Energy under Contract No. DE-AC02-05CH1123, by the National Energy Research Scientific Computing Center, a DOE Office of Science User Facility under the same contract; and by the U.S. National Science Foundation, Division of Astronomical Sciences under Contract No. AST-0950945 to NOAO.

This work has made use of data from the European Space Agency (ESA) mission {\it Gaia} (\url{https://www.cosmos.esa.int/gaia}), processed by the {\it Gaia} Data Processing and Analysis Consortium (DPAC, \url{https://www.cosmos.esa.int/web/gaia/dpac/consortium}).
Funding for the DPAC has been provided by national institutions, in particular the institutions participating in the {\it Gaia} Multilateral Agreement.

\par This work made use of Astropy:\footnote{\url{http://www.astropy.org}} a community-developed core Python package and an ecosystem of tools and resources for astronomy.

\par This work has made use of the Local Volume Database\footnote{\url{https://github.com/apace7/local_volume_database}}.

This manuscript has been authored by Fermi Research Alliance, LLC, under contract No.\ DE-AC02-07CH11359 with the US Department of Energy, Office of Science, Office of High Energy Physics. The United States Government retains and the publisher, by accepting the article for publication, acknowledges that the United States Government retains a non-exclusive, paid-up, irrevocable, worldwide license to publish or reproduce the published form of this manuscript, or allow others to do so, for United States Government purposes.

\facility{Blanco, \Gaia, Keck:II (DEIMOS)}

\software{\code{numpy} \citep{2011CSE....13b..22V,2020Natur.585..357H}, \code{scipy} \citep{2020NatMe..17..261V}, \emcee \citep{Foreman-Mackey:2013}, \healpix \citep{2005ApJ...622..759G},\footnote{\url{http://healpix.sourceforge.net}} \code{healpy} \citep{2019JOSS....4.1298Z} , \ugali \citep{Bechtol:2015} \footnote{\url{https://github.com/DarkEnergySurvey/ugali}}, \code{ChainConsumer} \citep{2019ascl.soft10017H}, \code{simple} \citep{Bechtol:2015,2015ApJ...813..109D}, \code{astropy} \citep{2013A&A...558A..33A,2018AJ....156..123A,2022ApJ...935..167A}, \code{Julia} \citep{2014arXiv1411.1607B}, \code{Korg} \citep{2024AJ....167...83W}}

\bibliography{main}

\begin{thebibliography}{}
\expandafter\ifx\csname natexlab\endcsname\relax\def\natexlab#1{#1}\fi
\providecommand{\url}[1]{\href{#1}{#1}}

\bibitem[{Abbott {et~al.}(2021)Abbott, Adamów, Aguena, Allam, Amon, Annis, Avila, Bacon, Banerji, Bechtol, Becker, Bernstein, Bertin, Bhargava, Bridle, Brooks, Burke, Rosell, Kind, Carretero, Castander, Cawthon, Chang, Choi, Conselice, Costanzi, Crocce, da~Costa, Davis, Vicente, DeRose, Desai, Diehl, Dietrich, Drlica-Wagner, Eckert, Elvin-Poole, Everett, Evrard, Ferrero, Ferté, Flaugher, Fosalba, Friedel, Frieman, García-Bellido, Gaztanaga, Gelman, Gerdes, Giannantonio, Gill, Gruen, Gruendl, Gschwend, Gutierrez, Hartley, Hinton, Hollowood, Honscheid, Huterer, James, Jeltema, Johnson, Kent, Kron, Kuehn, Kuropatkin, Lahav, Li, Lidman, Lin, MacCrann, Maia, Manning, Maloney, March, Marshall, Martini, Melchior, Menanteau, Miquel, Morgan, Myles, Neilsen, Ogando, Palmese, Paz-Chinchón, Petravick, Pieres, Plazas, Pond, Rodriguez-Monroy, Romer, Roodman, Rykoff, Sako, Sanchez, Santiago, Scarpine, Serrano, Sevilla-Noarbe, Smith, Smith, Soares-Santos, Suchyta, Swanson, Tarle, Thomas, To, Tremblay, Troxel, Tucker,
  Turner, Varga, Walker, Wechsler, Weller, Wester, Wilkinson, Yanny, Zhang, Nikutta, Fitzpatrick, Jacques, Scott, Olsen, Huang, Herrera, Juneau, Nidever, Weaver, Adean, Correia, de~Freitas, Freitas, Singulani, Vila-Verde, \& Server)}]{Abbott_2021}
Abbott, T. M.~C., Adamów, M., Aguena, M., {et~al.} 2021, The Astrophysical Journal Supplement Series, 255, 20.
\newblock \url{https://dx.doi.org/10.3847/1538-4365/ac00b3}

\bibitem[{{Ackermann} {et~al.}(2015){Ackermann}, {Albert}, {Anderson}, {Atwood}, {Baldini}, {Barbiellini}, {Bastieri}, {Bechtol}, {Bellazzini}, {Bissaldi}, {Blandford}, {Bloom}, {Bonino}, {Bottacini}, {Brandt}, {Bregeon}, {Bruel}, {Buehler}, {Caliandro}, {Cameron}, {Caputo}, {Caragiulo}, {Caraveo}, {Cecchi}, {Charles}, {Chekhtman}, {Chiang}, {Chiaro}, {Ciprini}, {Claus}, {Cohen-Tanugi}, {Conrad}, {Cuoco}, {Cutini}, {D'Ammando}, {de Angelis}, {de Palma}, {Desiante}, {Digel}, {Di Venere}, {Drell}, {Drlica-Wagner}, {Essig}, {Favuzzi}, {Fegan}, {Ferrara}, {Focke}, {Franckowiak}, {Fukazawa}, {Funk}, {Fusco}, {Gargano}, {Gasparrini}, {Giglietto}, {Giordano}, {Giroletti}, {Glanzman}, {Godfrey}, {Gomez-Vargas}, {Grenier}, {Guiriec}, {Gustafsson}, {Hays}, {Hewitt}, {Horan}, {Jogler}, {J{\'o}hannesson}, {Kuss}, {Larsson}, {Latronico}, {Li}, {Li}, {Llena Garde}, {Longo}, {Loparco}, {Lubrano}, {Malyshev}, {Mayer}, {Mazziotta}, {McEnery}, {Meyer}, {Michelson}, {Mizuno}, {Moiseev}, {Monzani}, {Morselli}, {Murgia}, {Nuss},
  {Ohsugi}, {Orienti}, {Orlando}, {Ormes}, {Paneque}, {Perkins}, {Pesce-Rollins}, {Piron}, {Pivato}, {Porter}, {Rain{\`o}}, {Rando}, {Razzano}, {Reimer}, {Reimer}, {Ritz}, {S{\'a}nchez-Conde}, {Schulz}, {Sehgal}, {Sgr{\`o}}, {Siskind}, {Spada}, {Spandre}, {Spinelli}, {Strigari}, {Tajima}, {Takahashi}, {Thayer}, {Tibaldo}, {Torres}, {Troja}, {Vianello}, {Werner}, {Winer}, {Wood}, {Wood}, {Zaharijas}, {Zimmer}, \& {Fermi-LAT Collaboration}}]{2015PhRvL.115w1301A}
{Ackermann}, M., {Albert}, A., {Anderson}, B., {et~al.} 2015, \prl, 115, 231301

\bibitem[{{Ahvazi} {et~al.}(2023){Ahvazi}, {Benson}, {Sales}, {Nadler}, {Weerasooriya}, {Du}, \& {Bovill}}]{2023arXiv230813599A}
{Ahvazi}, N., {Benson}, A., {Sales}, L.~V., {et~al.} 2023, arXiv e-prints, arXiv:2308.13599

\bibitem[{{Aoki} {et~al.}(2007){Aoki}, {Beers}, {Christlieb}, {Norris}, {Ryan}, \& {Tsangarides}}]{2007ApJ...655..492A}
{Aoki}, W., {Beers}, T.~C., {Christlieb}, N., {et~al.} 2007, \apj, 655, 492

\bibitem[{{Applebaum} {et~al.}(2021){Applebaum}, {Brooks}, {Christensen}, {Munshi}, {Quinn}, {Shen}, \& {Tremmel}}]{2021ApJ...906...96A}
{Applebaum}, E., {Brooks}, A.~M., {Christensen}, C.~R., {et~al.} 2021, \apj, 906, 96

\bibitem[{{Arentsen} {et~al.}(2022){Arentsen}, {Placco}, {Lee}, {Aguado}, {Martin}, {Starkenburg}, \& {Yoon}}]{2022MNRAS.515.4082A}
{Arentsen}, A., {Placco}, V.~M., {Lee}, Y.~S., {et~al.} 2022, \mnras, 515, 4082

\bibitem[{{Astropy Collaboration} {et~al.}(2013){Astropy Collaboration}, {Robitaille}, {Tollerud}, {Greenfield}, {Droettboom}, {Bray}, {Aldcroft}, {Davis}, {Ginsburg}, {Price-Whelan}, {Kerzendorf}, {Conley}, {Crighton}, {Barbary}, {Muna}, {Ferguson}, {Grollier}, {Parikh}, {Nair}, {Unther}, {Deil}, {Woillez}, {Conseil}, {Kramer}, {Turner}, {Singer}, {Fox}, {Weaver}, {Zabalza}, {Edwards}, {Azalee Bostroem}, {Burke}, {Casey}, {Crawford}, {Dencheva}, {Ely}, {Jenness}, {Labrie}, {Lim}, {Pierfederici}, {Pontzen}, {Ptak}, {Refsdal}, {Servillat}, \& {Streicher}}]{2013A&A...558A..33A}
{Astropy Collaboration}, {Robitaille}, T.~P., {Tollerud}, E.~J., {et~al.} 2013, \aap, 558, A33

\bibitem[{{Astropy Collaboration} {et~al.}(2018){Astropy Collaboration}, {Price-Whelan}, {Sip{\H{o}}cz}, {G{\"u}nther}, {Lim}, {Crawford}, {Conseil}, {Shupe}, {Craig}, {Dencheva}, {Ginsburg}, {VanderPlas}, {Bradley}, {P{\'e}rez-Su{\'a}rez}, {de Val-Borro}, {Aldcroft}, {Cruz}, {Robitaille}, {Tollerud}, {Ardelean}, {Babej}, {Bach}, {Bachetti}, {Bakanov}, {Bamford}, {Barentsen}, {Barmby}, {Baumbach}, {Berry}, {Biscani}, {Boquien}, {Bostroem}, {Bouma}, {Brammer}, {Bray}, {Breytenbach}, {Buddelmeijer}, {Burke}, {Calderone}, {Cano Rodr{\'\i}guez}, {Cara}, {Cardoso}, {Cheedella}, {Copin}, {Corrales}, {Crichton}, {D'Avella}, {Deil}, {Depagne}, {Dietrich}, {Donath}, {Droettboom}, {Earl}, {Erben}, {Fabbro}, {Ferreira}, {Finethy}, {Fox}, {Garrison}, {Gibbons}, {Goldstein}, {Gommers}, {Greco}, {Greenfield}, {Groener}, {Grollier}, {Hagen}, {Hirst}, {Homeier}, {Horton}, {Hosseinzadeh}, {Hu}, {Hunkeler}, {Ivezi{\'c}}, {Jain}, {Jenness}, {Kanarek}, {Kendrew}, {Kern}, {Kerzendorf}, {Khvalko}, {King}, {Kirkby}, {Kulkarni},
  {Kumar}, {Lee}, {Lenz}, {Littlefair}, {Ma}, {Macleod}, {Mastropietro}, {McCully}, {Montagnac}, {Morris}, {Mueller}, {Mumford}, {Muna}, {Murphy}, {Nelson}, {Nguyen}, {Ninan}, {N{\"o}the}, {Ogaz}, {Oh}, {Parejko}, {Parley}, {Pascual}, {Patil}, {Patil}, {Plunkett}, {Prochaska}, {Rastogi}, {Reddy Janga}, {Sabater}, {Sakurikar}, {Seifert}, {Sherbert}, {Sherwood-Taylor}, {Shih}, {Sick}, {Silbiger}, {Singanamalla}, {Singer}, {Sladen}, {Sooley}, {Sornarajah}, {Streicher}, {Teuben}, {Thomas}, {Tremblay}, {Turner}, {Terr{\'o}n}, {van Kerkwijk}, {de la Vega}, {Watkins}, {Weaver}, {Whitmore}, {Woillez}, {Zabalza}, \& {Astropy Contributors}}]{2018AJ....156..123A}
{Astropy Collaboration}, {Price-Whelan}, A.~M., {Sip{\H{o}}cz}, B.~M., {et~al.} 2018, \aj, 156, 123

\bibitem[{{Astropy Collaboration} {et~al.}(2022){Astropy Collaboration}, {Price-Whelan}, {Lim}, {Earl}, {Starkman}, {Bradley}, {Shupe}, {Patil}, {Corrales}, {Brasseur}, {N{\"o}the}, {Donath}, {Tollerud}, {Morris}, {Ginsburg}, {Vaher}, {Weaver}, {Tocknell}, {Jamieson}, {van Kerkwijk}, {Robitaille}, {Merry}, {Bachetti}, {G{\"u}nther}, {Aldcroft}, {Alvarado-Montes}, {Archibald}, {B{\'o}di}, {Bapat}, {Barentsen}, {Baz{\'a}n}, {Biswas}, {Boquien}, {Burke}, {Cara}, {Cara}, {Conroy}, {Conseil}, {Craig}, {Cross}, {Cruz}, {D'Eugenio}, {Dencheva}, {Devillepoix}, {Dietrich}, {Eigenbrot}, {Erben}, {Ferreira}, {Foreman-Mackey}, {Fox}, {Freij}, {Garg}, {Geda}, {Glattly}, {Gondhalekar}, {Gordon}, {Grant}, {Greenfield}, {Groener}, {Guest}, {Gurovich}, {Handberg}, {Hart}, {Hatfield-Dodds}, {Homeier}, {Hosseinzadeh}, {Jenness}, {Jones}, {Joseph}, {Kalmbach}, {Karamehmetoglu}, {Ka{\l}uszy{\'n}ski}, {Kelley}, {Kern}, {Kerzendorf}, {Koch}, {Kulumani}, {Lee}, {Ly}, {Ma}, {MacBride}, {Maljaars}, {Muna}, {Murphy}, {Norman},
  {O'Steen}, {Oman}, {Pacifici}, {Pascual}, {Pascual-Granado}, {Patil}, {Perren}, {Pickering}, {Rastogi}, {Roulston}, {Ryan}, {Rykoff}, {Sabater}, {Sakurikar}, {Salgado}, {Sanghi}, {Saunders}, {Savchenko}, {Schwardt}, {Seifert-Eckert}, {Shih}, {Jain}, {Shukla}, {Sick}, {Simpson}, {Singanamalla}, {Singer}, {Singhal}, {Sinha}, {Sip{\H{o}}cz}, {Spitler}, {Stansby}, {Streicher}, {{\v{S}}umak}, {Swinbank}, {Taranu}, {Tewary}, {Tremblay}, {Val-Borro}, {Van Kooten}, {Vasovi{\'c}}, {Verma}, {de Miranda Cardoso}, {Williams}, {Wilson}, {Winkel}, {Wood-Vasey}, {Xue}, {Yoachim}, {Zhang}, {Zonca}, \& {Astropy Project Contributors}}]{2022ApJ...935..167A}
{Astropy Collaboration}, {Price-Whelan}, A.~M., {Lim}, P.~L., {et~al.} 2022, \apj, 935, 167

\bibitem[{{Balbinot} {et~al.}(2013){Balbinot}, {Santiago}, {da Costa}, {Maia}, {Majewski}, {Nidever}, {Rocha-Pinto}, {Thomas}, {Wechsler}, \& {Yanny}}]{2013ApJ...767..101B}
{Balbinot}, E., {Santiago}, B.~X., {da Costa}, L., {et~al.} 2013, \apj, 767, 101

\bibitem[{{Battaglia} \& {Starkenburg}(2012)}]{2012A&A...539A.123B}
{Battaglia}, G., \& {Starkenburg}, E. 2012, \aap, 539, A123

\bibitem[{{Baumgardt} \& {Hilker}(2018)}]{2018MNRAS.478.1520B}
{Baumgardt}, H., \& {Hilker}, M. 2018, \mnras, 478, 1520

\bibitem[{{Baumgardt} {et~al.}(2020){Baumgardt}, {Sollima}, \& {Hilker}}]{2020PASA...37...46B}
{Baumgardt}, H., {Sollima}, A., \& {Hilker}, M. 2020, \pasa, 37, e046

\bibitem[{{Bechtol} {et~al.}(2015){Bechtol}, {Drlica-Wagner}, {Balbinot}, {Pieres}, {Simon}, {Yanny}, {Santiago}, {Wechsler}, {Frieman}, {Walker}, {Williams}, {Rozo}, {Rykoff}, {Queiroz}, {Luque}, {Benoit-L{\'e}vy}, {Tucker}, {Sevilla}, {Gruendl}, {da Costa}, {Fausti Neto}, {Maia}, {Abbott}, {Allam}, {Armstrong}, {Bauer}, {Bernstein}, {Bernstein}, {Bertin}, {Brooks}, {Buckley-Geer}, {Burke}, {Carnero Rosell}, {Castander}, {Covarrubias}, {D'Andrea}, {DePoy}, {Desai}, {Diehl}, {Eifler}, {Estrada}, {Evrard}, {Fernandez}, {Finley}, {Flaugher}, {Gaztanaga}, {Gerdes}, {Girardi}, {Gladders}, {Gruen}, {Gutierrez}, {Hao}, {Honscheid}, {Jain}, {James}, {Kent}, {Kron}, {Kuehn}, {Kuropatkin}, {Lahav}, {Li}, {Lin}, {Makler}, {March}, {Marshall}, {Martini}, {Merritt}, {Miller}, {Miquel}, {Mohr}, {Neilsen}, {Nichol}, {Nord}, {Ogando}, {Peoples}, {Petravick}, {Plazas}, {Romer}, {Roodman}, {Sako}, {Sanchez}, {Scarpine}, {Schubnell}, {Smith}, {Soares-Santos}, {Sobreira}, {Suchyta}, {Swanson}, {Tarle}, {Thaler}, {Thomas},
  {Wester}, {Zuntz}, \& {DES Collaboration}}]{Bechtol:2015}
{Bechtol}, K., {Drlica-Wagner}, A., {Balbinot}, E., {et~al.} 2015, \apj, 807, 50

\bibitem[{{Beers} \& {Christlieb}(2005)}]{2005ARA&A..43..531B}
{Beers}, T.~C., \& {Christlieb}, N. 2005, \araa, 43, 531

\bibitem[{{Beers} {et~al.}(1990){Beers}, {Preston}, {Shectman}, \& {Kage}}]{1990AJ....100..849B}
{Beers}, T.~C., {Preston}, G.~W., {Shectman}, S.~A., \& {Kage}, J.~A. 1990, \aj, 100, 849

\bibitem[{{Belokurov} {et~al.}(2006){Belokurov}, {Zucker}, {Evans}, {Wilkinson}, {Irwin}, {Hodgkin}, {Bramich}, {Irwin}, {Gilmore}, {Willman}, {Vidrih}, {Newberg}, {Wyse}, {Fellhauer}, {Hewett}, {Cole}, {Bell}, {Beers}, {Rockosi}, {Yanny}, {Grebel}, {Schneider}, {Lupton}, {Barentine}, {Brewington}, {Brinkmann}, {Harvanek}, {Kleinman}, {Krzesinski}, {Long}, {Nitta}, {Smith}, \& {Snedden}}]{2006ApJ...647L.111B}
{Belokurov}, V., {Zucker}, D.~B., {Evans}, N.~W., {et~al.} 2006, \apjl, 647, L111

\bibitem[{{Benitez-Llambay} \& {Frenk}(2020)}]{2020MNRAS.498.4887B}
{Benitez-Llambay}, A., \& {Frenk}, C. 2020, \mnras, 498, 4887

\bibitem[{{Besla} {et~al.}(2010){Besla}, {Kallivayalil}, {Hernquist}, {van der Marel}, {Cox}, \& {Kere{\v{s}}}}]{2010ApJ...721L..97B}
{Besla}, G., {Kallivayalil}, N., {Hernquist}, L., {et~al.} 2010, \apjl, 721, L97

\bibitem[{{Bezanson} {et~al.}(2014){Bezanson}, {Edelman}, {Karpinski}, \& {Shah}}]{2014arXiv1411.1607B}
{Bezanson}, J., {Edelman}, A., {Karpinski}, S., \& {Shah}, V.~B. 2014, arXiv e-prints, arXiv:1411.1607

\bibitem[{{Boddy} {et~al.}(2024){Boddy}, {Carter}, {Kumar}, {Rufino}, {Sandick}, \& {Tapia-Arellano}}]{2024PhRvD.109j3007B}
{Boddy}, K.~K., {Carter}, Z.~J., {Kumar}, J., {et~al.} 2024, \prd, 109, 103007

\bibitem[{{Bonnivard} {et~al.}(2015){Bonnivard}, {Combet}, {Daniel}, {Funk}, {Geringer-Sameth}, {Hinton}, {Maurin}, {Read}, {Sarkar}, {Walker}, \& {Wilkinson}}]{Bonnivard2015MNRAS.453..849B}
{Bonnivard}, V., {Combet}, C., {Daniel}, M., {et~al.} 2015, \mnras, 453, 849

\bibitem[{{Bovy}(2015)}]{2015ApJS..216...29B}
{Bovy}, J. 2015, \apjs, 216, 29

\bibitem[{{Bressan} {et~al.}(2012){Bressan}, {Marigo}, {Girardi}, {Salasnich}, {Dal Cero}, {Rubele}, \& {Nanni}}]{Bressan:2012}
{Bressan}, A., {Marigo}, P., {Girardi}, L., {et~al.} 2012, \mnras, 427, 127

\bibitem[{{Bruce} {et~al.}(2023){Bruce}, {Li}, {Pace}, {Heiger}, {Song}, \& {Simon}}]{2023ApJ...950..167B}
{Bruce}, J., {Li}, T.~S., {Pace}, A.~B., {et~al.} 2023, \apj, 950, 167

\bibitem[{{Bullock} \& {Boylan-Kolchin}(2017)}]{2017ARA&A..55..343B}
{Bullock}, J.~S., \& {Boylan-Kolchin}, M. 2017, \araa, 55, 343

\bibitem[{{Buttry} {et~al.}(2022){Buttry}, {Pace}, {Koposov}, {Walker}, {Caldwell}, {Kirby}, {Martin}, {Mateo}, {Olszewski}, {Starkenburg}, {Badenes}, \& {Daher}}]{2022MNRAS.514.1706B}
{Buttry}, R., {Pace}, A.~B., {Koposov}, S.~E., {et~al.} 2022, \mnras, 514, 1706

\bibitem[{{Cantu} {et~al.}(2021){Cantu}, {Pace}, {Marshall}, {Strigari}, {Crnojevic}, {Simon}, {Drlica-Wagner}, {Bechtol}, {Mart{\'\i}nez-V{\'a}zquez}, {Santiago}, {Amara}, {Stringer}, {Diehl}, {Aguena}, {Allam}, {Avila}, {Brooks}, {Carnero Rosell}, {Carrasco Kind}, {Carretero}, {Costanzi}, {Da Costa}, {De Vicente}, {Desai}, {Doel}, {Eifler}, {Everett}, {Frieman}, {Garc{\'\i}a-Bellido}, {Gaztanaga}, {Gruen}, {Gruendl}, {Gschwend}, {Gutierrez}, {Hinton}, {Hollowood}, {Honscheid}, {James}, {Kuehn}, {Maia}, {Menanteau}, {Miquel}, {Palmese}, {Paz-Chinch{\'o}n}, {Plazas}, {Sanchez}, {Scarpine}, {Schubnell}, {Serrano}, {Sevilla-Noarbe}, {Smith}, {Soares-Santos}, {Suchyta}, {Swanson}, {Tarle}, {Walker}, {Wilkinson}, \& {DES Collaboration}}]{2021ApJ...916...81C}
{Cantu}, S.~A., {Pace}, A.~B., {Marshall}, J., {et~al.} 2021, \apj, 916, 81

\bibitem[{{Carlin} {et~al.}(2009){Carlin}, {Grillmair}, {Mu{\~n}oz}, {Nidever}, \& {Majewski}}]{2009ApJ...702L...9C}
{Carlin}, J.~L., {Grillmair}, C.~J., {Mu{\~n}oz}, R.~R., {Nidever}, D.~L., \& {Majewski}, S.~R. 2009, \apjl, 702, L9

\bibitem[{{Carrera} {et~al.}(2013){Carrera}, {Pancino}, {Gallart}, \& {del Pino}}]{2013MNRAS.434.1681C}
{Carrera}, R., {Pancino}, E., {Gallart}, C., \& {del Pino}, A. 2013, \mnras, 434, 1681

\bibitem[{{Cerny} {et~al.}(2021{\natexlab{a}}){Cerny}, {Pace}, {Drlica-Wagner}, {Ferguson}, {Mau}, {Adam{\'o}w}, {Carlin}, {Choi}, {Erkal}, {Johnson}, {Li}, {Mart{\'\i}nez-V{\'a}zquez}, {Mutlu-Pakdil}, {Nidever}, {Olsen}, {Pieres}, {Tollerud}, {Simon}, {Vivas}, {James}, {Kuropatkin}, {Majewski}, {Mart{\'\i}nez-Delgado}, {Massana}, {Miller}, {Neilsen}, {No{\"e}l}, {Riley}, {Sand}, {Santana-Silva}, {Stringfellow}, {Tucker}, \& {Delve Collaboration}}]{2021ApJ...910...18C}
{Cerny}, W., {Pace}, A.~B., {Drlica-Wagner}, A., {et~al.} 2021{\natexlab{a}}, \apj, 910, 18

\bibitem[{{Cerny} {et~al.}(2021{\natexlab{b}}){Cerny}, {Pace}, {Drlica-Wagner}, {Koposov}, {Vivas}, {Mau}, {Riley}, {Bom}, {Carlin}, {Choi}, {Erkal}, {Ferguson}, {James}, {Li}, {Mart{\'\i}nez-Delgado}, {Mart{\'\i}nez-V{\'a}zquez}, {Munoz}, {Mutlu-Pakdil}, {Olsen}, {Pieres}, {Sakowska}, {Sand}, {Simon}, {Smercina}, {Stringfellow}, {Tollerud}, {Adam{\'o}w}, {Hernandez-Lang}, {Kuropatkin}, {Santana-Silva}, {Tucker}, {Zenteno}, \& {Delve Collaboration}}]{2021ApJ...920L..44C}
---. 2021{\natexlab{b}}, \apjl, 920, L44

\bibitem[{{Cerny} {et~al.}(2022){Cerny}, {Mart{\'\i}nez-V{\'a}zquez}, {Drlica-Wagner}, {Pace}, {Mutlu-Pakdil}, {Li}, {Riley}, {Crnojevi{\'c}}, {Bom}, {Carballo-Bello}, {Carlin}, {Chiti}, {Choi}, {Collins}, {Darragh-Ford}, {Ferguson}, {Geha}, {Mart{\'\i}nez-Delgado}, {Massana}, {Mau}, {Medina}, {Mu{\~n}oz}, {Nadler}, {Olsen}, {Pieres}, {Sakowska}, {Simon}, {Stringfellow}, {Vivas}, {Walker}, \& {Wechsler}}]{2022arXiv220912422C}
{Cerny}, W., {Mart{\'\i}nez-V{\'a}zquez}, C.~E., {Drlica-Wagner}, A., {et~al.} 2022, arXiv e-prints, arXiv:2209.12422

\bibitem[{{Cerny} {et~al.}(2023{\natexlab{a}}){Cerny}, {Mart{\'\i}nez-V{\'a}zquez}, {Drlica-Wagner}, {Pace}, {Mutlu-Pakdil}, {Li}, {Riley}, {Crnojevi{\'c}}, {Bom}, {Carballo-Bello}, {Carlin}, {Chiti}, {Choi}, {Collins}, {Darragh-Ford}, {Ferguson}, {Geha}, {Mart{\'\i}nez-Delgado}, {Massana}, {Mau}, {Medina}, {Mu{\~n}oz}, {Nadler}, {No{\"e}l}, {Olsen}, {Pieres}, {Sakowska}, {Simon}, {Stringfellow}, {Tollerud}, {Vivas}, {Walker}, {Wechsler}, \& {Delve Collaboration}}]{2023ApJ...953....1C}
---. 2023{\natexlab{a}}, \apj, 953, 1

\bibitem[{{Cerny} {et~al.}(2023{\natexlab{b}}){Cerny}, {Simon}, {Li}, {Drlica-Wagner}, {Pace}, {Mart{\'\i}nez-V{\'a}zquez}, {Riley}, {Mutlu-Pakdil}, {Mau}, {Ferguson}, {Erkal}, {Munoz}, {Bom}, {Carlin}, {Carollo}, {Choi}, {Ji}, {Manwadkar}, {Mart{\'\i}nez-Delgado}, {Miller}, {No{\"e}l}, {Sakowska}, {Sand}, {Stringfellow}, {Tollerud}, {Vivas}, {Carballo-Bello}, {Hernandez-Lang}, {James}, {Nidever}, {Castellon}, {Olsen}, {Zenteno}, \& {Delve Collaboration}}]{2023ApJ...942..111C}
{Cerny}, W., {Simon}, J.~D., {Li}, T.~S., {et~al.} 2023{\natexlab{b}}, \apj, 942, 111

\bibitem[{{Cerny} {et~al.}(2023{\natexlab{c}}){Cerny}, {Drlica-Wagner}, {Li}, {Pace}, {Olsen}, {No{\"e}l}, {van der Marel}, {Carlin}, {Choi}, {Erkal}, {Geha}, {James}, {Mart{\'\i}nez-V{\'a}zquez}, {Massana}, {Medina}, {Miller}, {Mutlu-Pakdil}, {Nidever}, {Sakowska}, {Stringfellow}, {Carballo-Bello}, {Ferguson}, {Kuropatkin}, {Mau}, {Tollerud}, {Vivas}, \& {Delve Collaboration}}]{2023ApJ...953L..21C}
{Cerny}, W., {Drlica-Wagner}, A., {Li}, T.~S., {et~al.} 2023{\natexlab{c}}, \apjl, 953, L21

\bibitem[{{Chandra} {et~al.}(2023){Chandra}, {Naidu}, {Conroy}, {Bonaca}, {Zaritsky}, {Cargile}, {Caldwell}, {Johnson}, {Han}, \& {Ting}}]{2023ApJ...956..110C}
{Chandra}, V., {Naidu}, R.~P., {Conroy}, C., {et~al.} 2023, \apj, 956, 110

\bibitem[{Chen {et~al.}(2015)Chen, Bressan, Girardi, Marigo, Kong, \& Lanza}]{10.1093/mnras/stv1281}
Chen, Y., Bressan, A., Girardi, L., {et~al.} 2015, Monthly Notices of the Royal Astronomical Society, 452, 1068.
\newblock \url{https://doi.org/10.1093/mnras/stv1281}

\bibitem[{{Chen} {et~al.}(2014){Chen}, {Girardi}, {Bressan}, {Marigo}, {Barbieri}, \& {Kong}}]{2014MNRAS.444.2525C}
{Chen}, Y., {Girardi}, L., {Bressan}, A., {et~al.} 2014, \mnras, 444, 2525

\bibitem[{{Chiti} {et~al.}(2022){Chiti}, {Simon}, {Frebel}, {Pace}, {Ji}, \& {Li}}]{2022ApJ...939...41C}
{Chiti}, A., {Simon}, J.~D., {Frebel}, A., {et~al.} 2022, \apj, 939, 41

\bibitem[{{Chiti} {et~al.}(2018){Chiti}, {Simon}, {Frebel}, {Thompson}, {Shectman}, {Mateo}, {Bailey}, {Crane}, \& {Walker}}]{2018ApJ...856..142C}
---. 2018, \apj, 856, 142

\bibitem[{{Chiti} {et~al.}(2021){Chiti}, {Frebel}, {Simon}, {Erkal}, {Chang}, {Necib}, {Ji}, {Jerjen}, {Kim}, \& {Norris}}]{2021NatAs...5..392C}
{Chiti}, A., {Frebel}, A., {Simon}, J.~D., {et~al.} 2021, Nature Astronomy, 5, 392

\bibitem[{{Chiti} {et~al.}(2023){Chiti}, {Frebel}, {Ji}, {Mardini}, {Ou}, {Simon}, {Jerjen}, {Kim}, \& {Norris}}]{2023AJ....165...55C}
{Chiti}, A., {Frebel}, A., {Ji}, A.~P., {et~al.} 2023, \aj, 165, 55

\bibitem[{{Choi} {et~al.}(2018){Choi}, {Nidever}, {Olsen}, {Besla}, {Blum}, {Zaritsky}, {Cioni}, {van der Marel}, {Bell}, {Johnson}, {Vivas}, {Walker}, {de Boer}, {No{\"e}l}, {Monachesi}, {Gallart}, {Monelli}, {Stringfellow}, {Massana}, {Martinez-Delgado}, \& {Mu{\~n}oz}}]{2018ApJ...869..125C}
{Choi}, Y., {Nidever}, D.~L., {Olsen}, K., {et~al.} 2018, \apj, 869, 125

\bibitem[{{Conn} {et~al.}(2018){Conn}, {Jerjen}, {Kim}, \& {Schirmer}}]{2018ApJ...852...68C}
{Conn}, B.~C., {Jerjen}, H., {Kim}, D., \& {Schirmer}, M. 2018, \apj, 852, 68

\bibitem[{{Correa Magnus} \& {Vasiliev}(2022)}]{2022MNRAS.511.2610C}
{Correa Magnus}, L., \& {Vasiliev}, E. 2022, \mnras, 511, 2610

\bibitem[{{Correnti} {et~al.}(2009){Correnti}, {Bellazzini}, \& {Ferraro}}]{2009MNRAS.397L..26C}
{Correnti}, M., {Bellazzini}, M., \& {Ferraro}, F.~R. 2009, \mnras, 397, L26

\bibitem[{{Crnojevi{\'c}} {et~al.}(2016){Crnojevi{\'c}}, {Sand}, {Zaritsky}, {Spekkens}, {Willman}, \& {Hargis}}]{2016ApJ...824L..14C}
{Crnojevi{\'c}}, D., {Sand}, D.~J., {Zaritsky}, D., {et~al.} 2016, \apjl, 824, L14

\bibitem[{{Dey} {et~al.}(2019){Dey}, {Schlegel}, {Lang}, {Blum}, {Burleigh}, {Fan}, {Findlay}, {Finkbeiner}, {Herrera}, {Juneau}, {Landriau}, {Levi}, {McGreer}, {Meisner}, {Myers}, {Moustakas}, {Nugent}, {Patej}, {Schlafly}, {Walker}, {Valdes}, {Weaver}, {Y{\`e}che}, {Zou}, {Zhou}, {Abareshi}, {Abbott}, {Abolfathi}, {Aguilera}, {Alam}, {Allen}, {Alvarez}, {Annis}, {Ansarinejad}, {Aubert}, {Beechert}, {Bell}, {BenZvi}, {Beutler}, {Bielby}, {Bolton}, {Brice{\~n}o}, {Buckley-Geer}, {Butler}, {Calamida}, {Carlberg}, {Carter}, {Casas}, {Castander}, {Choi}, {Comparat}, {Cukanovaite}, {Delubac}, {DeVries}, {Dey}, {Dhungana}, {Dickinson}, {Ding}, {Donaldson}, {Duan}, {Duckworth}, {Eftekharzadeh}, {Eisenstein}, {Etourneau}, {Fagrelius}, {Farihi}, {Fitzpatrick}, {Font-Ribera}, {Fulmer}, {G{\"a}nsicke}, {Gaztanaga}, {George}, {Gerdes}, {Gontcho}, {Gorgoni}, {Green}, {Guy}, {Harmer}, {Hernandez}, {Honscheid}, {Huang}, {James}, {Jannuzi}, {Jiang}, {Joyce}, {Karcher}, {Karkar}, {Kehoe}, {Kneib}, {Kueter-Young}, {Lan},
  {Lauer}, {Le Guillou}, {Le Van Suu}, {Lee}, {Lesser}, {Perreault Levasseur}, {Li}, {Mann}, {Marshall}, {Mart{\'\i}nez-V{\'a}zquez}, {Martini}, {du Mas des Bourboux}, {McManus}, {Meier}, {M{\'e}nard}, {Metcalfe}, {Mu{\~n}oz-Guti{\'e}rrez}, {Najita}, {Napier}, {Narayan}, {Newman}, {Nie}, {Nord}, {Norman}, {Olsen}, {Paat}, {Palanque-Delabrouille}, {Peng}, {Poppett}, {Poremba}, {Prakash}, {Rabinowitz}, {Raichoor}, {Rezaie}, {Robertson}, {Roe}, {Ross}, {Ross}, {Rudnick}, {Safonova}, {Saha}, {S{\'a}nchez}, {Savary}, {Schweiker}, {Scott}, {Seo}, {Shan}, {Silva}, {Slepian}, {Soto}, {Sprayberry}, {Staten}, {Stillman}, {Stupak}, {Summers}, {Sien Tie}, {Tirado}, {Vargas-Maga{\~n}a}, {Vivas}, {Wechsler}, {Williams}, {Yang}, {Yang}, {Yapici}, {Zaritsky}, {Zenteno}, {Zhang}, {Zhang}, {Zhou}, \& {Zhou}}]{2019AJ....157..168D}
{Dey}, A., {Schlegel}, D.~J., {Lang}, D., {et~al.} 2019, \aj, 157, 168

\bibitem[{{Drlica-Wagner} {et~al.}(2015){Drlica-Wagner}, {Bechtol}, {Rykoff}, {Luque}, {Queiroz}, {Mao}, {Wechsler}, {Simon}, {Santiago}, {Yanny}, {Balbinot}, {Dodelson}, {Fausti Neto}, {James}, {Li}, {Maia}, {Marshall}, {}, {Stringer}, {Walker}, {Abbott}, {Abdalla}, {Allam}, {Benoit-L{\'e}vy}, {Bernstein}, {Bertin}, {Brooks}, {Buckley-Geer}, {Burke}, {Carnero Rosell}, {Carrasco Kind}, {Carretero}, {Crocce}, {da Costa}, {Desai}, {Diehl}, {Dietrich}, {Doel}, {Eifler}, {Evrard}, {Finley}, {Flaugher}, {Fosalba}, {Frieman}, {Gaztanaga}, {Gerdes}, {Gruen}, {Gruendl}, {Gutierrez}, {Honscheid}, {Kuehn}, {Kuropatkin}, {Lahav}, {Martini}, {Miquel}, {Nord}, {Ogando}, {Plazas}, {Reil}, {Roodman}, {Sako}, {Sanchez}, {Scarpine}, {Schubnell}, {Sevilla-Noarbe}, {Smith}, {Soares-Santos}, {Sobreira}, {Suchyta}, {Swanson}, {Tarle}, {Tucker}, {Vikram}, {Wester}, {Zhang}, {Zuntz}, \& {DES Collaboration}}]{2015ApJ...813..109D}
{Drlica-Wagner}, A., {Bechtol}, K., {Rykoff}, E.~S., {et~al.} 2015, \apj, 813, 109

\bibitem[{{Drlica-Wagner} {et~al.}(2016){Drlica-Wagner}, {Bechtol}, {Allam}, {Tucker}, {Gruendl}, {Johnson}, {Walker}, {James}, {Nidever}, {Olsen}, {Wechsler}, {Cioni}, {Conn}, {Kuehn}, {Li}, {Mao}, {Martin}, {Neilsen}, {Noel}, {Pieres}, {Simon}, {Stringfellow}, {van der Marel}, \& {Yanny}}]{2016ApJ...833L...5D}
{Drlica-Wagner}, A., {Bechtol}, K., {Allam}, S., {et~al.} 2016, \apjl, 833, L5

\bibitem[{{Drlica-Wagner} {et~al.}(2020){Drlica-Wagner}, {Bechtol}, {Mau}, {McNanna}, {Nadler}, {Pace}, {Li}, {Pieres}, {Rozo}, {Simon}, {Walker}, {Wechsler}, {Abbott}, {Allam}, {Annis}, {Bertin}, {Brooks}, {Burke}, {Rosell}, {Carrasco Kind}, {Carretero}, {Costanzi}, {da Costa}, {De Vicente}, {Desai}, {Diehl}, {Doel}, {Eifler}, {Everett}, {Flaugher}, {Frieman}, {Garc{\'\i}a-Bellido}, {Gaztanaga}, {Gruen}, {Gruendl}, {Gschwend}, {Gutierrez}, {Honscheid}, {James}, {Krause}, {Kuehn}, {Kuropatkin}, {Lahav}, {Maia}, {Marshall}, {Melchior}, {Menanteau}, {Miquel}, {Palmese}, {Plazas}, {Sanchez}, {Scarpine}, {Schubnell}, {Serrano}, {Sevilla-Noarbe}, {Smith}, {Suchyta}, {Tarle}, \& {DES Collaboration}}]{2020ApJ...893...47D}
{Drlica-Wagner}, A., {Bechtol}, K., {Mau}, S., {et~al.} 2020, \apj, 893, 47

\bibitem[{{Drlica-Wagner} {et~al.}(2021){Drlica-Wagner}, {Carlin}, {Nidever}, {Ferguson}, {Kuropatkin}, {Adam{\'o}w}, {Cerny}, {Choi}, {Esteves}, {Mart{\'\i}nez-V{\'a}zquez}, {Mau}, {Miller}, {Mutlu-Pakdil}, {Neilsen}, {Olsen}, {Pace}, {Riley}, {Sakowska}, {Sand}, {Santana-Silva}, {Tollerud}, {Tucker}, {Vivas}, {Zaborowski}, {Zenteno}, {Abbott}, {Allam}, {Bechtol}, {Bell}, {Bell}, {Bilaji}, {Bom}, {Carballo-Bello}, {Crnojevi{\'c}}, {Cioni}, {Diaz-Ocampo}, {de Boer}, {Erkal}, {Gruendl}, {Hernandez-Lang}, {Hughes}, {James}, {Johnson}, {Li}, {Mao}, {Mart{\'\i}nez-Delgado}, {Massana}, {McNanna}, {Morgan}, {Nadler}, {No{\"e}l}, {Palmese}, {Peter}, {Rykoff}, {S{\'a}nchez}, {Shipp}, {Simon}, {Smercina}, {Soares-Santos}, {Stringfellow}, {Tavangar}, {van der Marel}, {Walker}, {Wechsler}, {Wu}, {Yanny}, {Fitzpatrick}, {Huang}, {Jacques}, {Nikutta}, {Scott}, \& {Astro Data Lab}}]{2021ApJS..256....2D}
{Drlica-Wagner}, A., {Carlin}, J.~L., {Nidever}, D.~L., {et~al.} 2021, \apjs, 256, 2

\bibitem[{{Drlica-Wagner} {et~al.}(2022){Drlica-Wagner}, {Ferguson}, {Adam{\'o}w}, {Aguena}, {Allam}, {Andrade-Oliveira}, {Bacon}, {Bechtol}, {Bell}, {Bertin}, {Bilaji}, {Bocquet}, {Bom}, {Brooks}, {Burke}, {Carballo-Bello}, {Carlin}, {Carnero Rosell}, {Carrasco Kind}, {Carretero}, {Castander}, {Cerny}, {Chang}, {Choi}, {Conselice}, {Costanzi}, {Crnojevi{\'c}}, {da Costa}, {de Vicente}, {Desai}, {Esteves}, {Everett}, {Ferrero}, {Fitzpatrick}, {Flaugher}, {Friedel}, {Frieman}, {Garc{\'\i}a-Bellido}, {Gatti}, {Gaztanaga}, {Gerdes}, {Gruen}, {Gruendl}, {Gschwend}, {Hartley}, {Hernandez-Lang}, {Hinton}, {Hollowood}, {Honscheid}, {Hughes}, {Jacques}, {James}, {Johnson}, {Kuehn}, {Kuropatkin}, {Lahav}, {Li}, {Lidman}, {Lin}, {March}, {Marshall}, {Mart{\'\i}nez-Delgado}, {Mart{\'\i}nez-V{\'a}zquez}, {Massana}, {Mau}, {McNanna}, {Melchior}, {Menanteau}, {Miller}, {Miquel}, {Mohr}, {Morgan}, {Mutlu-Pakdil}, {Mu{\~n}oz}, {Neilsen}, {Nidever}, {Nikutta}, {Nilo Castellon}, {No{\"e}l}, {Ogando}, {Olsen}, {Pace},
  {Palmese}, {Paz-Chinch{\'o}n}, {Pereira}, {Pieres}, {Plazas Malag{\'o}n}, {Prat}, {Riley}, {Rodriguez-Monroy}, {Romer}, {Roodman}, {Sako}, {Sakowska}, {Sanchez}, {S{\'a}nchez}, {Sand}, {Santana-Silva}, {Santiago}, {Schubnell}, {Serrano}, {Sevilla-Noarbe}, {Simon}, {Smith}, {Soares-Santos}, {Stringfellow}, {Suchyta}, {Suson}, {Tan}, {Tarle}, {Tavangar}, {Thomas}, {To}, {Tollerud}, {Troxel}, {Tucker}, {Varga}, {Vivas}, {Walker}, {Weller}, {Wilkinson}, {Wu}, {Yanny}, {Zaborowski}, {Zenteno}, {Delve Collaboration}, {Des Collaboration}, \& {Astro Data Lab}}]{2022ApJS..261...38D}
{Drlica-Wagner}, A., {Ferguson}, P.~S., {Adam{\'o}w}, M., {et~al.} 2022, \apjs, 261, 38

\bibitem[{{Efstathiou}(1992)}]{1992MNRAS.256P..43E}
{Efstathiou}, G. 1992, \mnras, 256, 43P

\bibitem[{{Esteban} {et~al.}(2023){Esteban}, {Peter}, \& {Kim}}]{2023arXiv230604674E}
{Esteban}, I., {Peter}, A. H.~G., \& {Kim}, S.~Y. 2023, arXiv e-prints, arXiv:2306.04674

\bibitem[{{Faber} {et~al.}(2003){Faber}, {Phillips}, {Kibrick}, {Alcott}, {Allen}, {Burrous}, {Cantrall}, {Clarke}, {Coil}, {Cowley}, {Davis}, {Deich}, {Dietsch}, {Gilmore}, {Harper}, {Hilyard}, {Lewis}, {McVeigh}, {Newman}, {Osborne}, {Schiavon}, {Stover}, {Tucker}, {Wallace}, {Wei}, {Wirth}, \& {Wright}}]{2003SPIE.4841.1657F}
{Faber}, S.~M., {Phillips}, A.~C., {Kibrick}, R.~I., {et~al.} 2003, in Society of Photo-Optical Instrumentation Engineers (SPIE) Conference Series, Vol. 4841, Instrument Design and Performance for Optical/Infrared Ground-based Telescopes, ed. M.~{Iye} \& A.~F.~M. {Moorwood}, 1657--1669

\bibitem[{{Fadely} {et~al.}(2011){Fadely}, {Willman}, {Geha}, {Walsh}, {Mu{\~n}oz}, {Jerjen}, {Vargas}, \& {Da Costa}}]{2011AJ....142...88F}
{Fadely}, R., {Willman}, B., {Geha}, M., {et~al.} 2011, \aj, 142, 88

\bibitem[{{Flaugher} {et~al.}(2015){Flaugher}, {Diehl}, {Honscheid}, {Abbott}, {Alvarez}, {Angstadt}, {Annis}, {Antonik}, {Ballester}, {Beaufore}, {Bernstein}, {Bernstein}, {Bigelow}, {Bonati}, {Boprie}, {Brooks}, {Buckley-Geer}, {Campa}, {Cardiel-Sas}, {Castander}, {Castilla}, {Cease}, {Cela-Ruiz}, {Chappa}, {Chi}, {Cooper}, {da Costa}, {Dede}, {Derylo}, {DePoy}, {de Vicente}, {Doel}, {Drlica-Wagner}, {Eiting}, {Elliott}, {Emes}, {Estrada}, {Fausti Neto}, {Finley}, {Flores}, {Frieman}, {Gerdes}, {Gladders}, {Gregory}, {Gutierrez}, {Hao}, {Holland}, {Holm}, {Huffman}, {Jackson}, {James}, {Jonas}, {Karcher}, {Karliner}, {Kent}, {Kessler}, {Kozlovsky}, {Kron}, {Kubik}, {Kuehn}, {Kuhlmann}, {Kuk}, {Lahav}, {Lathrop}, {Lee}, {Levi}, {Lewis}, {Li}, {Mandrichenko}, {Marshall}, {Martinez}, {Merritt}, {Miquel}, {Mu{\~n}oz}, {Neilsen}, {Nichol}, {Nord}, {Ogando}, {Olsen}, {Palaio}, {Patton}, {Peoples}, {Plazas}, {Rauch}, {Reil}, {Rheault}, {Roe}, {Rogers}, {Roodman}, {Sanchez}, {Scarpine}, {Schindler}, {Schmidt},
  {Schmitt}, {Schubnell}, {Schultz}, {Schurter}, {Scott}, {Serrano}, {Shaw}, {Smith}, {Soares-Santos}, {Stefanik}, {Stuermer}, {Suchyta}, {Sypniewski}, {Tarle}, {Thaler}, {Tighe}, {Tran}, {Tucker}, {Walker}, {Wang}, {Watson}, {Weaverdyck}, {Wester}, {Woods}, {Yanny}, \& {DES Collaboration}}]{2015AJ....150..150F}
{Flaugher}, B., {Diehl}, H.~T., {Honscheid}, K., {et~al.} 2015, \aj, 150, 150

\bibitem[{{Foreman-Mackey} {et~al.}(2013){Foreman-Mackey}, {Hogg}, {Lang}, \& {Goodman}}]{Foreman-Mackey:2013}
{Foreman-Mackey}, D., {Hogg}, D.~W., {Lang}, D., \& {Goodman}, J. 2013, \pasp, 125, 306

\bibitem[{{Fritz} {et~al.}(2019){Fritz}, {Carrera}, {Battaglia}, \& {Taibi}}]{2019A&A...623A.129F}
{Fritz}, T.~K., {Carrera}, R., {Battaglia}, G., \& {Taibi}, S. 2019, \aap, 623, A129

\bibitem[{{Fu} {et~al.}(2023){Fu}, {Weisz}, {Starkenburg}, {Martin}, {Savino}, {Boylan-Kolchin}, {C{\^o}t{\'e}}, {Dolphin}, {Ji}, {Longeard}, {Mateo}, {Patel}, \& {Sandford}}]{2023ApJ...958..167F}
{Fu}, S.~W., {Weisz}, D.~R., {Starkenburg}, E., {et~al.} 2023, \apj, 958, 167

\bibitem[{{Gaia Collaboration} {et~al.}(2016){Gaia Collaboration}, {Prusti}, {de Bruijne}, {Brown}, {Vallenari}, {Babusiaux}, {Bailer-Jones}, {Bastian}, {Biermann}, {Evans}, {Eyer}, {Jansen}, {Jordi}, {Klioner}, {Lammers}, {Lindegren}, {Luri}, {Mignard}, {Milligan}, {Panem}, {Poinsignon}, {Pourbaix}, {Randich}, {Sarri}, {Sartoretti}, {Siddiqui}, {Soubiran}, {Valette}, {van Leeuwen}, {Walton}, {Aerts}, {Arenou}, {Cropper}, {Drimmel}, {H{\o}g}, {Katz}, {Lattanzi}, {O'Mullane}, {Grebel}, {Holland}, {Huc}, {Passot}, {Bramante}, {Cacciari}, {Casta{\~n}eda}, {Chaoul}, {Cheek}, {De Angeli}, {Fabricius}, {Guerra}, {Hern{\'a}ndez}, {Jean-Antoine-Piccolo}, {Masana}, {Messineo}, {Mowlavi}, {Nienartowicz}, {Ord{\'o}{\~n}ez-Blanco}, {Panuzzo}, {Portell}, {Richards}, {Riello}, {Seabroke}, {Tanga}, {Th{\'e}venin}, {Torra}, {Els}, {Gracia-Abril}, {Comoretto}, {Garcia-Reinaldos}, {Lock}, {Mercier}, {Altmann}, {Andrae}, {Astraatmadja}, {Bellas-Velidis}, {Benson}, {Berthier}, {Blomme}, {Busso}, {Carry}, {Cellino}, {Clementini},
  {Cowell}, {Creevey}, {Cuypers}, {Davidson}, {De Ridder}, {de Torres}, {Delchambre}, {Dell'Oro}, {Ducourant}, {Fr{\'e}mat}, {Garc{\'\i}a-Torres}, {Gosset}, {Halbwachs}, {Hambly}, {Harrison}, {Hauser}, {Hestroffer}, {Hodgkin}, {Huckle}, {Hutton}, {Jasniewicz}, {Jordan}, {Kontizas}, {Korn}, {Lanzafame}, {Manteiga}, {Moitinho}, {Muinonen}, {Osinde}, {Pancino}, {Pauwels}, {Petit}, {Recio-Blanco}, {Robin}, {Sarro}, {Siopis}, {Smith}, {Smith}, {Sozzetti}, {Thuillot}, {van Reeven}, {Viala}, {Abbas}, {Abreu Aramburu}, {Accart}, {Aguado}, {Allan}, {Allasia}, {Altavilla}, {{\'A}lvarez}, {Alves}, {Anderson}, {Andrei}, {Anglada Varela}, {Antiche}, {Antoja}, {Ant{\'o}n}, {Arcay}, {Atzei}, {Ayache}, {Bach}, {Baker}, {Balaguer-N{\'u}{\~n}ez}, {Barache}, {Barata}, {Barbier}, {Barblan}, {Baroni}, {Barrado y Navascu{\'e}s}, {Barros}, {Barstow}, {Becciani}, {Bellazzini}, {Bellei}, {Bello Garc{\'\i}a}, {Belokurov}, {Bendjoya}, {Berihuete}, {Bianchi}, {Bienaym{\'e}}, {Billebaud}, {Blagorodnova}, {Blanco-Cuaresma}, {Boch},
  {Bombrun}, {Borrachero}, {Bouquillon}, {Bourda}, {Bouy}, {Bragaglia}, {Breddels}, {Brouillet}, {Br{\"u}semeister}, {Bucciarelli}, {Budnik}, {Burgess}, {Burgon}, {Burlacu}, {Busonero}, {Buzzi}, {Caffau}, {Cambras}, {Campbell}, {Cancelliere}, {Cantat-Gaudin}, {Carlucci}, {Carrasco}, {Castellani}, {Charlot}, {Charnas}, {Charvet}, {Chassat}, {Chiavassa}, {Clotet}, {Cocozza}, {Collins}, {Collins}, {Costigan}, {Crifo}, {Cross}, {Crosta}, {Crowley}, {Dafonte}, {Damerdji}, {Dapergolas}, {David}, {David}, {De Cat}, {de Felice}, {de Laverny}, {De Luise}, {De March}, {de Martino}, {de Souza}, {Debosscher}, {del Pozo}, {Delbo}, {Delgado}, {Delgado}, {di Marco}, {Di Matteo}, {Diakite}, {Distefano}, {Dolding}, {Dos Anjos}, {Drazinos}, {Dur{\'a}n}, {Dzigan}, {Ecale}, {Edvardsson}, {Enke}, {Erdmann}, {Escolar}, {Espina}, {Evans}, {Eynard Bontemps}, {Fabre}, {Fabrizio}, {Faigler}, {Falc{\~a}o}, {Farr{\`a}s Casas}, {Faye}, {Federici}, {Fedorets}, {Fern{\'a}ndez-Hern{\'a}ndez}, {Fernique}, {Fienga}, {Figueras}, {Filippi},
  {Findeisen}, {Fonti}, {Fouesneau}, {Fraile}, {Fraser}, {Fuchs}, {Furnell}, {Gai}, {Galleti}, {Galluccio}, {Garabato}, {Garc{\'\i}a-Sedano}, {Gar{\'e}}, {Garofalo}, {Garralda}, {Gavras}, {Gerssen}, {Geyer}, {Gilmore}, {Girona}, {Giuffrida}, {Gomes}, {Gonz{\'a}lez-Marcos}, {Gonz{\'a}lez-N{\'u}{\~n}ez}, {Gonz{\'a}lez-Vidal}, {Granvik}, {Guerrier}, {Guillout}, {Guiraud}, {G{\'u}rpide}, {Guti{\'e}rrez-S{\'a}nchez}, {Guy}, {Haigron}, {Hatzidimitriou}, {Haywood}, {Heiter}, {Helmi}, {Hobbs}, {Hofmann}, {Holl}, {Holland}, {Hunt}, {Hypki}, {Icardi}, {Irwin}, {Jevardat de Fombelle}, {Jofr{\'e}}, {Jonker}, {Jorissen}, {Julbe}, {Karampelas}, {Kochoska}, {Kohley}, {Kolenberg}, {Kontizas}, {Koposov}, {Kordopatis}, {Koubsky}, {Kowalczyk}, {Krone-Martins}, {Kudryashova}, {Kull}, {Bachchan}, {Lacoste-Seris}, {Lanza}, {Lavigne}, {Le Poncin-Lafitte}, {Lebreton}, {Lebzelter}, {Leccia}, {Leclerc}, {Lecoeur-Taibi}, {Lemaitre}, {Lenhardt}, {Leroux}, {Liao}, {Licata}, {Lindstr{\o}m}, {Lister}, {Livanou}, {Lobel}, {L{\"o}ffler},
  {L{\'o}pez}, {Lopez-Lozano}, {Lorenz}, {Loureiro}, {MacDonald}, {Magalh{\~a}es Fernandes}, {Managau}, {Mann}, {Mantelet}, {Marchal}, {Marchant}, {Marconi}, {Marie}, {Marinoni}, {Marrese}, {Marschalk{\'o}}, {Marshall}, {Mart{\'\i}n-Fleitas}, {Martino}, {Mary}, {Matijevi{\v{c}}}, {Mazeh}, {McMillan}, {Messina}, {Mestre}, {Michalik}, {Millar}, {Miranda}, {Molina}, {Molinaro}, {Molinaro}, {Moln{\'a}r}, {Moniez}, {Montegriffo}, {Monteiro}, {Mor}, {Mora}, {Morbidelli}, {Morel}, {Morgenthaler}, {Morley}, {Morris}, {Mulone}, {Muraveva}, {Musella}, {Narbonne}, {Nelemans}, {Nicastro}, {Noval}, {Ord{\'e}novic}, {Ordieres-Mer{\'e}}, {Osborne}, {Pagani}, {Pagano}, {Pailler}, {Palacin}, {Palaversa}, {Parsons}, {Paulsen}, {Pecoraro}, {Pedrosa}, {Pentik{\"a}inen}, {Pereira}, {Pichon}, {Piersimoni}, {Pineau}, {Plachy}, {Plum}, {Poujoulet}, {Pr{\v{s}}a}, {Pulone}, {Ragaini}, {Rago}, {Rambaux}, {Ramos-Lerate}, {Ranalli}, {Rauw}, {Read}, {Regibo}, {Renk}, {Reyl{\'e}}, {Ribeiro}, {Rimoldini}, {Ripepi}, {Riva}, {Rixon},
  {Roelens}, {Romero-G{\'o}mez}, {Rowell}, {Royer}, {Rudolph}, {Ruiz-Dern}, {Sadowski}, {Sagrist{\`a} Sell{\'e}s}, {Sahlmann}, {Salgado}, {Salguero}, {Sarasso}, {Savietto}, {Schnorhk}, {Schultheis}, {Sciacca}, {Segol}, {Segovia}, {Segransan}, {Serpell}, {Shih}, {Smareglia}, {Smart}, {Smith}, {Solano}, {Solitro}, {Sordo}, {Soria Nieto}, {Souchay}, {Spagna}, {Spoto}, {Stampa}, {Steele}, {Steidelm{\"u}ller}, {Stephenson}, {Stoev}, {Suess}, {S{\"u}veges}, {Surdej}, {Szabados}, {Szegedi-Elek}, {Tapiador}, {Taris}, {Tauran}, {Taylor}, {Teixeira}, {Terrett}, {Tingley}, {Trager}, {Turon}, {Ulla}, {Utrilla}, {Valentini}, {van Elteren}, {Van Hemelryck}, {van Leeuwen}, {Varadi}, {Vecchiato}, {Veljanoski}, {Via}, {Vicente}, {Vogt}, {Voss}, {Votruba}, {Voutsinas}, {Walmsley}, {Weiler}, {Weingrill}, {Werner}, {Wevers}, {Whitehead}, {Wyrzykowski}, {Yoldas}, {{\v{Z}}erjal}, {Zucker}, {Zurbach}, {Zwitter}, {Alecu}, {Allen}, {Allende Prieto}, {Amorim}, {Anglada-Escud{\'e}}, {Arsenijevic}, {Azaz}, {Balm}, {Beck}, {Bernstein},
  {Bigot}, {Bijaoui}, {Blasco}, {Bonfigli}, {Bono}, {Boudreault}, {Bressan}, {Brown}, {Brunet}, {Bunclark}, {Buonanno}, {Butkevich}, {Carret}, {Carrion}, {Chemin}, {Ch{\'e}reau}, {Corcione}, {Darmigny}, {de Boer}, {de Teodoro}, {de Zeeuw}, {Delle Luche}, {Domingues}, {Dubath}, {Fodor}, {Fr{\'e}zouls}, {Fries}, {Fustes}, {Fyfe}, {Gallardo}, {Gallegos}, {Gardiol}, {Gebran}, {Gomboc}, {G{\'o}mez}, {Grux}, {Gueguen}, {Heyrovsky}, {Hoar}, {Iannicola}, {Isasi Parache}, {Janotto}, {Joliet}, {Jonckheere}, {Keil}, {Kim}, {Klagyivik}, {Klar}, {Knude}, {Kochukhov}, {Kolka}, {Kos}, {Kutka}, {Lainey}, {LeBouquin}, {Liu}, {Loreggia}, {Makarov}, {Marseille}, {Martayan}, {Martinez-Rubi}, {Massart}, {Meynadier}, {Mignot}, {Munari}, {Nguyen}, {Nordlander}, {Ocvirk}, {O'Flaherty}, {Olias Sanz}, {Ortiz}, {Osorio}, {Oszkiewicz}, {Ouzounis}, {Palmer}, {Park}, {Pasquato}, {Peltzer}, {Peralta}, {P{\'e}turaud}, {Pieniluoma}, {Pigozzi}, {Poels}, {Prat}, {Prod'homme}, {Raison}, {Rebordao}, {Risquez}, {Rocca-Volmerange}, {Rosen},
  {Ruiz-Fuertes}, {Russo}, {Sembay}, {Serraller Vizcaino}, {Short}, {Siebert}, {Silva}, {Sinachopoulos}, {Slezak}, {Soffel}, {Sosnowska}, {Strai{\v{z}}ys}, {ter Linden}, {Terrell}, {Theil}, {Tiede}, {Troisi}, {Tsalmantza}, {Tur}, {Vaccari}, {Vachier}, {Valles}, {Van Hamme}, {Veltz}, {Virtanen}, {Wallut}, {Wichmann}, {Wilkinson}, {Ziaeepour}, \& {Zschocke}}]{2016A&A...595A...1G}
{Gaia Collaboration}, {Prusti}, T., {de Bruijne}, J.~H.~J., {et~al.} 2016, \aap, 595, A1

\bibitem[{{Gaia Collaboration} {et~al.}(2023){Gaia Collaboration}, {Vallenari}, {Brown}, {Prusti}, {de Bruijne}, {Arenou}, {Babusiaux}, {Biermann}, {Creevey}, {Ducourant}, {Evans}, {Eyer}, {Guerra}, {Hutton}, {Jordi}, {Klioner}, {Lammers}, {Lindegren}, {Luri}, {Mignard}, {Panem}, {Pourbaix}, {Randich}, {Sartoretti}, {Soubiran}, {Tanga}, {Walton}, {Bailer-Jones}, {Bastian}, {Drimmel}, {Jansen}, {Katz}, {Lattanzi}, {van Leeuwen}, {Bakker}, {Cacciari}, {Casta{\~n}eda}, {De Angeli}, {Fabricius}, {Fouesneau}, {Fr{\'e}mat}, {Galluccio}, {Guerrier}, {Heiter}, {Masana}, {Messineo}, {Mowlavi}, {Nicolas}, {Nienartowicz}, {Pailler}, {Panuzzo}, {Riclet}, {Roux}, {Seabroke}, {Sordo}, {Th{\'e}venin}, {Gracia-Abril}, {Portell}, {Teyssier}, {Altmann}, {Andrae}, {Audard}, {Bellas-Velidis}, {Benson}, {Berthier}, {Blomme}, {Burgess}, {Busonero}, {Busso}, {C{\'a}novas}, {Carry}, {Cellino}, {Cheek}, {Clementini}, {Damerdji}, {Davidson}, {de Teodoro}, {Nu{\~n}ez Campos}, {Delchambre}, {Dell'Oro}, {Esquej},
  {Fern{\'a}ndez-Hern{\'a}ndez}, {Fraile}, {Garabato}, {Garc{\'\i}a-Lario}, {Gosset}, {Haigron}, {Halbwachs}, {Hambly}, {Harrison}, {Hern{\'a}ndez}, {Hestroffer}, {Hodgkin}, {Holl}, {Jan{\ss}en}, {Jevardat de Fombelle}, {Jordan}, {Krone-Martins}, {Lanzafame}, {L{\"o}ffler}, {Marchal}, {Marrese}, {Moitinho}, {Muinonen}, {Osborne}, {Pancino}, {Pauwels}, {Recio-Blanco}, {Reyl{\'e}}, {Riello}, {Rimoldini}, {Roegiers}, {Rybizki}, {Sarro}, {Siopis}, {Smith}, {Sozzetti}, {Utrilla}, {van Leeuwen}, {Abbas}, {{\'A}brah{\'a}m}, {Abreu Aramburu}, {Aerts}, {Aguado}, {Ajaj}, {Aldea-Montero}, {Altavilla}, {{\'A}lvarez}, {Alves}, {Anders}, {Anderson}, {Anglada Varela}, {Antoja}, {Baines}, {Baker}, {Balaguer-N{\'u}{\~n}ez}, {Balbinot}, {Balog}, {Barache}, {Barbato}, {Barros}, {Barstow}, {Bartolom{\'e}}, {Bassilana}, {Bauchet}, {Becciani}, {Bellazzini}, {Berihuete}, {Bernet}, {Bertone}, {Bianchi}, {Binnenfeld}, {Blanco-Cuaresma}, {Blazere}, {Boch}, {Bombrun}, {Bossini}, {Bouquillon}, {Bragaglia}, {Bramante}, {Breedt},
  {Bressan}, {Brouillet}, {Brugaletta}, {Bucciarelli}, {Burlacu}, {Butkevich}, {Buzzi}, {Caffau}, {Cancelliere}, {Cantat-Gaudin}, {Carballo}, {Carlucci}, {Carnerero}, {Carrasco}, {Casamiquela}, {Castellani}, {Castro-Ginard}, {Chaoul}, {Charlot}, {Chemin}, {Chiaramida}, {Chiavassa}, {Chornay}, {Comoretto}, {Contursi}, {Cooper}, {Cornez}, {Cowell}, {Crifo}, {Cropper}, {Crosta}, {Crowley}, {Dafonte}, {Dapergolas}, {David}, {David}, {de Laverny}, {De Luise}, {De March}, {De Ridder}, {de Souza}, {de Torres}, {del Peloso}, {del Pozo}, {Delbo}, {Delgado}, {Delisle}, {Demouchy}, {Dharmawardena}, {Di Matteo}, {Diakite}, {Diener}, {Distefano}, {Dolding}, {Edvardsson}, {Enke}, {Fabre}, {Fabrizio}, {Faigler}, {Fedorets}, {Fernique}, {Fienga}, {Figueras}, {Fournier}, {Fouron}, {Fragkoudi}, {Gai}, {Garcia-Gutierrez}, {Garcia-Reinaldos}, {Garc{\'\i}a-Torres}, {Garofalo}, {Gavel}, {Gavras}, {Gerlach}, {Geyer}, {Giacobbe}, {Gilmore}, {Girona}, {Giuffrida}, {Gomel}, {Gomez}, {Gonz{\'a}lez-N{\'u}{\~n}ez},
  {Gonz{\'a}lez-Santamar{\'\i}a}, {Gonz{\'a}lez-Vidal}, {Granvik}, {Guillout}, {Guiraud}, {Guti{\'e}rrez-S{\'a}nchez}, {Guy}, {Hatzidimitriou}, {Hauser}, {Haywood}, {Helmer}, {Helmi}, {Sarmiento}, {Hidalgo}, {Hilger}, {H{\l}adczuk}, {Hobbs}, {Holland}, {Huckle}, {Jardine}, {Jasniewicz}, {Jean-Antoine Piccolo}, {Jim{\'e}nez-Arranz}, {Jorissen}, {Juaristi Campillo}, {Julbe}, {Karbevska}, {Kervella}, {Khanna}, {Kontizas}, {Kordopatis}, {Korn}, {K{\'o}sp{\'a}l}, {Kostrzewa-Rutkowska}, {Kruszy{\'n}ska}, {Kun}, {Laizeau}, {Lambert}, {Lanza}, {Lasne}, {Le Campion}, {Lebreton}, {Lebzelter}, {Leccia}, {Leclerc}, {Lecoeur-Taibi}, {Liao}, {Licata}, {Lindstr{\o}m}, {Lister}, {Livanou}, {Lobel}, {Lorca}, {Loup}, {Madrero Pardo}, {Magdaleno Romeo}, {Managau}, {Mann}, {Manteiga}, {Marchant}, {Marconi}, {Marcos}, {Marcos Santos}, {Mar{\'\i}n Pina}, {Marinoni}, {Marocco}, {Marshall}, {Martin Polo}, {Mart{\'\i}n-Fleitas}, {Marton}, {Mary}, {Masip}, {Massari}, {Mastrobuono-Battisti}, {Mazeh}, {McMillan}, {Messina}, {Michalik},
  {Millar}, {Mints}, {Molina}, {Molinaro}, {Moln{\'a}r}, {Monari}, {Mongui{\'o}}, {Montegriffo}, {Montero}, {Mor}, {Mora}, {Morbidelli}, {Morel}, {Morris}, {Muraveva}, {Murphy}, {Musella}, {Nagy}, {Noval}, {Oca{\~n}a}, {Ogden}, {Ordenovic}, {Osinde}, {Pagani}, {Pagano}, {Palaversa}, {Palicio}, {Pallas-Quintela}, {Panahi}, {Payne-Wardenaar}, {Pe{\~n}alosa Esteller}, {Penttil{\"a}}, {Pichon}, {Piersimoni}, {Pineau}, {Plachy}, {Plum}, {Poggio}, {Pr{\v{s}}a}, {Pulone}, {Racero}, {Ragaini}, {Rainer}, {Raiteri}, {Rambaux}, {Ramos}, {Ramos-Lerate}, {Re Fiorentin}, {Regibo}, {Richards}, {Rios Diaz}, {Ripepi}, {Riva}, {Rix}, {Rixon}, {Robichon}, {Robin}, {Robin}, {Roelens}, {Rogues}, {Rohrbasser}, {Romero-G{\'o}mez}, {Rowell}, {Royer}, {Ruz Mieres}, {Rybicki}, {Sadowski}, {S{\'a}ez N{\'u}{\~n}ez}, {Sagrist{\`a} Sell{\'e}s}, {Sahlmann}, {Salguero}, {Samaras}, {Sanchez Gimenez}, {Sanna}, {Santove{\~n}a}, {Sarasso}, {Schultheis}, {Sciacca}, {Segol}, {Segovia}, {S{\'e}gransan}, {Semeux}, {Shahaf}, {Siddiqui}, {Siebert},
  {Siltala}, {Silvelo}, {Slezak}, {Slezak}, {Smart}, {Snaith}, {Solano}, {Solitro}, {Souami}, {Souchay}, {Spagna}, {Spina}, {Spoto}, {Steele}, {Steidelm{\"u}ller}, {Stephenson}, {S{\"u}veges}, {Surdej}, {Szabados}, {Szegedi-Elek}, {Taris}, {Taylor}, {Teixeira}, {Tolomei}, {Tonello}, {Torra}, {Torra}, {Torralba Elipe}, {Trabucchi}, {Tsounis}, {Turon}, {Ulla}, {Unger}, {Vaillant}, {van Dillen}, {van Reeven}, {Vanel}, {Vecchiato}, {Viala}, {Vicente}, {Voutsinas}, {Weiler}, {Wevers}, {Wyrzykowski}, {Yoldas}, {Yvard}, {Zhao}, {Zorec}, {Zucker}, \& {Zwitter}}]{2023A&A...674A...1G}
{Gaia Collaboration}, {Vallenari}, A., {Brown}, A.~G.~A., {et~al.} 2023, \aap, 674, A1

\bibitem[{{Gatto} {et~al.}(2022){Gatto}, {Ripepi}, {Bellazzini}, {Dall'ora}, {Tosi}, {Tortora}, {Cignoni}, {Cioni}, {Cusano}, {Longo}, {Marconi}, {Musella}, {Schipani}, \& {Spavone}}]{2022ApJ...929L..21G}
{Gatto}, M., {Ripepi}, V., {Bellazzini}, M., {et~al.} 2022, \apjl, 929, L21

\bibitem[{{Gatto} {et~al.}(2023){Gatto}, {Bellazzini}, {Tortora}, {Ripepi}, {Dall'Ora}, {Cignoni}, {Kuijken}, {Hildebrandt}, {Zhang}, {de Jong}, \& {Napolitano}}]{2023arXiv231106037G}
{Gatto}, M., {Bellazzini}, M., {Tortora}, C., {et~al.} 2023, arXiv e-prints, arXiv:2311.06037

\bibitem[{{Geringer-Sameth} {et~al.}(2015{\natexlab{a}}){Geringer-Sameth}, {Koushiappas}, \& {Walker}}]{GeringerSameth2015ApJ...801...74G}
{Geringer-Sameth}, A., {Koushiappas}, S.~M., \& {Walker}, M. 2015{\natexlab{a}}, \apj, 801, 74

\bibitem[{{Geringer-Sameth} {et~al.}(2015{\natexlab{b}}){Geringer-Sameth}, {Koushiappas}, \& {Walker}}]{2015PhRvD..91h3535G}
{Geringer-Sameth}, A., {Koushiappas}, S.~M., \& {Walker}, M.~G. 2015{\natexlab{b}}, \prd, 91, 083535

\bibitem[{{Gieles} {et~al.}(2021){Gieles}, {Erkal}, {Antonini}, {Balbinot}, \& {Pe{\~n}arrubia}}]{2021NatAs...5..957G}
{Gieles}, M., {Erkal}, D., {Antonini}, F., {Balbinot}, E., \& {Pe{\~n}arrubia}, J. 2021, Nature Astronomy, 5, 957

\bibitem[{{G{\'o}rski} {et~al.}(2005){G{\'o}rski}, {Hivon}, {Banday}, {Wandelt}, {Hansen}, {Reinecke}, \& {Bartelmann}}]{2005ApJ...622..759G}
{G{\'o}rski}, K.~M., {Hivon}, E., {Banday}, A.~J., {et~al.} 2005, \apj, 622, 759

\bibitem[{{Griffen} {et~al.}(2016){Griffen}, {Ji}, {Dooley}, {G{\'o}mez}, {Vogelsberger}, {O'Shea}, \& {Frebel}}]{2016ApJ...818...10G}
{Griffen}, B.~F., {Ji}, A.~P., {Dooley}, G.~A., {et~al.} 2016, \apj, 818, 10

\bibitem[{{Gullikson} {et~al.}(2014){Gullikson}, {Dodson-Robinson}, \& {Kraus}}]{2014AJ....148...53G}
{Gullikson}, K., {Dodson-Robinson}, S., \& {Kraus}, A. 2014, \aj, 148, 53

\bibitem[{{Hammer} {et~al.}(2021){Hammer}, {Wang}, {Pawlowski}, {Yang}, {Bonifacio}, {Li}, {Babusiaux}, \& {Arenou}}]{2021ApJ...922...93H}
{Hammer}, F., {Wang}, J., {Pawlowski}, M.~S., {et~al.} 2021, \apj, 922, 93

\bibitem[{{Hamren} {et~al.}(2013){Hamren}, {Smith}, {Guhathakurta}, {Dolphin}, {Weisz}, {Rajan}, \& {Grillmair}}]{2013AJ....146..116H}
{Hamren}, K.~M., {Smith}, G.~H., {Guhathakurta}, P., {et~al.} 2013, \aj, 146, 116

\bibitem[{{Hansen} {et~al.}(2024){Hansen}, {Simon}, {Li}, {Sharkey}, {Ji}, {Thompson}, {Reggiani}, \& {Galarza}}]{2024ApJ...968...21H}
{Hansen}, T.~T., {Simon}, J.~D., {Li}, T.~S., {et~al.} 2024, \apj, 968, 21

\bibitem[{{Hargis} {et~al.}(2014){Hargis}, {Willman}, \& {Peter}}]{2014ApJ...795L..13H}
{Hargis}, J.~R., {Willman}, B., \& {Peter}, A. H.~G. 2014, \apjl, 795, L13

\bibitem[{{Harris} {et~al.}(2020){Harris}, {Millman}, {van der Walt}, {Gommers}, {Virtanen}, {Cournapeau}, {Wieser}, {Taylor}, {Berg}, {Smith}, {Kern}, {Picus}, {Hoyer}, {van Kerkwijk}, {Brett}, {Haldane}, {del R{\'\i}o}, {Wiebe}, {Peterson}, {G{\'e}rard-Marchant}, {Sheppard}, {Reddy}, {Weckesser}, {Abbasi}, {Gohlke}, \& {Oliphant}}]{2020Natur.585..357H}
{Harris}, C.~R., {Millman}, K.~J., {van der Walt}, S.~J., {et~al.} 2020, \nat, 585, 357

\bibitem[{{Harris}(1996)}]{1996AJ....112.1487H}
{Harris}, W.~E. 1996, \aj, 112, 1487

\bibitem[{Hartley {et~al.}(2021)Hartley, Choi, Amon, Gruendl, Sheldon, Harrison, Bernstein, Sevilla-Noarbe, Yanny, Eckert, Diehl, Alarcon, Banerji, Bechtol, Buchs, Cantu, Conselice, Cordero, Davis, Davis, Dodelson, Drlica-Wagner, Everett, Ferté, Gruen, Honscheid, Jarvis, Johnson, Kokron, MacCrann, Myles, Pace, Palmese, Paz-Chinchón, Pereira, Plazas, Prat, Rodriguez-Monroy, Rykoff, Samuroff, Sánchez, Secco, Tarsitano, Tong, Troxel, Vasquez, Wang, Zhou, Abbott, Aguena, Allam, Annis, Bacon, Bertin, Bhargava, Brooks, Burke, Carnero Rosell, Carrasco Kind, Carretero, Castander, Costanzi, Crocce, da Costa, De Vicente, DeRose, Desai, Dietrich, Eifler, Elvin-Poole, Ferrero, Flaugher, Fosalba, García-Bellido, Gaztanaga, Gerdes, Gschwend, Gutierrez, Hinton, Hollowood, Huterer, James, Kent, Krause, Kuehn, Kuropatkin, Lahav, Lin, Maia, March, Marshall, Martini, Melchior, Menanteau, Miquel, Mohr, Morgan, Neilsen, Ogando, Pandey, Romer, Roodman, Sako, Sanchez, Scarpine, Serrano, Smith, Soares-Santos, Suchyta,
  Swanson, Tarle, Thomas, To, Varga, Walker, Wester, Wilkinson, Zuntz, \& Collaboration)}]{10.1093/mnras/stab3055}
Hartley, W.~G., Choi, A., Amon, A., {et~al.} 2021, Monthly Notices of the Royal Astronomical Society, 509, 3547.
\newblock \url{https://doi.org/10.1093/mnras/stab3055}

\bibitem[{{Heiger} {et~al.}(2024){Heiger}, {Li}, {Pace}, {Simon}, {Ji}, {Chiti}, {Bom}, {Carballo-Bello}, {Carlin}, {Cerny}, {Choi}, {Drlica-Wagner}, {James}, {Mart{\'\i}nez-V{\'a}zquez}, {Medina}, {Mutlu-Pakdil}, {Navabi}, {No{\"e}l}, {Sakowska}, {Stringfellow}, \& {DELVE Collaboration}}]{2024ApJ...961..234H}
{Heiger}, M.~E., {Li}, T.~S., {Pace}, A.~B., {et~al.} 2024, \apj, 961, 234

\bibitem[{{Hinton}(2019)}]{2019ascl.soft10017H}
{Hinton}, S.~R. 2019, {ChainConsumer: Corner plots, LaTeX tables and plotting walks}, Astrophysics Source Code Library, record ascl:1910.017, , , ascl:1910.017

\bibitem[{{Homma} {et~al.}(2018){Homma}, {Chiba}, {Okamoto}, {Komiyama}, {Tanaka}, {Tanaka}, {Ishigaki}, {Hayashi}, {Arimoto}, {Garmilla}, {Lupton}, {Strauss}, {Miyazaki}, {Wang}, \& {Murayama}}]{2018PASJ...70S..18H}
{Homma}, D., {Chiba}, M., {Okamoto}, S., {et~al.} 2018, \pasj, 70, S18

\bibitem[{{Homma} {et~al.}(2019){Homma}, {Chiba}, {Komiyama}, {Tanaka}, {Okamoto}, {Tanaka}, {Ishigaki}, {Hayashi}, {Arimoto}, {Carlsten}, {Lupton}, {Strauss}, {Miyazaki}, {Torrealba}, {Wang}, \& {Murayama}}]{2019PASJ...71...94H}
{Homma}, D., {Chiba}, M., {Komiyama}, Y., {et~al.} 2019, \pasj, 71, 94

\bibitem[{{Homma} {et~al.}(2023){Homma}, {Chiba}, {Komiyama}, {Tanaka}, {Okamoto}, {Tanaka}, {Ishigaki}, {Hayashi}, {Arimoto}, {Lupton}, {Strauss}, {Miyazaki}, {Wang}, \& {Murayama}}]{2023arXiv231105439H}
---. 2023, arXiv e-prints, arXiv:2311.05439

\bibitem[{{Hou} {et~al.}(2014){Hou}, {Parker}, \& {Harris}}]{2014MNRAS.442..406H}
{Hou}, A., {Parker}, L.~C., \& {Harris}, W.~E. 2014, \mnras, 442, 406

\bibitem[{{Husser} {et~al.}(2013){Husser}, {Wende-von Berg}, {Dreizler}, {Homeier}, {Reiners}, {Barman}, \& {Hauschildt}}]{2013A&A...553A...6H}
{Husser}, T.~O., {Wende-von Berg}, S., {Dreizler}, S., {et~al.} 2013, \aap, 553, A6

\bibitem[{{Jenkins} {et~al.}(2021){Jenkins}, {Li}, {Pace}, {Ji}, {Koposov}, \& {Mutlu-Pakdil}}]{2021ApJ...920...92J}
{Jenkins}, S.~A., {Li}, T.~S., {Pace}, A.~B., {et~al.} 2021, \apj, 920, 92

\bibitem[{{Jeon} {et~al.}(2021){Jeon}, {Bromm}, {Besla}, {Yoon}, \& {Choi}}]{2021MNRAS.502....1J}
{Jeon}, M., {Bromm}, V., {Besla}, G., {Yoon}, J., \& {Choi}, Y. 2021, \mnras, 502, 1

\bibitem[{{Jethwa} {et~al.}(2018){Jethwa}, {Erkal}, \& {Belokurov}}]{2018MNRAS.473.2060J}
{Jethwa}, P., {Erkal}, D., \& {Belokurov}, V. 2018, \mnras, 473, 2060

\bibitem[{{Ji} {et~al.}(2019){Ji}, {Simon}, {Frebel}, {Venn}, \& {Hansen}}]{2019ApJ...870...83J}
{Ji}, A.~P., {Simon}, J.~D., {Frebel}, A., {Venn}, K.~A., \& {Hansen}, T.~T. 2019, \apj, 870, 83

\bibitem[{{Ji} {et~al.}(2021){Ji}, {Koposov}, {Li}, {Erkal}, {Pace}, {Simon}, {Belokurov}, {Cullinane}, {Da Costa}, {Kuehn}, {Lewis}, {Mackey}, {Shipp}, {Simpson}, {Zucker}, {Hansen}, {Bland-Hawthorn}, \& {S5 Collaboration}}]{2021ApJ...921...32J}
{Ji}, A.~P., {Koposov}, S.~E., {Li}, T.~S., {et~al.} 2021, \apj, 921, 32

\bibitem[{{Jones} {et~al.}(2023){Jones}, {Mutlu-Pakdil}, {Sand}, {Donnerstein}, {Crnojevi{\'c}}, {Bennet}, {Fielder}, {Karunakaran}, {Spekkens}, {Strader}, {Urquhart}, \& {Zaritsky}}]{2023ApJ...957L...5J}
{Jones}, M.~G., {Mutlu-Pakdil}, B., {Sand}, D.~J., {et~al.} 2023, \apjl, 957, L5

\bibitem[{{Joshi} {et~al.}(2019){Joshi}, {Parker}, {Wadsley}, \& {Keller}}]{2019MNRAS.483..235J}
{Joshi}, G.~D., {Parker}, L.~C., {Wadsley}, J., \& {Keller}, B.~W. 2019, \mnras, 483, 235

\bibitem[{{Kallivayalil} {et~al.}(2013){Kallivayalil}, {van der Marel}, {Besla}, {Anderson}, \& {Alcock}}]{2013ApJ...764..161K}
{Kallivayalil}, N., {van der Marel}, R.~P., {Besla}, G., {Anderson}, J., \& {Alcock}, C. 2013, \apj, 764, 161

\bibitem[{Kauffmann {et~al.}(1993)Kauffmann, White, \& Guiderdoni}]{10.1093/mnras/264.1.201}
Kauffmann, G., White, S. D.~M., \& Guiderdoni, B. 1993, Monthly Notices of the Royal Astronomical Society, 264, 201.
\newblock \url{https://doi.org/10.1093/mnras/264.1.201}

\bibitem[{{Kelson}(2003)}]{2003PASP..115..688K}
{Kelson}, D.~D. 2003, \pasp, 115, 688

\bibitem[{{Kelson} {et~al.}(2000){Kelson}, {Illingworth}, {van Dokkum}, \& {Franx}}]{2000ApJ...531..159K}
{Kelson}, D.~D., {Illingworth}, G.~D., {van Dokkum}, P.~G., \& {Franx}, M. 2000, \apj, 531, 159

\bibitem[{{Kim} \& {Jerjen}(2015{\natexlab{a}})}]{2015ApJ...808L..39K}
{Kim}, D., \& {Jerjen}, H. 2015{\natexlab{a}}, \apjl, 808, L39

\bibitem[{{Kim} \& {Jerjen}(2015{\natexlab{b}})}]{2015ApJ...799...73K}
---. 2015{\natexlab{b}}, \apj, 799, 73

\bibitem[{{Kim} {et~al.}(2016{\natexlab{a}}){Kim}, {Jerjen}, {Mackey}, {Da Costa}, \& {Milone}}]{2016ApJ...820..119K}
{Kim}, D., {Jerjen}, H., {Mackey}, D., {Da Costa}, G.~S., \& {Milone}, A.~P. 2016{\natexlab{a}}, \apj, 820, 119

\bibitem[{{Kim} {et~al.}(2015){Kim}, {Jerjen}, {Milone}, {Mackey}, \& {Da Costa}}]{2015ApJ...803...63K}
{Kim}, D., {Jerjen}, H., {Milone}, A.~P., {Mackey}, D., \& {Da Costa}, G.~S. 2015, \apj, 803, 63

\bibitem[{{Kim} {et~al.}(2016{\natexlab{b}}){Kim}, {Jerjen}, {Geha}, {Chiti}, {Milone}, {Da Costa}, {Mackey}, {Frebel}, \& {Conn}}]{2016ApJ...833...16K}
{Kim}, D., {Jerjen}, H., {Geha}, M., {et~al.} 2016{\natexlab{b}}, \apj, 833, 16

\bibitem[{{Kim} {et~al.}(2018){Kim}, {Peter}, \& {Hargis}}]{2018PhRvL.121u1302K}
{Kim}, S.~Y., {Peter}, A. H.~G., \& {Hargis}, J.~R. 2018, \prl, 121, 211302

\bibitem[{{Kirby} {et~al.}(2013{\natexlab{a}}){Kirby}, {Boylan-Kolchin}, {Cohen}, {Geha}, {Bullock}, \& {Kaplinghat}}]{2013ApJ...770...16K}
{Kirby}, E.~N., {Boylan-Kolchin}, M., {Cohen}, J.~G., {et~al.} 2013{\natexlab{a}}, \apj, 770, 16

\bibitem[{{Kirby} {et~al.}(2013{\natexlab{b}}){Kirby}, {Cohen}, {Guhathakurta}, {Cheng}, {Bullock}, \& {Gallazzi}}]{2013ApJ...779..102K}
{Kirby}, E.~N., {Cohen}, J.~G., {Guhathakurta}, P., {et~al.} 2013{\natexlab{b}}, \apj, 779, 102

\bibitem[{{Kirby} {et~al.}(2017){Kirby}, {Cohen}, {Simon}, {Guhathakurta}, {Thygesen}, \& {Duggan}}]{2017ApJ...838...83K}
{Kirby}, E.~N., {Cohen}, J.~G., {Simon}, J.~D., {et~al.} 2017, \apj, 838, 83

\bibitem[{{Kirby} {et~al.}(2015){Kirby}, {Simon}, \& {Cohen}}]{2015ApJ...810...56K}
{Kirby}, E.~N., {Simon}, J.~D., \& {Cohen}, J.~G. 2015, \apj, 810, 56

\bibitem[{{Kleyna} {et~al.}(2005){Kleyna}, {Wilkinson}, {Evans}, \& {Gilmore}}]{2005ApJ...630L.141K}
{Kleyna}, J.~T., {Wilkinson}, M.~I., {Evans}, N.~W., \& {Gilmore}, G. 2005, \apjl, 630, L141

\bibitem[{{Klypin} {et~al.}(1999){Klypin}, {Kravtsov}, {Valenzuela}, \& {Prada}}]{1999ApJ...522...82K}
{Klypin}, A., {Kravtsov}, A.~V., {Valenzuela}, O., \& {Prada}, F. 1999, \apj, 522, 82

\bibitem[{{Kobulnicky} {et~al.}(2005){Kobulnicky}, {Monson}, {Buckalew}, {Darnel}, {Uzpen}, {Meade}, {Babler}, {Indebetouw}, {Whitney}, {Watson}, {Churchwell}, {Wolfire}, {Wolff}, {Clemens}, {Shah}, {Bania}, {Benjamin}, {Cohen}, {Dickey}, {Jackson}, {Marston}, {Mathis}, {Mercer}, {Stauffer}, {Stolovy}, {Norris}, {Kutyrev}, {Canterna}, \& {Pierce}}]{2005AJ....129..239K}
{Kobulnicky}, H.~A., {Monson}, A.~J., {Buckalew}, B.~A., {et~al.} 2005, \aj, 129, 239

\bibitem[{{Koposov} {et~al.}(2015{\natexlab{a}}){Koposov}, {Belokurov}, {Torrealba}, \& {Evans}}]{2015ApJ...805..130K}
{Koposov}, S.~E., {Belokurov}, V., {Torrealba}, G., \& {Evans}, N.~W. 2015{\natexlab{a}}, \apj, 805, 130

\bibitem[{{Koposov} {et~al.}(2009){Koposov}, {Yoo}, {Rix}, {Weinberg}, {Macci{\`o}}, \& {Escud{\'e}}}]{2009ApJ...696.2179K}
{Koposov}, S.~E., {Yoo}, J., {Rix}, H.-W., {et~al.} 2009, \apj, 696, 2179

\bibitem[{{Koposov} {et~al.}(2011){Koposov}, {Gilmore}, {Walker}, {Belokurov}, {Evans}, {Fellhauer}, {Gieren}, {Geisler}, {Monaco}, {Norris}, {Okamoto}, {Pe{\~n}arrubia}, {Wilkinson}, {Wyse}, \& {Zucker}}]{2011ApJ...736..146K}
{Koposov}, S.~E., {Gilmore}, G., {Walker}, M.~G., {et~al.} 2011, \apj, 736, 146

\bibitem[{{Koposov} {et~al.}(2015{\natexlab{b}}){Koposov}, {Casey}, {Belokurov}, {Lewis}, {Gilmore}, {Worley}, {Hourihane}, {Randich}, {Bensby}, {Bragaglia}, {Bergemann}, {Carraro}, {Costado}, {Flaccomio}, {Francois}, {Heiter}, {Hill}, {Jofre}, {Lando}, {Lanzafame}, {de Laverny}, {Monaco}, {Morbidelli}, {Sbordone}, {Mikolaitis}, \& {Ryde}}]{2015ApJ...811...62K}
{Koposov}, S.~E., {Casey}, A.~R., {Belokurov}, V., {et~al.} 2015{\natexlab{b}}, \apj, 811, 62

\bibitem[{{Koposov} {et~al.}(2018){Koposov}, {Walker}, {Belokurov}, {Casey}, {Geringer-Sameth}, {Mackey}, {Da Costa}, {Erkal}, {Jethwa}, {Mateo}, {Olszewski}, \& {Bailey}}]{2018MNRAS.479.5343K}
{Koposov}, S.~E., {Walker}, M.~G., {Belokurov}, V., {et~al.} 2018, \mnras, 479, 5343

\bibitem[{{Kurtev} {et~al.}(2008){Kurtev}, {Ivanov}, {Borissova}, \& {Ortolani}}]{2008A&A...489..583K}
{Kurtev}, R., {Ivanov}, V.~D., {Borissova}, J., \& {Ortolani}, S. 2008, \aap, 489, 583

\bibitem[{{Leanza} {et~al.}(2024){Leanza}, {Pallanca}, {Ferraro}, {Lanzoni}, {Vesperini}, {Cadelano}, {Origlia}, {Fanelli}, {Dalessandro}, \& {Valenti}}]{2024arXiv240513558L}
{Leanza}, S., {Pallanca}, C., {Ferraro}, F.~R., {et~al.} 2024, arXiv e-prints, arXiv:2405.13558

\bibitem[{{Lee} {et~al.}(2013){Lee}, {Beers}, {Masseron}, {Plez}, {Rockosi}, {Sobeck}, {Yanny}, {Lucatello}, {Sivarani}, {Placco}, \& {Carollo}}]{2013AJ....146..132L}
{Lee}, Y.~S., {Beers}, T.~C., {Masseron}, T., {et~al.} 2013, \aj, 146, 132

\bibitem[{{Li} {et~al.}(2017){Li}, {Simon}, {Drlica-Wagner}, {Bechtol}, {Wang}, {Garc{\'\i}a-Bellido}, {Frieman}, {Marshall}, {James}, {Strigari}, {Pace}, {Balbinot}, {Zhang}, {Abbott}, {Allam}, {Benoit-L{\'e}vy}, {Bernstein}, {Bertin}, {Brooks}, {Burke}, {Carnero Rosell}, {Carrasco Kind}, {Carretero}, {Cunha}, {D'Andrea}, {da Costa}, {DePoy}, {Desai}, {Diehl}, {Eifler}, {Flaugher}, {Goldstein}, {Gruen}, {Gruendl}, {Gschwend}, {Gutierrez}, {Krause}, {Kuehn}, {Lin}, {Maia}, {March}, {Menanteau}, {Miquel}, {Plazas}, {Romer}, {Sanchez}, {Santiago}, {Schubnell}, {Sevilla-Noarbe}, {Smith}, {Sobreira}, {Suchyta}, {Tarle}, {Thomas}, {Tucker}, {Walker}, {Wechsler}, {Wester}, {Yanny}, \& {DES Collaboration}}]{2017ApJ...838....8L}
{Li}, T.~S., {Simon}, J.~D., {Drlica-Wagner}, A., {et~al.} 2017, \apj, 838, 8

\bibitem[{{Li} {et~al.}(2018{\natexlab{a}}){Li}, {Simon}, {Kuehn}, {Pace}, {Erkal}, {Bechtol}, {Yanny}, {Drlica-Wagner}, {Marshall}, {Lidman}, {Balbinot}, {Carollo}, {Jenkins}, {Mart{\'\i}nez-V{\'a}zquez}, {Shipp}, {Stringer}, {Vivas}, {Walker}, {Wechsler}, {Abdalla}, {Allam}, {Annis}, {Avila}, {Bertin}, {Brooks}, {Buckley-Geer}, {Burke}, {Carnero Rosell}, {Carrasco Kind}, {Carretero}, {Cunha}, {D'Andrea}, {da Costa}, {Davis}, {De Vicente}, {Doel}, {Eifler}, {Evrard}, {Flaugher}, {Frieman}, {Garc{\'\i}a-Bellido}, {Gaztanaga}, {Gerdes}, {Gruen}, {Gruendl}, {Gschwend}, {Gutierrez}, {Hartley}, {Hollowood}, {Honscheid}, {James}, {Krause}, {Maia}, {March}, {Menanteau}, {Miquel}, {Plazas}, {Sanchez}, {Santiago}, {Scarpine}, {Schindler}, {Schubnell}, {Sevilla-Noarbe}, {Smith}, {Smith}, {Soares-Santos}, {Sobreira}, {Suchyta}, {Swanson}, {Tarle}, {Tucker}, \& {DES Collaboration}}]{2018ApJ...866...22L}
{Li}, T.~S., {Simon}, J.~D., {Kuehn}, K., {et~al.} 2018{\natexlab{a}}, \apj, 866, 22

\bibitem[{{Li} {et~al.}(2018{\natexlab{b}}){Li}, {Simon}, {Pace}, {Torrealba}, {Kuehn}, {Drlica-Wagner}, {Bechtol}, {Vivas}, {van der Marel}, {Wood}, {Yanny}, {Belokurov}, {Jethwa}, {Zucker}, {Lewis}, {Kron}, {Nidever}, {S{\'a}nchez-Conde}, {Ji}, {Conn}, {James}, {Martin}, {Martinez-Delgado}, {No{\"e}l}, \& {MagLiteS Collaboration}}]{2018ApJ...857..145L}
{Li}, T.~S., {Simon}, J.~D., {Pace}, A.~B., {et~al.} 2018{\natexlab{b}}, \apj, 857, 145

\bibitem[{{Lindegren} {et~al.}(2021){Lindegren}, {Klioner}, {Hern{\'a}ndez}, {Bombrun}, {Ramos-Lerate}, {Steidelm{\"u}ller}, {Bastian}, {Biermann}, {de Torres}, {Gerlach}, {Geyer}, {Hilger}, {Hobbs}, {Lammers}, {McMillan}, {Stephenson}, {Casta{\~n}eda}, {Davidson}, {Fabricius}, {Gracia-Abril}, {Portell}, {Rowell}, {Teyssier}, {Torra}, {Bartolom{\'e}}, {Clotet}, {Garralda}, {Gonz{\'a}lez-Vidal}, {Torra}, {Abbas}, {Altmann}, {Anglada Varela}, {Balaguer-N{\'u}{\~n}ez}, {Balog}, {Barache}, {Becciani}, {Bernet}, {Bertone}, {Bianchi}, {Bouquillon}, {Brown}, {Bucciarelli}, {Busonero}, {Butkevich}, {Buzzi}, {Cancelliere}, {Carlucci}, {Charlot}, {Cioni}, {Crosta}, {Crowley}, {del Peloso}, {del Pozo}, {Drimmel}, {Esquej}, {Fienga}, {Fraile}, {Gai}, {Garcia-Reinaldos}, {Guerra}, {Hambly}, {Hauser}, {Jan{\ss}en}, {Jordan}, {Kostrzewa-Rutkowska}, {Lattanzi}, {Liao}, {Licata}, {Lister}, {L{\"o}ffler}, {Marchant}, {Masip}, {Mignard}, {Mints}, {Molina}, {Mora}, {Morbidelli}, {Murphy}, {Pagani}, {Panuzzo}, {Pe{\~n}alosa
  Esteller}, {Poggio}, {Re Fiorentin}, {Riva}, {Sagrist{\`a} Sell{\'e}s}, {Sanchez Gimenez}, {Sarasso}, {Sciacca}, {Siddiqui}, {Smart}, {Souami}, {Spagna}, {Steele}, {Taris}, {Utrilla}, {van Reeven}, \& {Vecchiato}}]{2021A&A...649A...2L}
{Lindegren}, L., {Klioner}, S.~A., {Hern{\'a}ndez}, J., {et~al.} 2021, \aap, 649, A2

\bibitem[{{Longeard} {et~al.}(2019){Longeard}, {Martin}, {Ibata}, {Collins}, {Laevens}, {Bell}, \& {Mackey}}]{2019MNRAS.490.1498L}
{Longeard}, N., {Martin}, N., {Ibata}, R.~A., {et~al.} 2019, \mnras, 490, 1498

\bibitem[{{Longeard} {et~al.}(2018){Longeard}, {Martin}, {Starkenburg}, {Ibata}, {Collins}, {Geha}, {Laevens}, {Rich}, {Aguado}, {Arentsen}, {Carlberg}, {C{\^o}t{\'e}}, {Hill}, {Jablonka}, {Gonz{\'a}lez Hern{\'a}ndez}, {Navarro}, {S{\'a}nchez-Janssen}, {Tolstoy}, {Venn}, \& {Youakim}}]{2018MNRAS.480.2609L}
{Longeard}, N., {Martin}, N., {Starkenburg}, E., {et~al.} 2018, \mnras, 480, 2609

\bibitem[{{Lucatello} {et~al.}(2006){Lucatello}, {Beers}, {Christlieb}, {Barklem}, {Rossi}, {Marsteller}, {Sivarani}, \& {Lee}}]{2006ApJ...652L..37L}
{Lucatello}, S., {Beers}, T.~C., {Christlieb}, N., {et~al.} 2006, \apjl, 652, L37

\bibitem[{{Luque} {et~al.}(2018){Luque}, {Santiago}, {Pieres}, {Marshall}, {Pace}, {Kron}, {Drlica-Wagner}, {Queiroz}, {Balbinot}, {dal Ponte}, {Fausti Neto}, {da Costa}, {Maia}, {Walker}, {Abdalla}, {Allam}, {Annis}, {Bechtol}, {Benoit-L{\'e}vy}, {Bertin}, {Brooks}, {Carnero Rosell}, {Carrasco Kind}, {Carretero}, {Crocce}, {Davis}, {Doel}, {Eifler}, {Flaugher}, {Garc{\'\i}a-Bellido}, {Gerdes}, {Gruen}, {Gruendl}, {Gutierrez}, {Honscheid}, {James}, {Kuehn}, {Kuropatkin}, {Miquel}, {Nichol}, {Plazas}, {Sanchez}, {Scarpine}, {Schindler}, {Sevilla-Noarbe}, {Smith}, {Soares-Santos}, {Sobreira}, {Suchyta}, {Tarle}, \& {Thomas}}]{2018MNRAS.478.2006L}
{Luque}, E., {Santiago}, B., {Pieres}, A., {et~al.} 2018, \mnras, 478, 2006

\bibitem[{Macciò {et~al.}(2010)Macciò, Kang, Fontanot, Somerville, Koposov, \& Monaco}]{10.1111/j.1365-2966.2009.16031.x}
Macciò, A.~V., Kang, X., Fontanot, F., {et~al.} 2010, Monthly Notices of the Royal Astronomical Society, 402, 1995.
\newblock \url{https://doi.org/10.1111/j.1365-2966.2009.16031.x}

\bibitem[{{Manwadkar} \& {Kravtsov}(2022)}]{2022MNRAS.516.3944M}
{Manwadkar}, V., \& {Kravtsov}, A.~V. 2022, \mnras, 516, 3944

\bibitem[{{Martin} {et~al.}(2008){Martin}, {de Jong}, \& {Rix}}]{2008ApJ...684.1075M}
{Martin}, N.~F., {de Jong}, J. T.~A., \& {Rix}, H.-W. 2008, \apj, 684, 1075

\bibitem[{{Martin} {et~al.}(2007){Martin}, {Ibata}, {Chapman}, {Irwin}, \& {Lewis}}]{2007MNRAS.380..281M}
{Martin}, N.~F., {Ibata}, R.~A., {Chapman}, S.~C., {Irwin}, M., \& {Lewis}, G.~F. 2007, \mnras, 380, 281

\bibitem[{{Martin} {et~al.}(2016){Martin}, {Jungbluth}, {Nidever}, {Bell}, {Besla}, {Blum}, {Cioni}, {Conn}, {Kaleida}, {Gallart}, {Jin}, {Majewski}, {Martinez-Delgado}, {Monachesi}, {Mu{\~n}oz}, {No{\"e}l}, {Olsen}, {Stringfellow}, {van der Marel}, {Vivas}, {Walker}, \& {Zaritsky}}]{2016ApJ...830L..10M}
{Martin}, N.~F., {Jungbluth}, V., {Nidever}, D.~L., {et~al.} 2016, \apjl, 830, L10

\bibitem[{{Mart{\'\i}nez-Delgado} {et~al.}(2022){Mart{\'\i}nez-Delgado}, {Karim}, {Charles}, {Boschin}, {Monelli}, {Collins}, {Donatiello}, \& {Alfaro}}]{2022MNRAS.509...16M}
{Mart{\'\i}nez-Delgado}, D., {Karim}, N., {Charles}, E. J.~E., {et~al.} 2022, \mnras, 509, 16

\bibitem[{{Mateo} {et~al.}(2008){Mateo}, {Olszewski}, \& {Walker}}]{2008ApJ...675..201M}
{Mateo}, M., {Olszewski}, E.~W., \& {Walker}, M.~G. 2008, \apj, 675, 201

\bibitem[{{Mau} {et~al.}(2020){Mau}, {Cerny}, {Pace}, {Choi}, {Drlica-Wagner}, {Santana-Silva}, {Riley}, {Erkal}, {Stringfellow}, {Adam{\'o}w}, {Carlin}, {Gruendl}, {Hernandez-Lang}, {Kuropatkin}, {Li}, {Mart{\'\i}nez-V{\'a}zquez}, {Morganson}, {Mutlu-Pakdil}, {Neilsen}, {Nidever}, {Olsen}, {Sand}, {Tollerud}, {Tucker}, {Yanny}, {Zenteno}, {Allam}, {Barkhouse}, {Bechtol}, {Bell}, {Balaji}, {Crnojevi{\'c}}, {Esteves}, {Ferguson}, {Gallart}, {Hughes}, {James}, {Jethwa}, {Johnson}, {Kuehn}, {Majewski}, {Mao}, {Massana}, {McNanna}, {Monachesi}, {Nadler}, {No{\"e}l}, {Palmese}, {Paz-Chinchon}, {Pieres}, {Sanchez}, {Shipp}, {Simon}, {Soares-Santos}, {Tavangar}, {van der Marel}, {Vivas}, {Walker}, \& {Wechsler}}]{2020ApJ...890..136M}
{Mau}, S., {Cerny}, W., {Pace}, A.~B., {et~al.} 2020, \apj, 890, 136

\bibitem[{{Mau} {et~al.}(2022){Mau}, {Nadler}, {Wechsler}, {Drlica-Wagner}, {Bechtol}, {Green}, {Huterer}, {Li}, {Mao}, {Mart{\'\i}nez-V{\'a}zquez}, {McNanna}, {Mutlu-Pakdil}, {Pace}, {Peter}, {Riley}, {Strigari}, {Wang}, {Aguena}, {Allam}, {Annis}, {Bacon}, {Bertin}, {Bocquet}, {Brooks}, {Burke}, {Carnero Rosell}, {Carrasco Kind}, {Carretero}, {Costanzi}, {Crocce}, {Pereira}, {Davis}, {De Vicente}, {Desai}, {Doel}, {Ferrero}, {Flaugher}, {Frieman}, {Garc{\'\i}a-Bellido}, {Gatti}, {Giannini}, {Gruen}, {Gruendl}, {Gschwend}, {Gutierrez}, {Hinton}, {Hollowood}, {Honscheid}, {James}, {Kuehn}, {Lahav}, {Maia}, {Marshall}, {Miquel}, {Mohr}, {Morgan}, {Ogando}, {Paz-Chinch{\'o}n}, {Pieres}, {Rodriguez-Monroy}, {Sanchez}, {Scarpine}, {Serrano}, {Sevilla-Noarbe}, {Suchyta}, {Tarle}, {To}, {Tucker}, {Weller}, \& {DES Collaboration}}]{2022ApJ...932..128M}
{Mau}, S., {Nadler}, E.~O., {Wechsler}, R.~H., {et~al.} 2022, \apj, 932, 128

\bibitem[{{McConnachie} \& {C{\^o}t{\'e}}(2010)}]{2010ApJ...722L.209M}
{McConnachie}, A.~W., \& {C{\^o}t{\'e}}, P. 2010, \apjl, 722, L209

\bibitem[{{McDaniel} {et~al.}(2023){McDaniel}, {Ajello}, {Karwin}, {Di Mauro}, {Drlica-Wagner}, \& {Sanchez-Conde}}]{2023arXiv231104982M}
{McDaniel}, A., {Ajello}, M., {Karwin}, C.~M., {et~al.} 2023, arXiv e-prints, arXiv:2311.04982

\bibitem[{McGee {et~al.}(2009)McGee, Balogh, Bower, Font, \& McCarthy}]{10.1111/j.1365-2966.2009.15507.x}
McGee, S.~L., Balogh, M.~L., Bower, R.~G., Font, A.~S., \& McCarthy, I.~G. 2009, Monthly Notices of the Royal Astronomical Society, 400, 937.
\newblock \url{https://doi.org/10.1111/j.1365-2966.2009.15507.x}

\bibitem[{{McMillan}(2017)}]{2017MNRAS.465...76M}
{McMillan}, P.~J. 2017, \mnras, 465, 76

\bibitem[{{McQuinn} {et~al.}(2023{\natexlab{a}}){McQuinn}, {Mao}, {Buckley}, {Shih}, {Cohen}, \& {Dolphin}}]{PegasusW}
{McQuinn}, K. B.~W., {Mao}, Y.-Y., {Buckley}, M.~R., {et~al.} 2023{\natexlab{a}}, \apj, 944, 14

\bibitem[{{McQuinn} {et~al.}(2023{\natexlab{b}}){McQuinn}, {Mao}, {Cohen}, {Shih}, {Buckley}, \& {Dolphin}}]{2023arXiv230708738M}
{McQuinn}, K. B.~W., {Mao}, Y.-Y., {Cohen}, R.~E., {et~al.} 2023{\natexlab{b}}, arXiv e-prints, arXiv:2307.08738

\bibitem[{{Minor} {et~al.}(2010){Minor}, {Martinez}, {Bullock}, {Kaplinghat}, \& {Trainor}}]{2010ApJ...721.1142M}
{Minor}, Q.~E., {Martinez}, G., {Bullock}, J., {Kaplinghat}, M., \& {Trainor}, R. 2010, \apj, 721, 1142

\bibitem[{{Moore} {et~al.}(1999){Moore}, {Ghigna}, {Governato}, {Lake}, {Quinn}, {Stadel}, \& {Tozzi}}]{1999ApJ...524L..19M}
{Moore}, B., {Ghigna}, S., {Governato}, F., {et~al.} 1999, \apjl, 524, L19

\bibitem[{{Moskowitz} \& {Walker}(2020)}]{2020ApJ...892...27M}
{Moskowitz}, A.~G., \& {Walker}, M.~G. 2020, \apj, 892, 27

\bibitem[{{Mu{\~n}oz} {et~al.}(2006){Mu{\~n}oz}, {Carlin}, {Frinchaboy}, {Nidever}, {Majewski}, \& {Patterson}}]{2006ApJ...650L..51M}
{Mu{\~n}oz}, R.~R., {Carlin}, J.~L., {Frinchaboy}, P.~M., {et~al.} 2006, \apjl, 650, L51

\bibitem[{{Mu{\~n}oz} {et~al.}(2018){Mu{\~n}oz}, {C{\^o}t{\'e}}, {Santana}, {Geha}, {Simon}, {Oyarz{\'u}n}, {Stetson}, \& {Djorgovski}}]{2018ApJ...860...66M}
{Mu{\~n}oz}, R.~R., {C{\^o}t{\'e}}, P., {Santana}, F.~A., {et~al.} 2018, \apj, 860, 66

\bibitem[{{Munshi} {et~al.}(2019){Munshi}, {Brooks}, {Christensen}, {Applebaum}, {Holley-Bockelmann}, {Quinn}, \& {Wadsley}}]{2019ApJ...874...40M}
{Munshi}, F., {Brooks}, A.~M., {Christensen}, C., {et~al.} 2019, \apj, 874, 40

\bibitem[{{Mutlu-Pakdil} {et~al.}(2018){Mutlu-Pakdil}, {Sand}, {Carlin}, {Spekkens}, {Caldwell}, {Crnojevi{\'c}}, {Hughes}, {Willman}, \& {Zaritsky}}]{2018ApJ...863...25M}
{Mutlu-Pakdil}, B., {Sand}, D.~J., {Carlin}, J.~L., {et~al.} 2018, \apj, 863, 25

\bibitem[{{Mutlu-Pakdil} {et~al.}(2022){Mutlu-Pakdil}, {Sand}, {Crnojevi{\'c}}, {Jones}, {Caldwell}, {Guhathakurta}, {Seth}, {Simon}, {Spekkens}, {Strader}, \& {Toloba}}]{2022ApJ...926...77M}
{Mutlu-Pakdil}, B., {Sand}, D.~J., {Crnojevi{\'c}}, D., {et~al.} 2022, \apj, 926, 77

\bibitem[{{Nadler} {et~al.}(2024){Nadler}, {Gluscevic}, {Driskell}, {Wechsler}, {Moustakas}, {Benson}, \& {Mao}}]{2024arXiv240110318N}
{Nadler}, E.~O., {Gluscevic}, V., {Driskell}, T., {et~al.} 2024, arXiv e-prints, arXiv:2401.10318

\bibitem[{{Nadler} {et~al.}(2020){Nadler}, {Wechsler}, {Bechtol}, {Mao}, {Green}, {Drlica-Wagner}, {McNanna}, {Mau}, {Pace}, {Simon}, {Kravtsov}, {Dodelson}, {Li}, {Riley}, {Wang}, {Abbott}, {Aguena}, {Allam}, {Annis}, {Avila}, {Bernstein}, {Bertin}, {Brooks}, {Burke}, {Rosell}, {Kind}, {Carretero}, {Costanzi}, {da Costa}, {De Vicente}, {Desai}, {Evrard}, {Flaugher}, {Fosalba}, {Frieman}, {Garc{\'\i}a-Bellido}, {Gaztanaga}, {Gerdes}, {Gruen}, {Gschwend}, {Gutierrez}, {Hartley}, {Hinton}, {Honscheid}, {Krause}, {Kuehn}, {Kuropatkin}, {Lahav}, {Maia}, {Marshall}, {Menanteau}, {Miquel}, {Palmese}, {Paz-Chinch{\'o}n}, {Plazas}, {Romer}, {Sanchez}, {Santiago}, {Scarpine}, {Serrano}, {Smith}, {Soares-Santos}, {Suchyta}, {Tarle}, {Thomas}, {Varga}, {Walker}, \& {DES Collaboration}}]{2020ApJ...893...48N}
{Nadler}, E.~O., {Wechsler}, R.~H., {Bechtol}, K., {et~al.} 2020, \apj, 893, 48

\bibitem[{{Nadler} {et~al.}(2021){Nadler}, {Drlica-Wagner}, {Bechtol}, {Mau}, {Wechsler}, {Gluscevic}, {Boddy}, {Pace}, {Li}, {McNanna}, {Riley}, {Garc{\'\i}a-Bellido}, {Mao}, {Green}, {Burke}, {Peter}, {Jain}, {Abbott}, {Aguena}, {Allam}, {Annis}, {Avila}, {Brooks}, {Carrasco Kind}, {Carretero}, {Costanzi}, {da Costa}, {De Vicente}, {Desai}, {Diehl}, {Doel}, {Everett}, {Evrard}, {Flaugher}, {Frieman}, {Gerdes}, {Gruen}, {Gruendl}, {Gschwend}, {Gutierrez}, {Hinton}, {Honscheid}, {Huterer}, {James}, {Krause}, {Kuehn}, {Kuropatkin}, {Lahav}, {Maia}, {Marshall}, {Menanteau}, {Miquel}, {Palmese}, {Paz-Chinch{\'o}n}, {Plazas}, {Romer}, {Sanchez}, {Scarpine}, {Serrano}, {Sevilla-Noarbe}, {Smith}, {Soares-Santos}, {Suchyta}, {Swanson}, {Tarle}, {Tucker}, {Walker}, {Wester}, \& {DES Collaboration}}]{2021PhRvL.126i1101N}
{Nadler}, E.~O., {Drlica-Wagner}, A., {Bechtol}, K., {et~al.} 2021, \prl, 126, 091101

\bibitem[{{Navarro} {et~al.}(1997){Navarro}, {Frenk}, \& {White}}]{1997ApJ...490..493N}
{Navarro}, J.~F., {Frenk}, C.~S., \& {White}, S. D.~M. 1997, \apj, 490, 493

\bibitem[{{Nebrin} {et~al.}(2023){Nebrin}, {Giri}, \& {Mellema}}]{2023MNRAS.524.2290N}
{Nebrin}, O., {Giri}, S.~K., \& {Mellema}, G. 2023, \mnras, 524, 2290

\bibitem[{{Newton} {et~al.}(2018){Newton}, {Cautun}, {Jenkins}, {Frenk}, \& {Helly}}]{2018MNRAS.479.2853N}
{Newton}, O., {Cautun}, M., {Jenkins}, A., {Frenk}, C.~S., \& {Helly}, J.~C. 2018, \mnras, 479, 2853

\bibitem[{{Newton} {et~al.}(2021){Newton}, {Leo}, {Cautun}, {Jenkins}, {Frenk}, {Lovell}, {Helly}, {Benson}, \& {Cole}}]{2021JCAP...08..062N}
{Newton}, O., {Leo}, M., {Cautun}, M., {et~al.} 2021, \jcap, 2021, 062

\bibitem[{{Pace} \& {Li}(2019)}]{2019ApJ...875...77P}
{Pace}, A.~B., \& {Li}, T.~S. 2019, \apj, 875, 77

\bibitem[{Pace \& Strigari(2019)}]{10.1093/mnras/sty2839}
Pace, A.~B., \& Strigari, L.~E. 2019, Monthly Notices of the Royal Astronomical Society, 482, 3480.
\newblock \url{https://doi.org/10.1093/mnras/sty2839}

\bibitem[{{Pace} {et~al.}(2020){Pace}, {Kaplinghat}, {Kirby}, {Simon}, {Tollerud}, {Mu{\~n}oz}, {C{\^o}t{\'e}}, {Djorgovski}, \& {Geha}}]{2020MNRAS.495.3022P}
{Pace}, A.~B., {Kaplinghat}, M., {Kirby}, E., {et~al.} 2020, \mnras, 495, 3022

\bibitem[{{Pallanca} {et~al.}(2023){Pallanca}, {Leanza}, {Ferraro}, {Lanzoni}, {Dalessandro}, {Cadelano}, {Vesperini}, {Origlia}, {Mucciarelli}, {Valenti}, \& {Miola}}]{2023ApJ...950..138P}
{Pallanca}, C., {Leanza}, S., {Ferraro}, F.~R., {et~al.} 2023, \apj, 950, 138

\bibitem[{{Pe{\~n}arrubia} {et~al.}(2008){Pe{\~n}arrubia}, {Navarro}, \& {McConnachie}}]{2008ApJ...673..226P}
{Pe{\~n}arrubia}, J., {Navarro}, J.~F., \& {McConnachie}, A.~W. 2008, \apj, 673, 226

\bibitem[{{Pianta} {et~al.}(2022){Pianta}, {Capuzzo-Dolcetta}, \& {Carraro}}]{2022ApJ...939....3P}
{Pianta}, C., {Capuzzo-Dolcetta}, R., \& {Carraro}, G. 2022, \apj, 939, 3

\bibitem[{{Placco} {et~al.}(2014){Placco}, {Frebel}, {Beers}, \& {Stancliffe}}]{2014ApJ...797...21P}
{Placco}, V.~M., {Frebel}, A., {Beers}, T.~C., \& {Stancliffe}, R.~J. 2014, \apj, 797, 21

\bibitem[{{Plummer}(1911)}]{1911MNRAS..71..460P}
{Plummer}, H.~C. 1911, \mnras, 71, 460

\bibitem[{{Press} \& {Schechter}(1974)}]{1974ApJ...187..425P}
{Press}, W.~H., \& {Schechter}, P. 1974, \apj, 187, 425

\bibitem[{{Prochaska} {et~al.}(2020){Prochaska}, {Hennawi}, {Westfall}, {Cooke}, {Wang}, {Hsyu}, {Davies}, {Farina}, \& {Pelliccia}}]{2020JOSS....5.2308P}
{Prochaska}, J., {Hennawi}, J., {Westfall}, K., {et~al.} 2020, The Journal of Open Source Software, 5, 2308

\bibitem[{{Richstein} {et~al.}(2022){Richstein}, {Patel}, {Kallivayalil}, {Simon}, {Zivick}, {Tollerud}, {Fritz}, {Warfield}, {Besla}, {van der Marel}, {Wetzel}, {Choi}, {Deason}, {Geha}, {Guhathakurta}, {Jeon}, {Kirby}, {Libralato}, {Sacchi}, \& {Sohn}}]{2022ApJ...933..217R}
{Richstein}, H., {Patel}, E., {Kallivayalil}, N., {et~al.} 2022, \apj, 933, 217

\bibitem[{{Richstein} {et~al.}(2024){Richstein}, {Kallivayalil}, {Simon}, {Garling}, {Wetzel}, {Warfield}, {van der Marel}, {Jeon}, {Rose}, {Torrey}, {Engelhardt}, {Besla}, {Choi}, {Geha}, {Guhathakurta}, {Kirby}, {Patel}, {Sacchi}, \& {Sohn}}]{2024ApJ...967...72R}
{Richstein}, H., {Kallivayalil}, N., {Simon}, J.~D., {et~al.} 2024, \apj, 967, 72

\bibitem[{{Ricotti} {et~al.}(2016){Ricotti}, {Parry}, \& {Gnedin}}]{2016ApJ...831..204R}
{Ricotti}, M., {Parry}, O.~H., \& {Gnedin}, N.~Y. 2016, \apj, 831, 204

\bibitem[{{Rosenberg} {et~al.}(1998){Rosenberg}, {Saviane}, {Piotto}, {Aparicio}, \& {Zaggia}}]{1998AJ....115..648R}
{Rosenberg}, A., {Saviane}, I., {Piotto}, G., {Aparicio}, A., \& {Zaggia}, S.~R. 1998, \aj, 115, 648

\bibitem[{{Rybizki} {et~al.}(2022){Rybizki}, {Green}, {Rix}, {El-Badry}, {Demleitner}, {Zari}, {Udalski}, {Smart}, \& {Gould}}]{2022MNRAS.510.2597R}
{Rybizki}, J., {Green}, G.~M., {Rix}, H.-W., {et~al.} 2022, \mnras, 510, 2597

\bibitem[{{Sales} {et~al.}(2022){Sales}, {Wetzel}, \& {Fattahi}}]{2022NatAs...6..897S}
{Sales}, L.~V., {Wetzel}, A., \& {Fattahi}, A. 2022, Nature Astronomy, 6, 897

\bibitem[{{Sand} {et~al.}(2022){Sand}, {Mutlu-Pakdil}, {Jones}, {Karunakaran}, {Wang}, {Yang}, {Chiti}, {Bennet}, {Crnojevi{\'c}}, \& {Spekkens}}]{2022ApJ...935L..17S}
{Sand}, D.~J., {Mutlu-Pakdil}, B., {Jones}, M.~G., {et~al.} 2022, \apjl, 935, L17

\bibitem[{{Schlafly} \& {Finkbeiner}(2011)}]{2011ApJ...737..103S}
{Schlafly}, E.~F., \& {Finkbeiner}, D.~P. 2011, \apj, 737, 103

\bibitem[{{Schlegel} {et~al.}(1998){Schlegel}, {Finkbeiner}, \& {Davis}}]{Schlegel:1998}
{Schlegel}, D.~J., {Finkbeiner}, D.~P., \& {Davis}, M. 1998, \apj, 500, 525

\bibitem[{{Sch{\"o}nrich} {et~al.}(2010){Sch{\"o}nrich}, {Binney}, \& {Dehnen}}]{2010MNRAS.403.1829S}
{Sch{\"o}nrich}, R., {Binney}, J., \& {Dehnen}, W. 2010, \mnras, 403, 1829

\bibitem[{{Sevilla-Noarbe} {et~al.}(2021){Sevilla-Noarbe}, {Bechtol}, {Carrasco Kind}, {Carnero Rosell}, {Becker}, {Drlica-Wagner}, {Gruendl}, {Rykoff}, {Sheldon}, {Yanny}, {Alarcon}, {Allam}, {Amon}, {Benoit-L{\'e}vy}, {Bernstein}, {Bertin}, {Burke}, {Carretero}, {Choi}, {Diehl}, {Everett}, {Flaugher}, {Gaztanaga}, {Gschwend}, {Harrison}, {Hartley}, {Hoyle}, {Jarvis}, {Johnson}, {Kessler}, {Kron}, {Kuropatkin}, {Leistedt}, {Li}, {Menanteau}, {Morganson}, {Ogando}, {Palmese}, {Paz-Chinch{\'o}n}, {Pieres}, {Pond}, {Rodriguez-Monroy}, {Smith}, {Stringer}, {Troxel}, {Tucker}, {de Vicente}, {Wester}, {Zhang}, {Abbott}, {Aguena}, {Annis}, {Avila}, {Bhargava}, {Bridle}, {Brooks}, {Brout}, {Castander}, {Cawthon}, {Chang}, {Conselice}, {Costanzi}, {Crocce}, {da Costa}, {Pereira}, {Davis}, {Desai}, {Dietrich}, {Doel}, {Eckert}, {Evrard}, {Ferrero}, {Fosalba}, {Garc{\'\i}a-Bellido}, {Gerdes}, {Giannantonio}, {Gruen}, {Gutierrez}, {Hinton}, {Hollowood}, {Honscheid}, {Huff}, {Huterer}, {James}, {Jeltema}, {Kuehn},
  {Lahav}, {Lidman}, {Lima}, {Lin}, {Maia}, {Marshall}, {Martini}, {Melchior}, {Miquel}, {Mohr}, {Morgan}, {Neilsen}, {Plazas}, {Romer}, {Roodman}, {Sanchez}, {Scarpine}, {Schubnell}, {Serrano}, {Smith}, {Suchyta}, {Tarle}, {Thomas}, {To}, {Varga}, {Wechsler}, {Weller}, {Wilkinson}, \& {DES Collaboration}}]{2021ApJS..254...24S}
{Sevilla-Noarbe}, I., {Bechtol}, K., {Carrasco Kind}, M., {et~al.} 2021, \apjs, 254, 24

\bibitem[{{Simon} {et~al.}(2019){Simon}, {Bechtol}, {Drlica-Wagner}, {Geha}, {Gluscevic}, {Ji}, {Kirby}, {Li}, {Nadler}, {Pace}, {Peter}, \& {Wechsler}}]{2019BAAS...51c.409S}
{Simon}, J., {Bechtol}, K., {Drlica-Wagner}, A., {et~al.} 2019, \baas, 51, 409

\bibitem[{{Simon}(2018)}]{2018ApJ...863...89S}
{Simon}, J.~D. 2018, \apj, 863, 89

\bibitem[{{Simon}(2019)}]{2019ARA&A..57..375S}
---. 2019, \araa, 57, 375

\bibitem[{{Simon} \& {Geha}(2007)}]{2007ApJ...670..313S}
{Simon}, J.~D., \& {Geha}, M. 2007, \apj, 670, 313

\bibitem[{{Simon} {et~al.}(2011){Simon}, {Geha}, {Minor}, {Martinez}, {Kirby}, {Bullock}, {Kaplinghat}, {Strigari}, {Willman}, {Choi}, {Tollerud}, \& {Wolf}}]{2011ApJ...733...46S}
{Simon}, J.~D., {Geha}, M., {Minor}, Q.~E., {et~al.} 2011, \apj, 733, 46

\bibitem[{{Simon} {et~al.}(2015){Simon}, {Drlica-Wagner}, {Li}, {Nord}, {Geha}, {Bechtol}, {Balbinot}, {Buckley-Geer}, {Lin}, {Marshall}, {Santiago}, {Strigari}, {Wang}, {Wechsler}, {Yanny}, {Abbott}, {Bauer}, {Bernstein}, {Bertin}, {Brooks}, {Burke}, {Capozzi}, {Carnero Rosell}, {Carrasco Kind}, {D'Andrea}, {da Costa}, {DePoy}, {Desai}, {Diehl}, {Dodelson}, {Cunha}, {Estrada}, {Evrard}, {Fausti Neto}, {Fernandez}, {Finley}, {Flaugher}, {Frieman}, {Gaztanaga}, {Gerdes}, {Gruen}, {Gruendl}, {Honscheid}, {James}, {Kent}, {Kuehn}, {Kuropatkin}, {Lahav}, {Maia}, {March}, {Martini}, {Miller}, {Miquel}, {Ogando}, {Romer}, {Roodman}, {Rykoff}, {Sako}, {Sanchez}, {Schubnell}, {Sevilla}, {Smith}, {Soares-Santos}, {Sobreira}, {Suchyta}, {Swanson}, {Tarle}, {Thaler}, {Tucker}, {Vikram}, {Walker}, {Wester}, \& {DES Collaboration}}]{2015ApJ...808...95S}
{Simon}, J.~D., {Drlica-Wagner}, A., {Li}, T.~S., {et~al.} 2015, \apj, 808, 95

\bibitem[{{Simon} {et~al.}(2017){Simon}, {Li}, {Drlica-Wagner}, {Bechtol}, {Marshall}, {James}, {Wang}, {Strigari}, {Balbinot}, {Kuehn}, {Walker}, {Abbott}, {Allam}, {Annis}, {Benoit-L{\'e}vy}, {Brooks}, {Buckley-Geer}, {Burke}, {Carnero Rosell}, {Carrasco Kind}, {Carretero}, {Cunha}, {D'Andrea}, {da Costa}, {DePoy}, {Desai}, {Doel}, {Fernandez}, {Flaugher}, {Frieman}, {Garc{\'\i}a-Bellido}, {Gaztanaga}, {Goldstein}, {Gruen}, {Gutierrez}, {Kuropatkin}, {Maia}, {Martini}, {Menanteau}, {Miller}, {Miquel}, {Neilsen}, {Nord}, {Ogando}, {Plazas}, {Romer}, {Rykoff}, {Sanchez}, {Santiago}, {Scarpine}, {Schubnell}, {Sevilla-Noarbe}, {Smith}, {Sobreira}, {Suchyta}, {Swanson}, {Tarle}, {Whiteway}, {Yanny}, \& {DES Collaboration}}]{2017ApJ...838...11S}
{Simon}, J.~D., {Li}, T.~S., {Drlica-Wagner}, A., {et~al.} 2017, \apj, 838, 11

\bibitem[{{Simon} {et~al.}(2020){Simon}, {Li}, {Erkal}, {Pace}, {Drlica-Wagner}, {James}, {Marshall}, {Bechtol}, {Hansen}, {Kuehn}, {Lidman}, {Allam}, {Annis}, {Avila}, {Bertin}, {Brooks}, {Burke}, {Rosell}, {Carrasco Kind}, {Carretero}, {da Costa}, {De Vicente}, {Desai}, {Doel}, {Eifler}, {Everett}, {Fosalba}, {Frieman}, {Garc{\'\i}a-Bellido}, {Gaztanaga}, {Gerdes}, {Gruen}, {Gruendl}, {Gschwend}, {Gutierrez}, {Hollowood}, {Honscheid}, {Krause}, {Kuropatkin}, {MacCrann}, {Maia}, {March}, {Miquel}, {Palmese}, {Paz-Chinch{\'o}n}, {Plazas}, {Reil}, {Roodman}, {Sanchez}, {Santiago}, {Scarpine}, {Schubnell}, {Serrano}, {Smith}, {Suchyta}, {Tarle}, {Walker}, \& {DES Collaboration}}]{2020ApJ...892..137S}
{Simon}, J.~D., {Li}, T.~S., {Erkal}, D., {et~al.} 2020, \apj, 892, 137

\bibitem[{{Smith} {et~al.}(2023{\natexlab{a}}){Smith}, {Jensen}, {Roediger}, {Sestito}, {Hayes}, {McConnachie}, {Cuillandre}, {Gwyn}, {Magnier}, {Chambers}, {Hammer}, {Hudson}, {Martin}, {Navarro}, \& {Scott}}]{2023AJ....166...76S}
{Smith}, S. E.~T., {Jensen}, J., {Roediger}, J., {et~al.} 2023{\natexlab{a}}, \aj, 166, 76

\bibitem[{{Smith} {et~al.}(2023{\natexlab{b}}){Smith}, {Cerny}, {Hayes}, {Sestito}, {Jensen}, {McConnachie}, {Geha}, {Navarro}, {Li}, {Cuillandre}, {Errani}, {Chambers}, {Gwyn}, {Hammer}, {Hudson}, {Magnier}, \& {Martin}}]{2023arXiv231110147S}
{Smith}, S. E.~T., {Cerny}, W., {Hayes}, C.~R., {et~al.} 2023{\natexlab{b}}, arXiv e-prints, arXiv:2311.10147

\bibitem[{{Smith} {et~al.}(2024){Smith}, {Cerny}, {Hayes}, {Sestito}, {Jensen}, {McConnachie}, {Geha}, {Navarro}, {Li}, {Cuillandre}, {Errani}, {Chambers}, {Gwyn}, {Hammer}, {Hudson}, {Magnier}, \& {Martin}}]{2024ApJ...961...92S}
---. 2024, \apj, 961, 92

\bibitem[{{Sohn} {et~al.}(2007){Sohn}, {Majewski}, {Mu{\~n}oz}, {Kunkel}, {Johnston}, {Ostheimer}, {Guhathakurta}, {Patterson}, {Siegel}, \& {Cooper}}]{2007ApJ...663..960S}
{Sohn}, S.~T., {Majewski}, S.~R., {Mu{\~n}oz}, R.~R., {et~al.} 2007, \apj, 663, 960

\bibitem[{{Somerville}(2002)}]{2002ApJ...572L..23S}
{Somerville}, R.~S. 2002, \apjl, 572, L23

\bibitem[{{Spencer} {et~al.}(2018){Spencer}, {Mateo}, {Olszewski}, {Walker}, {McConnachie}, \& {Kirby}}]{2018AJ....156..257S}
{Spencer}, M.~E., {Mateo}, M., {Olszewski}, E.~W., {et~al.} 2018, \aj, 156, 257

\bibitem[{{Spencer} {et~al.}(2017){Spencer}, {Mateo}, {Walker}, \& {Olszewski}}]{2017ApJ...836..202S}
{Spencer}, M.~E., {Mateo}, M., {Walker}, M.~G., \& {Olszewski}, E.~W. 2017, \apj, 836, 202

\bibitem[{{Springel} {et~al.}(2008){Springel}, {Wang}, {Vogelsberger}, {Ludlow}, {Jenkins}, {Helmi}, {Navarro}, {Frenk}, \& {White}}]{2008MNRAS.391.1685S}
{Springel}, V., {Wang}, J., {Vogelsberger}, M., {et~al.} 2008, \mnras, 391, 1685

\bibitem[{{Strader} \& {Kobulnicky}(2008)}]{2008AJ....136.2102S}
{Strader}, J., \& {Kobulnicky}, H.~A. 2008, \aj, 136, 2102

\bibitem[{{Tan} {et~al.}(2024){Tan}, {Cerny}, {Drlica-Wagner}, {Pace}, {Geha}, {Ji}, {Li}, {Adam{\'o}w}, {Anbajagane}, {Bom}, {Carballo-Bello}, {Carlin}, {Chang}, {Choi}, {Collins}, {Doliva-Dolinsky}, {Ferguson}, {Gruendl}, {James}, {Limberg}, {Navabi}, {Mart{\'\i}nez-Delgado}, {Mart{\'\i}nez-V{\'a}zquez}, {Medina}, {Mutlu-Pakdil}, {Nidever}, {No{\"e}l}, {Riley}, {Sakowska}, {Sand}, {Sharp}, {Stringfellow}, {Tolley}, \& {Vivas}}]{2024arXiv240800865T}
{Tan}, C.~Y., {Cerny}, W., {Drlica-Wagner}, A., {et~al.} 2024, arXiv e-prints, arXiv:2408.00865

\bibitem[{{Ting} {et~al.}(2019){Ting}, {Conroy}, {Rix}, \& {Cargile}}]{2019ApJ...879...69T}
{Ting}, Y.-S., {Conroy}, C., {Rix}, H.-W., \& {Cargile}, P. 2019, \apj, 879, 69

\bibitem[{{Torrealba} {et~al.}(2019){Torrealba}, {Belokurov}, \& {Koposov}}]{2019MNRAS.484.2181T}
{Torrealba}, G., {Belokurov}, V., \& {Koposov}, S.~E. 2019, \mnras, 484, 2181

\bibitem[{{Torrealba} {et~al.}(2016{\natexlab{a}}){Torrealba}, {Koposov}, {Belokurov}, \& {Irwin}}]{2016MNRAS.459.2370T}
{Torrealba}, G., {Koposov}, S.~E., {Belokurov}, V., \& {Irwin}, M. 2016{\natexlab{a}}, \mnras, 459, 2370

\bibitem[{{Torrealba} {et~al.}(2016{\natexlab{b}}){Torrealba}, {Koposov}, {Belokurov}, {Irwin}, {Collins}, {Spencer}, {Ibata}, {Mateo}, {Bonaca}, \& {Jethwa}}]{2016MNRAS.463..712T}
{Torrealba}, G., {Koposov}, S.~E., {Belokurov}, V., {et~al.} 2016{\natexlab{b}}, \mnras, 463, 712

\bibitem[{{Torrealba} {et~al.}(2018){Torrealba}, {Belokurov}, {Koposov}, {Bechtol}, {Drlica-Wagner}, {Olsen}, {Vivas}, {Yanny}, {Jethwa}, {Walker}, {Li}, {Allam}, {Conn}, {Gallart}, {Gruendl}, {James}, {Johnson}, {Kuehn}, {Kuropatkin}, {Martin}, {Martinez-Delgado}, {Nidever}, {No{\"e}l}, {Simon}, {Stringfellow}, \& {Tucker}}]{2018MNRAS.475.5085T}
{Torrealba}, G., {Belokurov}, V., {Koposov}, S.~E., {et~al.} 2018, \mnras, 475, 5085

\bibitem[{{van der Walt} {et~al.}(2011){van der Walt}, {Colbert}, \& {Varoquaux}}]{2011CSE....13b..22V}
{van der Walt}, S., {Colbert}, S.~C., \& {Varoquaux}, G. 2011, Computing in Science and Engineering, 13, 22

\bibitem[{Virtanen {et~al.}(2020)Virtanen, Gommers, Oliphant, Haberland, Reddy, Cournapeau, Burovski, Peterson, Weckesser, Bright, {van der Walt}, Brett, Wilson, Millman, Mayorov, Nelson, Jones, Kern, Larson, Carey, Polat, Feng, Moore, {VanderPlas}, Laxalde, Perktold, Cimrman, Henriksen, Quintero, Harris, Archibald, Ribeiro, Pedregosa, {van Mulbregt}, \& {SciPy 1.0 Contributors}}]{2020SciPy-NMeth}
Virtanen, P., Gommers, R., Oliphant, T.~E., {et~al.} 2020, Nature Methods, 17, 261

\bibitem[{{Virtanen} {et~al.}(2020){Virtanen}, {Gommers}, {Oliphant}, {Haberland}, {Reddy}, {Cournapeau}, {Burovski}, {Peterson}, {Weckesser}, {Bright}, {van der Walt}, {Brett}, {Wilson}, {Millman}, {Mayorov}, {Nelson}, {Jones}, {Kern}, {Larson}, {Carey}, {Polat}, {Feng}, {Moore}, {VanderPlas}, {Laxalde}, {Perktold}, {Cimrman}, {Henriksen}, {Quintero}, {Harris}, {Archibald}, {Ribeiro}, {Pedregosa}, {van Mulbregt}, \& {SciPy 1. 0 Contributors}}]{2020NatMe..17..261V}
{Virtanen}, P., {Gommers}, R., {Oliphant}, T.~E., {et~al.} 2020, Nature Methods, 17, 261

\bibitem[{{Walker} {et~al.}(2009){Walker}, {Mateo}, \& {Olszewski}}]{2009AJ....137.3100W}
{Walker}, M.~G., {Mateo}, M., \& {Olszewski}, E.~W. 2009, \aj, 137, 3100

\bibitem[{{Walker} {et~al.}(2015{\natexlab{a}}){Walker}, {Mateo}, {Olszewski}, {Bailey}, {Koposov}, {Belokurov}, \& {Evans}}]{2015ApJ...808..108W}
{Walker}, M.~G., {Mateo}, M., {Olszewski}, E.~W., {et~al.} 2015{\natexlab{a}}, \apj, 808, 108

\bibitem[{{Walker} {et~al.}(2006){Walker}, {Mateo}, {Olszewski}, {Bernstein}, {Wang}, \& {Woodroofe}}]{2006AJ....131.2114W}
---. 2006, \aj, 131, 2114

\bibitem[{{Walker} {et~al.}(2015{\natexlab{b}}){Walker}, {Olszewski}, \& {Mateo}}]{2015MNRAS.448.2717W}
{Walker}, M.~G., {Olszewski}, E.~W., \& {Mateo}, M. 2015{\natexlab{b}}, \mnras, 448, 2717

\bibitem[{{Wang} {et~al.}(2019){Wang}, {de Boer}, {Pieres}, {Li}, {Drlica-Wagner}, {Koposov}, {Vivas}, {Pace}, {Santiago}, {Walker}, {Tucker}, {Strigari}, {Marshall}, {Yanny}, {DePoy}, {Bechtol}, {Roodman}, {Abbott}, {Abdalla}, {Allam}, {Annis}, {Avila}, {Bertin}, {Brooks}, {Burke}, {Carnero Rosell}, {Carrasco Kind}, {Cunha}, {D'Andrea}, {da Costa}, {De Vicente}, {Desai}, {Eifler}, {Estrada}, {Flaugher}, {Frieman}, {Garc{\'\i}a-Bellido}, {Gerdes}, {Gruen}, {Gruendl}, {Gutierrez}, {Hollowood}, {Honscheid}, {James}, {Kuehn}, {Kuropatkin}, {Lahav}, {Maia}, {Miquel}, {Sanchez}, {Scarpine}, {Sevilla-Noarbe}, {Smith}, {Smith}, {Sobreira}, {Suchyta}, {Swanson}, {Tarle}, \& {DES Collaboration}}]{2019ApJ...881..118W}
{Wang}, M.~Y., {de Boer}, T., {Pieres}, A., {et~al.} 2019, \apj, 881, 118

\bibitem[{{Wang} {et~al.}(2023){Wang}, {Zhu}, {Jing}, {Grand}, {Li}, {Fu}, {Li}, {Han}, {Li}, {Feng}, \& {Frenk}}]{2023ApJ...956...91W}
{Wang}, W., {Zhu}, L., {Jing}, Y., {et~al.} 2023, \apj, 956, 91

\bibitem[{{Weerasooriya} {et~al.}(2023){Weerasooriya}, {Bovill}, {Benson}, {Musick}, \& {Ricotti}}]{2023ApJ...948...87W}
{Weerasooriya}, S., {Bovill}, M.~S., {Benson}, A., {Musick}, A.~M., \& {Ricotti}, M. 2023, \apj, 948, 87

\bibitem[{{Weisz} {et~al.}(2016){Weisz}, {Koposov}, {Dolphin}, {Belokurov}, {Gieles}, {Mateo}, {Olszewski}, {Sills}, \& {Walker}}]{2016ApJ...822...32W}
{Weisz}, D.~R., {Koposov}, S.~E., {Dolphin}, A.~E., {et~al.} 2016, \apj, 822, 32

\bibitem[{{Wetzel} {et~al.}(2015){Wetzel}, {Deason}, \& {Garrison-Kimmel}}]{2015ApJ...807...49W}
{Wetzel}, A.~R., {Deason}, A.~J., \& {Garrison-Kimmel}, S. 2015, \apj, 807, 49

\bibitem[{{Wetzel} {et~al.}(2013){Wetzel}, {Tinker}, {Conroy}, \& {van den Bosch}}]{2013MNRAS.432..336W}
{Wetzel}, A.~R., {Tinker}, J.~L., {Conroy}, C., \& {van den Bosch}, F.~C. 2013, \mnras, 432, 336

\bibitem[{{Wheeler} {et~al.}(2023){Wheeler}, {Abruzzo}, {Casey}, \& {Ness}}]{2023AJ....165...68W}
{Wheeler}, A.~J., {Abruzzo}, M.~W., {Casey}, A.~R., \& {Ness}, M.~K. 2023, \aj, 165, 68

\bibitem[{{Wheeler} {et~al.}(2024){Wheeler}, {Casey}, \& {Abruzzo}}]{2024AJ....167...83W}
{Wheeler}, A.~J., {Casey}, A.~R., \& {Abruzzo}, M.~W. 2024, \aj, 167, 83

\bibitem[{{Wheeler} {et~al.}(2015){Wheeler}, {O{\~n}orbe}, {Bullock}, {Boylan-Kolchin}, {Elbert}, {Garrison-Kimmel}, {Hopkins}, \& {Kere{\v{s}}}}]{2015MNRAS.453.1305W}
{Wheeler}, C., {O{\~n}orbe}, J., {Bullock}, J.~S., {et~al.} 2015, \mnras, 453, 1305

\bibitem[{{White} \& {Frenk}(1991)}]{1991ApJ...379...52W}
{White}, S. D.~M., \& {Frenk}, C.~S. 1991, \apj, 379, 52

\bibitem[{{White} \& {Rees}(1978)}]{1978MNRAS.183..341W}
{White}, S.~D.~M., \& {Rees}, M.~J. 1978, \mnras, 183, 341

\bibitem[{{Willman} {et~al.}(2011){Willman}, {Geha}, {Strader}, {Strigari}, {Simon}, {Kirby}, {Ho}, \& {Warres}}]{2011AJ....142..128W}
{Willman}, B., {Geha}, M., {Strader}, J., {et~al.} 2011, \aj, 142, 128

\bibitem[{{Willman} \& {Strader}(2012)}]{2012AJ....144...76W}
{Willman}, B., \& {Strader}, J. 2012, \aj, 144, 76

\bibitem[{{Willman} {et~al.}(2005{\natexlab{a}}){Willman}, {Blanton}, {West}, {Dalcanton}, {Hogg}, {Schneider}, {Wherry}, {Yanny}, \& {Brinkmann}}]{2005AJ....129.2692W}
{Willman}, B., {Blanton}, M.~R., {West}, A.~A., {et~al.} 2005{\natexlab{a}}, \aj, 129, 2692

\bibitem[{{Willman} {et~al.}(2005{\natexlab{b}}){Willman}, {Dalcanton}, {Martinez-Delgado}, {West}, {Blanton}, {Hogg}, {Barentine}, {Brewington}, {Harvanek}, {Kleinman}, {Krzesinski}, {Long}, {Neilsen}, {Nitta}, \& {Snedden}}]{2005ApJ...626L..85W}
{Willman}, B., {Dalcanton}, J.~J., {Martinez-Delgado}, D., {et~al.} 2005{\natexlab{b}}, \apjl, 626, L85

\bibitem[{{Wolf} {et~al.}(2010){Wolf}, {Martinez}, {Bullock}, {Kaplinghat}, {Geha}, {Mu{\~n}oz}, {Simon}, \& {Avedo}}]{2010MNRAS.406.1220W}
{Wolf}, J., {Martinez}, G.~D., {Bullock}, J.~S., {et~al.} 2010, \mnras, 406, 1220

\bibitem[{{Zonca} {et~al.}(2019){Zonca}, {Singer}, {Lenz}, {Reinecke}, {Rosset}, {Hivon}, \& {Gorski}}]{2019JOSS....4.1298Z}
{Zonca}, A., {Singer}, L., {Lenz}, D., {et~al.} 2019, The Journal of Open Source Software, 4, 1298

\bibitem[{{Zucker} {et~al.}(2006){Zucker}, {Belokurov}, {Evans}, {Kleyna}, {Irwin}, {Wilkinson}, {Fellhauer}, {Bramich}, {Gilmore}, {Newberg}, {Yanny}, {Smith}, {Hewett}, {Bell}, {Rix}, {Gnedin}, {Vidrih}, {Wyse}, {Willman}, {Grebel}, {Schneider}, {Beers}, {Kniazev}, {Barentine}, {Brewington}, {Brinkmann}, {Harvanek}, {Kleinman}, {Krzesinski}, {Long}, {Nitta}, \& {Snedden}}]{2006ApJ...650L..41Z}
{Zucker}, D.~B., {Belokurov}, V., {Evans}, N.~W., {et~al.} 2006, \apjl, 650, L41

\end{thebibliography}
\appendix

\section{References for Data in Figure 5}
\label{sec:references}
\figref{pop} presents a comparison between Aquarius~III's properties and those of the Milky Way's globular clusters, satellite dwarf galaxies, and ultra-faint compact satellites (which may include both star clusters and dwarf galaxies). The measurements in this figure were adopted from the Local Volume Database (A. Pace et al., in prep). The underlying individual references are as follows:
\par The dwarf galaxy measurements reported in the Local Volume Database version used here
were compiled from individual studies including \citet{2007ApJ...670..313S}, \citet{2008ApJ...675..201M},  \citet{2009MNRAS.397L..26C}, \citet{2009AJ....137.3100W}, \citet{2009ApJ...702L...9C}, \citet{2011AJ....142..128W}, \citet{2011ApJ...733...46S}, \citet{2011ApJ...736..146K},  \citet{2013ApJ...770...16K}, \citet{2015ApJ...810...56K},  \citet{2015ApJ...808...95S}, \citet{2015ApJ...813..109D}, \citet{2015MNRAS.448.2717W}, \citet{2015ApJ...808..108W}, 
\citet{2015ApJ...805..130K}, \citet{2015ApJ...811...62K}, \citet{2016ApJ...824L..14C}, \citet{2016ApJ...833L...5D}, \citet{2016ApJ...833...16K}, \citet{2016MNRAS.463..712T},   \citet{2016MNRAS.459.2370T}, \citet{2017ApJ...838...83K},   \citet{2017ApJ...836..202S}, \citet{2017ApJ...838...11S}, \citet{2017ApJ...838....8L}, \citet{2018ApJ...869..125C}, \citet{2018PASJ...70S..18H}, \citet{2018MNRAS.480.2609L},  \citet{2018ApJ...857..145L}, \citet{2018MNRAS.479.5343K},  \citet{2018AJ....156..257S}, \citet{2018MNRAS.475.5085T},  \citet{2018ApJ...863...25M}, \citet{2018ApJ...860...66M}, \citet{2019PASJ...71...94H}, \citet{2019A&A...623A.129F}, \citet{2019ARA&A..57..375S}, \citet{2019ApJ...881..118W}, \citet{2020MNRAS.495.3022P}, \citet{2020ApJ...892...27M},  \citet{2020ApJ...890..136M}, \citet{2020ApJ...892..137S}, \citet{2021ApJ...921...32J}, \citet{2021NatAs...5..392C}, \citet{2021ApJ...916...81C}, \citet{2021ApJ...910...18C}, \citet{2021ApJ...920...92J}, \citet{2021ApJ...920L..44C}, \citet{2022ApJ...939...41C}, \citet{2022ApJ...933..217R}, \citet{2023ApJ...942..111C}, \citet{2023AJ....166...76S}, \citet{2023arXiv231105439H}, \citet{2023ApJ...953....1C}, \citet{2023ApJ...950..167B}, \citet{2023AJ....165...55C}, \citet{2024ApJ...968...21H}, \citet{2024ApJ...961..234H}, \citet{2024ApJ...967...72R}.

\par Globular cluster measurements were primarily drawn from the compilations of \citet{2018MNRAS.478.1520B} for structural parameters and \citet{2020PASA...37...46B} for absolute magnitudes. Other measurements for select individual GCs were taken from \citet[][2010 edition]{1996AJ....112.1487H}, \citet{1998AJ....115..648R}, \citet{2005AJ....129..239K}, \citet{2008AJ....136.2102S}, \citet{2008A&A...489..583K}, \citet{2013AJ....146..116H}, \citet{2016ApJ...822...32W}, \citet{2021NatAs...5..957G}, \citet{2023ApJ...950..138P}, \citet{2024arXiv240513558L}, and \citet{2024ApJ...967...72R}.
\par Data for the ultra-faint compact satellites (roughly defined here as halo systems with $M_V \gtrsim -3.5$, $r_{1/2} < 15$~pc) were taken from \citet{2011AJ....142...88F}, \citet{2013ApJ...767..101B},  \citet{2015ApJ...799...73K}, \citet{2015ApJ...803...63K}, \citet{2016ApJ...820..119K}, \citet{2016ApJ...830L..10M}, \citet{2018MNRAS.478.2006L}, \citet{2018ApJ...852...68C}, \citet{2018ApJ...860...66M}, \citet{2019MNRAS.484.2181T}, \citet{2019PASJ...71...94H}, \citet{2019MNRAS.490.1498L}, \citet{2020ApJ...890..136M}, \citet{2021ApJ...910...18C}, \citet{2022ApJ...929L..21G}, \citet{2023ApJ...953....1C}, \citet{2023ApJ...953L..21C}, and \citet{2024ApJ...961...92S}.

\end{document}